\documentclass[acmsmall]{acmart}

\AtBeginDocument{%
  \providecommand\BibTeX{{%
    \normalfont B\kern-0.5em{\scshape i\kern-0.25em b}\kern-0.8em\TeX}}}

\setcopyright{acmcopyright}
\copyrightyear{2022}
\acmYear{2022}
\acmDOI{nn.nnnn/nnnnnnn.nnnnnnn}

\acmConference[CSCW]{}{2022}{Virtual}
\acmBooktitle{}

\usepackage{csquotes}
\usepackage{float}
\usepackage{subcaption}
\usepackage{soul}
\usepackage{url}
\usepackage{cleveref} 
\usepackage[labelfont=bf]{caption}
\usepackage{array}
\usepackage{multirow}
\usepackage{subcaption}    
\usepackage{enumitem}    
\usepackage{graphicx}
\usepackage{caption}
\usepackage{tabularx}
\definecolor{labelpurple}{HTML}{cfcfea}
\definecolor{labelpink}{HTML}{eadada}
\definecolor{labelblue}{HTML}{bfcce1}

\definecolor{labelgreen}{HTML}{bdd5d2}
\definecolor{labellightgrey}{HTML}{F0EFEF}
\definecolor{labelsky}{HTML}{DBE9F0}

\usepackage{blindtext,graphicx}
\usepackage[absolute]{textpos}
\begin{document}
\begin{textblock}{4}(0.5,1)
\noindent\textcolor{red} {To cite: Prerna Juneja and Tanushree Mitra. 2022. Human and technological infrastructures of fact-checking. In Proceedings of the 2022 Conference On Computer
Supported Cooperative Work And Social Computing (CSCW '22). Proceedings of the ACM on Human-Computer Interaction.}
\end{textblock}

\title{Human and technological infrastructures of fact-checking}




\author{Prerna Juneja}
\affiliation{%
 \institution{University of Washington}
 \city{Seattle}
 \state{Washington}
 \country{USA}}
\email{prerna79@uw.edu}

\author{Tanushree Mitra}
\affiliation{%
 \institution{University of Washington}
 \city{Seattle}
 \state{Washington}
 \country{USA}}
\email{tanmit@uw.edu}
\renewcommand{\shortauthors}{Anonymous, et al.}

\begin{abstract}

Increasing demands for fact-checking have led to a growing interest in developing systems and tools to automate the fact-checking process.  However,  such systems are limited in practice because their system design often does not take into account how fact-checking is done in the real world and ignores the insights and needs  of various stakeholder groups core to the fact-checking process. This paper  unpacks the fact-checking process by revealing the  infrastructures---both human and technological---that support  and shape fact-checking work. We interviewed {26} participants belonging to {16} fact-checking teams and organizations with representation from 4 continents. Through these interviews, 
we describe the human infrastructure of fact-checking by identifying and presenting, in-depth, the roles of six primary stakeholder groups,  1) Editors, 2) External fact-checkers, 3) In-house fact-checkers, 4) Investigators and researchers, 5) Social media managers, and 6) Advocators.  Our findings highlight that the fact-checking process is a collaborative effort among various stakeholder groups and associated technological and informational infrastructures. By rendering visibility to the infrastructures, we reveal how fact-checking has evolved to include both \textit{short-term claims centric} and \textit{long-term advocacy centric} fact-checking. 
Our work also identifies key social and technical needs and challenges faced by each  stakeholder group. Based on our findings, we suggest that improving the quality of fact-checking requires systematic changes in the civic,  informational, and technological contexts.

\end{abstract}




\maketitle

\section{Introduction}


\begin{small}
\begin{displayquote}
\enquote{[There is] potentially an issue with.. the academic community as well, at least around fact checking.. There is loads of research done but I'm not sure [how] it makes an impact on how we do fact checking and how we report fact checking.} - \emph{Fact-checking advocator from Full Fact} \cite{fullfact} 
\end{displayquote}
\end{small}

\begin{small}
\begin{displayquote}
\enquote{Duke reporters lab runs \emph{Tech \& Check alerts} \cite{TechChec3:online} I'd say maybe 10 times in the last few years has it helped us identify something to fact check. Lately I haven't found it quite as useful for alerting me to something that is worth seeing.} - \emph{Fact-checking editor from The Washington Post} \cite{wp}
\end{displayquote}
\end{small}

With the increase in scale and diffusion of online misinformation, efforts to develop scalable technological systems for fact-checking online information have also increased.
However, such systems are limited in practice {\mbox{\cite{graves2018understanding}}}, primarily for two reasons. First, their design is treated as a technical solution to what is often seen as a purely technological problem. But fact-checking is a complex socially-situated technical phenomena involving collaboration among multiple stakeholder groups at various stages of the process. Yet, current automated fact-checking systems rarely take into account the insights and needs of ``the human''---stakeholder groups who are central to this  process. Second, automated fact-checking systems are limited in their applicability. 
For instance, most systems are either restricted to verifying claims about very specific public statistics by matching them against official figures (e.g., unemployment rate, inflation rate, etc.) or they are limited to identifying simple declarative claims to debunk \cite{graves2018understanding}. 
Hence, automated fact-checking solutions fail to generalize to real-world fact-checking scenarios \cite{graves2018understanding}. In other words, the rigidity of a purely technical system lacks the social flexibility necessary to support an inherent socio-technical process. 

In the last decade, HCI and CSCW communities have developed a better understanding of the gap between the social and the technical \mbox{\cite{girardin2007towards,dwyer2007task,ackerman2000intellectual,tuffley2009mind}} and thus, are well positioned to develop an understanding of the socio-technical mechanisms underlying fact-checking.  Yet, till date, we know very little about how fact-checking is done in practice and what we could do to socially and technically  support the fact-checking process. In this paper, we elucidate how fact-checking is practiced by laying bare the human  and technological infrastructures that facilitate and shape the fact-checking process in a fact-checking team/organization. 
We attempt to foreground the social by revealing the  synergistic collaboration that occurs among human infrastructure {---various stakeholder groups that work together to accomplish fact-checking work}. We  provide visibility to   the stakeholder groups'  roles,   needs and activities, many of which often remain  invisible to the external world. We also highlight the technological  infrastructure {---the tools, technology, processes and policies---}that supports and enables the work of the stakeholder groups. The foregrounding of the infrastructures supporting the fact-checking  work helps us unravel the technical, policy and information barriers to fact-checking. Our hope is that by considering both the human and technological infrastructures underlying the fact-checking process, we might narrow the ``design-reality gap'' \mbox{\cite{heeks2003most}}---the gap that exists between the needs of stakeholder groups involved in fact-checking  and design of technical systems for fact-checking. Overall, in this work we answer the following research questions.

\begin{itemize}[leftmargin=*]
\item[] \indent \textbf{RQ1:} What are the various infrastructures supporting fact-checking work?

\begin{itemize}
\item[] \indent \textbf{RQ1a {Human infrastructure:}} Who are the various stakeholder groups involved in the fact-checking process? What roles do they play? How do they evaluate priorities and collaborate together to make decisions?
\item[] \indent \textbf{RQ2b Technological infrastructure:} How do tools,  technology and policy support stakeholder groups in performing their roles?
\end{itemize}

\item[] \indent \textbf{RQ2 {Barriers to fact-checking:}} What are the various {needs and challenges of stakeholder groups involved in the fact-checking process? }
\end{itemize}

To answer the research questions, we adopt a multi-stakeholder approach and perform semi-structured interviews with 26 participants belonging to 16 fact-checking organizations  or fact-checking teams within publication houses. We began our study by interviewing fact-checkers and editors---the stakeholder groups identified in previous works \mbox{\cite{doi:10.1080/17512786.2019.1643767,amazeen2013critical,graves2017anatomy}}. We discovered the existence of other stakeholder groups through these interviews  and expanded our recruitment by reaching out to them using convenience \cite{etikan2016comparison} and snowball sampling \cite{goodman1961snowball}.  Our participants had diverse representations from 4 continents---North America, Europe, Asia and Africa. We intentionally sampled participants widely across fact-checking teams, organizations, and countries 
to capture the practices and challenges emerging in this space. Our work  aims at 
uncovering the possible human and technological infrastructures supporting the fact-checking work in  teams/organizations across the regions instead of  capturing the  variability of fact-checking process across regions.

{Our findings reveal the existence of   six distinct stakeholder groups involved in the fact-checking process and the various roles performed by them.} The  identified stakeholder groups are: (1) \textit{Editors} who are responsible for overseeing the fact-checking process, including planning what topics to target and ensuring the integrity of the fact-checks produced, (2) \textit{External fact-checkers} who are responsible for monitoring the external world (social media platforms, presidential speeches, etc.), investigating dubious claims and writing fact-checks,  (3) \textit{In-house fact-checkers} who are responsible for fixing incorrect claims present in the news stories or articles produced internally in the media/news publication house, (4) \textit{Investigators and researchers} who conduct in-depth investigation  and data analysis of persistently circulating disinformation campaigns  (e.g. {investigating coordinated campaigns that used anti-Ruto hashtags\footnote{William Ruto is the current Deputy President of the Republic of Kenya. In May 2020, several anti-Ruto hashtags (e.g. \#RutoMustGo, \#RutoWantedToKillUhuru, etc.) began trending on Twitter in an attempt to discredit the Deputy President.} on Twitter to spread misinformation\footnote{https://investigate.africa/opt-report-post/}}), (5) \textit{Social media managers} who distribute fact-checks across multiple social media platforms and strategize on ways to increase audience engagement with the fact-checks, and (6) \textit{Advocators} who spearhead initiatives to improve policies around availability of information and statistics in their countries to improve the quality of fact-checking. 
 
By studying the roles performed by the stakeholder groups (\emph{human infrastructure}), we establish how fact-checking has evolved from a process to debunk individual pieces of misinformation (\textit{short-term claims centric} fact-checking) to a multi-step long-term campaign involving research, policy, and advocacy work (\textit{long-term advocacy centric} fact-checking). We find that   stakeholder groups mediate their roles via different tools (\emph{technological infrastructure}) ranging from third party social media monitoring tools (e.g. BuzzSumo \cite{BuzzSumo25:online}), public databases, process management tools (e.g. Trello \cite{Trello12:online}), color coding schemes, to training and educational workshops.  Our interviews reveal that fact-checkers are skeptical of using fully automated AI based tools. They desire algorithm explainability and involvement of humans in the decision making process  as key values in the systems they would use. We also identify several technical, policy and informational challenges. For example,  there are limited tools to monitor and flag content on private messaging platforms and investigate false claims in videos and content in local regional languages. We also find that in some countries, information to investigate claims from government and civic bodies is either unavailable,  difficult to obtain, or is not updated periodically.



Overall, our work makes the following contributions to the CSCW/HCI community:
\setlist{nolistsep}
\begin{itemize}

\item Our study identifies the human infrastructure supporting the fact-checking process. We present the six key stakeholder groups involved in fact-checking  and the roles performed by them. Through their roles, we show  the evolvement of fact-checking process from \textit{short-term claims centric} to  \textit{long-term advocacy centric} work.

\item We  reveal the technological infrastructure  that supports and
facilitates the role of each of the stakeholder groups identified in the study.

\item We offer a comprehensive  understanding of complexities and peculiarities of the current practices of fact-checking and 
provide a holistic view of the collaborative effort among various stakeholder groups, technological infrastructure and the social-informational context in which the fact-checking takes place.

\item We present the  technical, policy and informational  challenges that  stakeholder groups face while performing their roles and state the implications of our study on the design of new tools and systems for the fact-checking process. We also highlight how in addition to the need for developing new tools and  improving current platform design and affordances, there is an exigency to improve the availability and quality of data and statistics globally.
\end{itemize}

\section{Related Work}

We start by presenting the definition, origin and evolution of  fact-checking  (Section \mbox{\ref{definitions}}). Next, we situate our analysis of the fact-checking process with the existing literature on human-technological infrastructure (Section \mbox{\ref{human_and_tech}}) and invisible work (Section \mbox{\ref{invisible}}). Finally, we  present the current landscape of computational research in  fact-checking . We show how previous work engages with the shortcoming of the fact-checking practices and tools in a limited manner and describe how our work addresses this gap (Section \mbox{\ref{computational_research}}).




\subsection{Fact-checking: Definition, Origin, and Evolution}\label{definitions}
The American Press Institute defines the process of fact-checking as ``re-reporting and researching the purported facts in published/recorded statements made by politicians and anyone whose words impact others’ lives and livelihoods.'' \cite{americanpressinstitute}. One of the early examples of fact-checking emerging as an integral part of journalism was when the \textit{Time} magazine set up a separate research department to objectively verify every printed word before releasing the publication, a phenomenon now known as ante hoc, internal, or in-house fact-checking \cite{cjr}. The last decade also witnessed the emergence of post-hoc or external fact-checking which consists of publishing an evidence based analysis of claims made in any public text (e.g., news report, political speech, social media posts, etc.)  after it is released to the world \cite{graves2019fact}. Today, fact-checking has emerged both as a principal part of news reporting as well as a separate entity \cite{van2020factcorp}. According to Duke Reporters’ Lab, by 2019 there were around 188 active fact-checking initiatives  spread across 60 countries \cite{poynter1}. 
These initiatives have incorporated a range of methodologies and data-driven journalistic practices to not only hold disinformation spreading individuals and organizations accountable  through their fact-checks,  but to also disseminate fact-checks in such a way that increases engagement with the public \cite{doi:10.1080/21670811.2018.1493940,Ballotpedia}.
With the aim of bringing together these fact-checking initiatives and in order to promote common fact-checking standards through a code of principles, the International Fact-checking Network (IFCN) was established in 2015 \cite{poynter2}. Major social media companies such as Facebook and Google have since then partnered with IFCN signatories to debunk false claims surfacing on their platforms \cite{graves2020discipline}. While existing work describes the evolution of fact-checking from journalism to external fact-checking \cite{graves2013deciding,graves2012fact}, there are still gaps in understanding how fact-checking is actually practiced, the identity and role of the participating stakeholders and the various collaborations and partnerships occurring in the process.  Our work deep dives into answering these missing aspects of the fact-checking phenomenon.

\subsection{Human and Technological Infrastructures}\label{human_and_tech}

The foundational work  of studying infrastructure as a subject could be credited to Star and Ruhleder \mbox{\cite{star1996steps}}. The authors consider
infrastructure  as ``something that emerges for people in practice,
connected to activities and structures''
\mbox{\cite{star1996steps}}. Rather than being a thing to use, Star and Ruhleder refer to infrastructure as a relational concept \mbox{\cite{star1996steps}}. Scholars have since advocated for broadening the understanding of infrastructure  by also including social practices, processes and flow of information \mbox{\cite{sambasivan2010human}} and have called for investigating the  complexities and particularities
of infrastructures in practice \mbox{\cite{CASS2018160,verdezoto2021invisible,karasti2018studying}}. In response to this call, several research studies in Computer Supported Cooperative Work (CSCW) and
related fields, such as Human Computer Interaction (HCI) have  examined the  infrastructures---both human and technological---supporting the diverse socio-technical systems \mbox{\cite{dye2018paquete,jack2017infrastructure,nguyen2016infrastructural,lee2006human}} in various contexts such as, in health-care \mbox{\cite{pendse2020like,tang2015restructuring,ferguson2005impact}}, e-governance \mbox{\cite{chaudhuri2019paradoxes}}, crisis situations \mbox{\cite{mark2009repairing}}, etc. For our study, we first focus on human infrastructure which Lee et al   define  as ``organizations and actors that must be brought into alignment in order for work to be accomplished'' \mbox{\cite{lee2006human}}. Scholars have used this  concept of human infrastructure  to denote the human partnerships that are necessary for a successful socio-technical system \mbox{\cite{sambasivan2010human}}. 
Drawing on such scholarly work \mbox{\cite{dye2018paquete,lee2006human,sambasivan2010human}},  we use the analytical lens of human infrastructure to ``magnify the social'' by rendering visibility to the stakeholder groups who collaborate to enable the fact-checking work. Highlighting the human infrastructure also allows us to focus on how collaboration and coordination is accomplished in  socio-technical systems  \mbox{\cite{lee2006human}}. Sustaining online collaboration among various groups in an organization can be challenging \mbox{\cite{morgan2013project}}. Thus, within CSCW, a lot of attention has also been paid on examining collaborative \mbox{\cite{jirotka2006collaboration,lampinen2016cscw,lee2006human}} and coordination efforts \mbox{\cite{morgan2013project,lundberg1999understanding,grinter1995using,wasson1998identifying}} in  socio technical systems, determining ways to foster collaboration and coordination \mbox{\cite{gao2011promoting,rummel2002promoting,mota2011fostering}} and designing cooperative work tools \mbox{\cite{he2006method,HCARSTENSEN1995327,fish1988quilt,karp2017hapteq}}. We complement these prior studies on by establishing fact-checking as a distributed problem that requires collaboration and coordination of the human infrastructure supporting the fact-checking process.



{In a socio-technical system, human infrastructure does not exist in a vacuum \mbox{\cite{tang2015restructuring}}. It is intertwined with the technological infrastructure which is the software, hardware and processes supporting the human actors in performing their roles \mbox{\cite{tang2015restructuring,rasmussen2007human}}. Scholars argue that technology and human actors are mutually constituting; one mediates the other \mbox{\cite{tang2015restructuring,robinson2015examining}}. Our work also borrows the concept of technological infrastructure to  shed light on how use of various tools facilitates the enactment of the stakeholder groups' roles in the fact-checking process.}

\subsection{Invisible Work of Fact-checking} \label{invisible}
Within CSCW, a lot of attention has also been paid on highlighting the invisible or the overlooked work in a process or within an organization \mbox{\cite{sawyer2006always,dye2018paquete,abraham2013re,ferreira2011user}}. Invisible work can include situations where person performing the work is visible but some of the work they perform is ``functionally invisible or taken for
granted'' \mbox{\cite{stisen2016accounting}}. Such work remains hidden in the background but is essential for collective functioning of a workplace \mbox{\cite{nardi1999web}}. It often includes informal work practices such as  informal conversations, operational and maintenance work, etc \mbox{\cite{nardi1999web,verdezoto2021invisible,unruh2008invisible}}. There are also situations where   the person performing the work itself is invisible, such as, service, design or domestic work \mbox{\cite{star1999layers,markussen1996politics}}. In certain complex environments (e.g. hospitals), both visible and invisible work practices can take place simultaneously \mbox{\cite{stisen2016accounting}}.  In similar vein, fact-checking is a complex ecosystem that includes somewhat \emph{visible} editorial and investigative work and \emph{invisible} advocacy, policy and  research work.  Most of the prior research has looked at fact-checking  as a process to debunk misinformative claims (\textit{short-term claims centric} fact-checking) \mbox{\cite{doi:10.1080/17512786.2019.1643767,amazeen2013critical,graves2019fact,graves2017anatomy}}. Scholars have mostly engaged with the role of  stakeholder groups such as fact-checkers and editors in supporting the fact-checking work \mbox{\cite{doi:10.1080/17512786.2019.1643767,graves2016rise,amazeen2013critical}}. We add on to the existing literature by not only expanding on the previously reported roles of fact-checkers and editors but also identifying other stakeholder groups, such as  investigators and researchers, and advocators  
whose roles remain  invisible and unexplored by prior research.   We shed light on the invisible work that fact-checking organizations are 
doing to improve the availability and quality of information in their country 
(\textit{long-term advocacy centric} fact-checking).

\subsection{Current Landscape of Computational Research in Fact-checking }\label{computational_research}
Past quantitative and computational research studies on fact-checking have primarily focused on automating multiple stages of the  fact-checking process \mbox{\cite{hassan2017toward,shao2016hoaxy,ghenai2017catching,cerone2020watch,kartal2020too,bondielli2019survey}}, determining the perception and believability of fact checks \mbox{\cite{fridkin2015liar,agadjanian2019counting,oeldorf2020ineffectiveness}} and  construction of fact-check databases \mbox{\cite{wang2017liar,thorne2018fever,kotonya2020explainable}}.
Scholars have adopted several   approaches to determine the veracity of content, such  as use of knowledge graphs \mbox{\cite{shiralkar2017finding}}, crowd-sourcing \mbox{\cite{hassan2019examining,kim2018leveraging}}, deep learning models \cite{karadzhov2017fully},   natural language processing techniques coupled with supervised learning techniques \cite{hassan2017toward} and combination of human knowledge and AI \cite{nguyen2018believe}. Work in the field of multimedia forensics has also led to the development of content verification tools especially for image and video verification such as Tineye, InVID etc. \cite{teyssou2017invid}. Despite the plethora of automated systems and tools available for fact-checking, {our understanding of their usefulness  in practise is limited. } Furthermore,
{there is a dearth of scholarly work that engages with the limitations of the current fact-checking tools and practices (\mbox{\cite{graves2018understanding,dias2020researching}}} are a few exceptions). Our paper addresses this gap by 
interviewing the various stakeholder groups involved in the fact-checking process to understand the 
technological infrastructure supporting their work 
including the tools that are actually used by the stakeholders in practice, limitations of the current tools and the  challenges faced by multiple stakeholder groups. 

 {The remainder of our paper is organized as follows. We begin by describing our study method and interview protocol in Section \mbox{\ref{method}}}. Then, we present a brief overview of \textit{short-term claims centric} and \textit{long-term advocacy centric} fact-checking  in Section \mbox{\ref{first}}. Next, we answer RQ1 by describing in detail the 
 human and technological infrastructures supporting both types of fact-checking
 in Section \mbox{\ref{labor-short-term}} and Section \mbox{\ref{labor-long-term}}.  We answer RQ2 by identifying the needs and challenges faced by the stakeholder groups in Section \ref{challenges}. Finally, we discuss the implications  of this study in Section \ref{sec:discussion}. 

\section{Method} \label{method}
To better understand how fact-checking is practised in real-world, we conducted semi-structured interviews
with  six stakeholder groups (N={26}):  (1) Editors (2) External fact-checkers (3) In-house fact-checkers (4) Investigators and researchers (5) Social media managers, and (6) Advocators. All interviews were conducted with the approval of the Institutional Review Board.  {We started by interviewing fact-checkers and editors employed in  fact-checking organizations and  publication houses. The qualitative analysis of the initial interviews (with P3, P4 and P5) as well as conversations with our contacts in the fact-checking organizations during the initial recruitment} gave us a new perspective on fact-checking revealing the complex workflows that include several other stakeholder groups (apart from editors and fact-checkers) who work together to achieve the end goal of fact-checking. We then expanded our recruitment to interview people belonging to these other stakeholder groups.

\subsection{Participant Sampling Technique}
We adopted convenience \cite{etikan2016comparison} and snowball sampling \cite{goodman1961snowball} to recruit our subjects. First, we employed convenience sampling to identify fact-checkers via Twitter search. We  sent out personal recruitment messages to those Twitter users whose Twitter bio revealed them to be fact-checkers and whose accounts allowed direct messaging. The second author had established collaboration with a few fact-checking organizations. We also reached out to individuals working in these organizations. Next, we used snowball sampling for recruitment.  We requested the individuals who participated in our study to further connect us with other individuals belonging to different stakeholder groups in their or other organizations.  
We interviewed a total 26 participants from  16 fact-checking teams/organizations including a Pulitzer Prize winning editor and journalist. 
{Tables \mbox{\ref{participants_details}} and \mbox{\ref{organization_details}} list our participants' demographics, experience in current role, stakeholder groups that we studied, participating organizations and the continents we covered by interviewing participants based in those continents. } {Note that the names of the roles of various stakeholder groups were not always self-reported (with an exception of fact-checkers, news editors and copy editors), but were rather found during qualitative analysis. Therefore, these roles might not match with participants' designation in the fact-checking team or organization. For example, the professional designations of two participants performing advocacy work in their respective fact-checking organizations were \textit{Head of Public Policy \& Institutional Development} and \textit{Partnerships manager}.} Several of our participants' roles were fluid and overlapping, i.e in some fact-checking teams/organizations, at times a single person enacted several  roles. {For example, participant  P12 performs both editorial and advocacy work.} In some cases, a participant provided us with insights about more than one {role} since they either led  or managed the entire fact-checking team or  work(ed) closely with the other stakeholder groups  and thus, were aware of the various roles involved in the fact-checking pipeline.  {For example, we interviewed an advocator working at Meedan, an organization that provides institutional and programmatic  support to partner  organizations doing fact checking work. 
Their job at Meedan allows them to work closely with people performing various roles at the fact-checking organizations.   Thus, the participant was  able to describe the responsibilities and tasks conducted by various stakeholder groups such as fact-checkers, advocators and investigators. }

\begin{table}[]
{\sffamily \scriptsize
\begin{tabular}{lll|lll|lll|lll}
\textbf{P\#} & \textbf{Gender} & \textbf{Exp.(yrs)} & \textbf{P\#} & \textbf{Gender} & \textbf{Exp.(yrs)} & \textbf{P\#} & \textbf{Gender} & \textbf{Exp.(yrs)} & \textbf{P\#} & \textbf{Gender} & \textbf{Exp.(yrs)} \\ \hline
P1 & Female & 0.67 & P8 & Male & 2 & P15 & Female & 2 & P22 & Male & 3 \\
P2 & Male & 0.21 & P9 & Female & 1 & P16 & Female & 0.5 & P23 & Male & 1 \\
P3 & Female & 2.5 & P10 & Male & 0.42 & P17 & Male & 0.83 & P24 & Male & 15 \\
P4 & Male & 2.25 & P11 & Female & 2 & P18 & Male & 10 & P25 & Female & 17 \\
P5 & Male & 3 & P12 & Male & 4 & P19 & Male & 4 & P26 & Female & 1 \\
P6 & Female & 0.75 & P13 & Female & 2.5 & P20 & Female & 7 &  &  &  \\
P7 & Female & 0.92 & P14 & Male & 0.5 & P21 & Male & 0.75 &  &  &  \\ \hline
\end{tabular}
}
\caption{{Table showing list of participants with their gender and experience (in years) in their current role. Some participants have been associated with  fact-checking work for a longer duration. We only report their experience in the current role in the organization.}}
\label{participants_details}
\vspace{-15pt}
\end{table}

\begin{table}[]
\centering
\footnotesize
\begin{tabular}{p{0.25\linewidth}p{0.46\linewidth}p{0.13\linewidth}}
 \textbf{Stakeholder group} & \textbf{Organization}  & \textbf{Continent}  \\
\hline
\begin{tabular}[t]{@{}l@{}}1) News desk editors and \\  copy editors\\ 2) External fact-checkers\\ 3) Social media managers\\ 4) Investigators and researchers\\ 5) Advocators\\ 6) In house fact-checkers \end{tabular}

& 
Pesacheck \cite{pesa}, Meedan \cite{meedan}, First Check \cite{dl}, Full Fact \cite{fullfact}, The African Network of Centers for Investigative Reporting's Investigative Lab  \cite{fnc}, AFP \cite{afp}, Africa Check \cite{africacheck}, The New Republic \cite{nr}, The Quint \cite{quint}, Al Jazeera \cite{aj}, The Washington Post \cite{wp}, DPA \cite{dpaen59}, Maldita \cite{ortada}, India Today \cite{FactChec52}, Der Spiegel \cite{DERSPIEG52:online}, Fine Tip Research \& Editing, Freelance

& Africa, Asia, Europe, North America \\
\hline
\end{tabular}
\caption{{Table showing the stakeholder groups identified in the study, the participating organizations and the continents we covered through the interviews. In the organization column, freelance refers to no association with a particular fact-checking organization/team. We aggregated the roles of stakeholders and their association with fact-checking organization/team to ensure anonymity as in some cases knowledge of network affiliation and role could potentially reveal the identities of  few participants. Note that the participants that we interviewed sometimes provided insights about more than one role.}}
\label{organization_details}
\vspace{-15pt}
\end{table}

\subsection{Interview Protocol and Data Analysis}
{All interviews were conducted between November 2020 to September, 2021.} {We designed a generic semi-structured interview script for our study that contained a set of broad questions about participant's and their organization's role.
}
{Based on participants' responses to these  questions, we inquired  them about the specific details of their roles. Thus, recruiting and interviewing different stakeholder groups did not require us to file any changes with our university's Institutional Review  Board.}
We first asked participants to describe their role within their organization, and the function of the organization itself. We encouraged
{participants to share their screen and describe various aspects of their work using real-world examples. We also asked the participants to demonstrate the tools they use wherever applicable.}
We probed them about the role of technology in their day-to-day work and how the affordances provided by online platforms facilitates or impede their work.  {To get insights about how fact-checking is practised in the participant's team/organization, we asked them to describe all the steps involved in the fact-checking pipeline. We inquired} about the other stakeholder groups working in their team/organization who also contribute towards the fact-checking process and how these various groups collaborate with each other. We also discussed  various challenges participants face in their job.  The interviews lasted between 60 to 125 minutes and averaged over 90 minutes. All interviews were recorded via {Zoom} or {Google Meet}.
{The first author transcribed all the video and audio recordings.  Then, two authors independently went through the transcripts and observation notes taken during the interviews and thematically analyzed them  using a
mixture of deductive and inductive coding scheme \mbox{\cite{braun2006using}}. First, both authors conducted a deductive scan where the transcripts were coded for categories: fact-checking process, stakeholder groups,  participant's role, decision making, use of tools, collaboration within stakeholder groups and challenges faced by the participant in performing their role. Then within each deductive code, inductive coding was conducted. Both authors read the transcripts multiple times to determine the codes.  Then, the two authors compared and contrasted their codes with each other to refine the codes and resolve the inconsistencies. After several rounds of discussions, both authors converged on a final set of themes.  We present these themes with respect to the research questions in the following sections.
}




 \begin{figure*}
  \centering
      \includegraphics[width=0.9\textwidth,keepaspectratio]{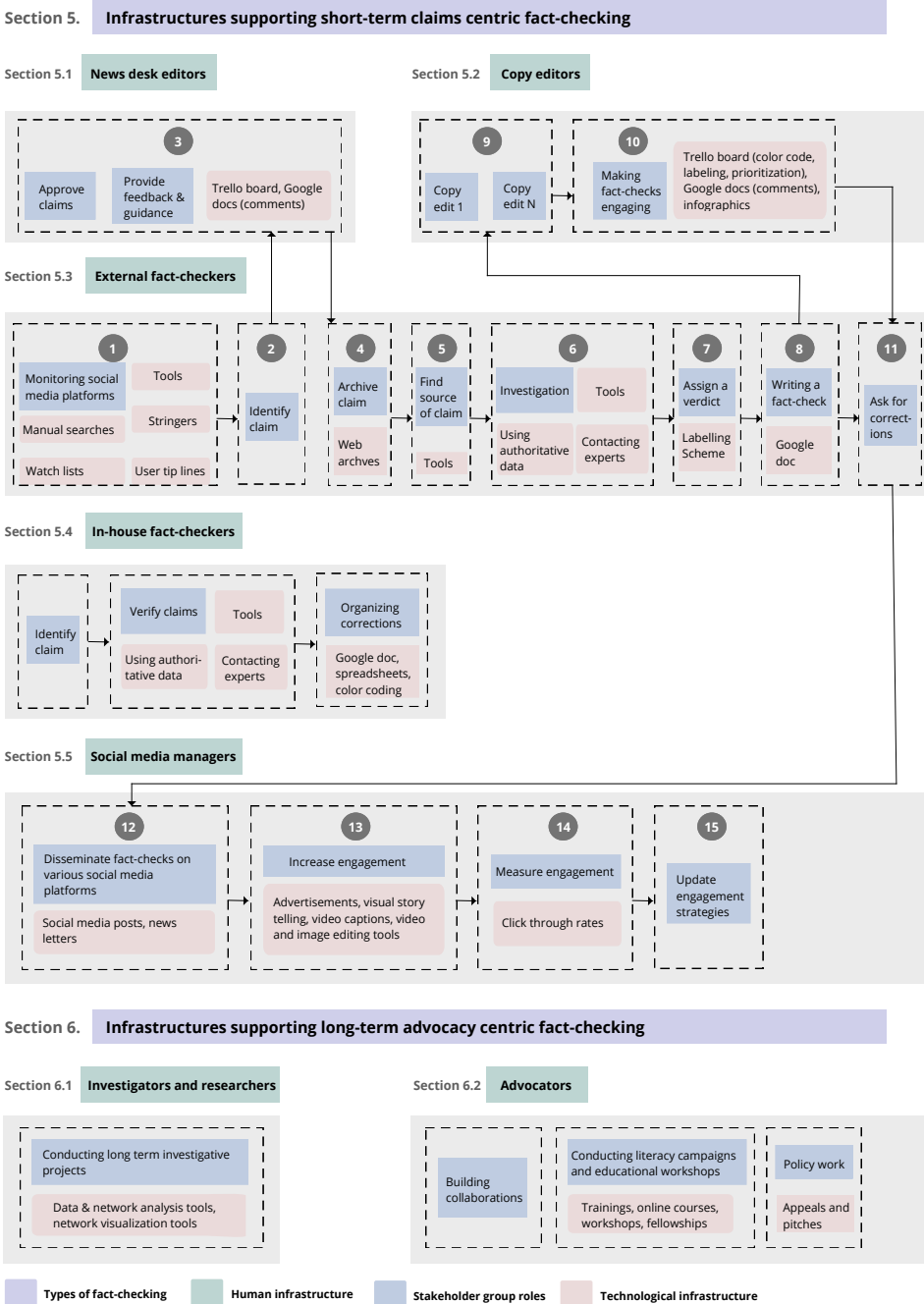}
  \caption{{Figure presenting the  ecosystem of  fact-checking, {the whole or part of which  could exist in a fact-checking organization or a news publication house.} \mbox{\fcolorbox{labelpurple}{labelpurple}{\rule{0pt}{4pt}\rule{4pt}{0pt}}} indicates the two types of fact-checking (\textit{short-term claims centric} and \textit{long-term advocacy centric} fact-checking) introduced in the study,  \mbox{\fcolorbox{labelgreen}{labelgreen}{\rule{0pt}{4pt}\rule{4pt}{0pt}}} presents the  stakeholder groups involved in the fact-checking process (human infrastructure), \mbox{\fcolorbox{labelblue}{labelblue}{\rule{0pt}{4pt}\rule{4pt}{0pt}}} shows  work done by the stakeholder groups as  part of their role, and \mbox{\fcolorbox{labelpink}{labelpink}{\rule{0pt}{4pt}\rule{4pt}{0pt}}}  specifies the tools stakeholders use to mediate their roles (technological infrastructure). The numbers indicate the sequence in which various roles are performed.} }
  \label{factcheckdiag}
  \vspace{-0.4cm}
\end{figure*}


\section{Types of Fact-checking: Short-term Claims and Long-term Advocacy} \label{first}
Our study aims to identify the  infrastructures---both human and technological---supporting the fact-checking work. { We identify and present the human infrastructure by elucidating the role of  six stakeholder groups that need to come in alignment to accomplish fact-checking. The six stakeholder groups are}  editors (news desk and copy editors), external fact-checkers, in-house fact-checkers, social media managers, investigators and researchers, and advocators. We show how these stakeholder groups' roles are  supported by technological infrastructure. Through the study of the fact-checking infrastructures, we establish that
 fact-checking exists as both \textit{short-term claims centric} and \textit{long-term advocacy centric} fact-checking. In this section, we first provide an overview of the two types of fact-checking before deep diving into the infrastructures supporting them in Section \ref{labor-short-term} and Section \ref{labor-long-term}. Figure \mbox{\ref{factcheckdiag}} provides an overview of the fact-checking ecosystem including the stakeholder groups, their roles and various tools that support them in performing their roles.


\subsection{Short-term Claims Centric Fact-checking} 

\textit{Short-term claims centric} fact-checking aims at informing the public by debunking  misleading claims circulating on the online platforms. It begins with fact-checkers continuously monitoring the online spaces for potentially misleading content.
They  identify the exact claim(s) that they want to fact-check and make a pitch to the editorial team about how they plan to debunk that claim  {\mbox{(Section \ref{fact-checkers-role})}}.
After gaining editor's approval  {\mbox{(Section \ref{editorial_checklist}})}, they archive the content, find the  source of the claim, and investigate the claim by  using online tools, consulting experts  and employing authoritative publicly available evidence (e,g, statistics on unemployment, census data, etc.). Based on their investigation, fact-checkers assign a label indicating the veracity of the claim and then 
write a report declaring all the sources  they gathered {\mbox{(Section \ref{fact-checkers-role}})}.
The written report goes through a rigorous copy editing pipeline to ensure integrity of the fact-check {\mbox{(Section \ref{copy_editors})}}.
If the claim is false, fact-checking organizations  reach out to person/organization who made the claim for correction(s) {\mbox{(Section \ref{fact-checkers-role}})}. 
Finally, social media engagement team publishes the fact-check story/document on  the social media pages of the organization and adopts several strategies to increase audience's engagement with the published fact-check { (Section \mbox{\ref{engagement_team}})}. {Our work also examines the role of in-house fact-checkers doing \textit{short-term claims centric} fact-checking in news and media publication houses. In house fact-checkers  assist reporters/journalists in verifying and validating their articles by ensuring that the  facts and quotes present in the article are correct and backed by authoritative sources {(Section \mbox{\ref{inhouse}})}. }

\subsection{Long-term Advocacy Centric Fact-checking} 

 \begin{small}
\begin{displayquote}
\enquote{Fact checking is more than publishing fact checks. In order to change the information ecosystem.., we need to spot patterns and try and do something about those patterns.. We try to influence policymakers, information producers and media to raise their standards and improve the quality of information and public debate. [Our role is] more akin to a campaigning organization. 
 } - P9
\end{displayquote}
\end{small}

We find that the work of the  fact-checking organizations is more than a one-off engagement with misleading claims and fact-checks (\textit{short-term claims centric} fact-checking). Most organizations also perform   \textit{long-term advocacy centric}  fact-checking. They  run several investigative projects to  study the misinformation ecosystem in their country {(Section \mbox{\ref{inv}})}.  They are also actively involved in advocacy and policy work  where they try to influence civic bodies and policy makers to improve the quality of data, organize workshops to train organizations and journalists to do fact-checking  and work towards forming coalitions among various fact-checking organizations and internet companies {(Section \mbox{\ref{advocacy}})}.  
To demystify the  fact-checking process, in the following sections, we present in detail the roles  performed by the stakeholder groups involved in both types of fact-checking  along with the collaborations occurring in process. For each stakeholder group, we also discuss how technological  infrastructure supports the enactment of their roles.

\section{Infrastructures Supporting Short-term Claims Centric Fact-checks } \label{labor-short-term}
\textit{Short-term claims centric} fact-checking is supported by five  stakeholder groups---news desk editors, copy editors, external and in-house fact-checkers, and social media managers. We present the roles played, activities performed,  decisions made, and tools used by them. 

\subsection{News Desk Editors---Approving Claims and Guiding Fact-checkers}  \label{editorial_checklist}

{News desk editors are one of the most critical stakeholder groups supporting the \textit{short-term claims centric} fact-checking.} They decide what their team/organization is going to fact check. 
They approve or reject the claims pitched by the fact-checkers and guide them in their work.

\subsubsection{Approving claims to fact-check} 
News desk editors are looking for newsworthy claims that impact a lot of people.  
First criteria of approving a claim  is the popularity and reach of the  person making the claim since it increases the chance of the claim  spreading far and wide (``\textit{it should be newsworthy, said by an important person,... if a popular public figure makes a claim, a lot of people are likely to be exposed to that claim, given the the bully pulpit that public figures have''---P18}).
Second criteria includes number of people likely to be misled based on the context surrounding the claim. For example, a claim touching upon a communal angle is likely to impact a lot of people in a country that has ``\textit{many illiterate people..who process information based on their communal experience} (P12)''.
Third, news editors  approve content that is gaining a lot of traction on social media platforms by accumulating engagement in the form of likes, comments and shares. 
Content that has received less attention from the public is not fact-checked from the fear of amplifying the false information by inadvertently bringing attention to the false claims (\textit{``Are people believing this or are they taking it just as a joke? Does it just have one share, meaning if I  fact check it, it will just amplify the fake news and not really give the correct information''---P11}).
Fourth, opinion pieces and claims that cannot be verified using sources are not fact-checked (``\textit{we check things that have facts,  that can be verified using records and information. We don't want a situation where we see this person says this and this person says this, that will just be hearsay''---P3}). Fifth,  fact-checkers and news desk editors consider  several stages of harm  that false information is likely to cause---physical, mental, social or emotional and prioritize the claims that are likely going to cause maximum damage to the public by affecting their  health and well being.

\begin{small}
\begin{displayquote}
\enquote{We have stages of harm that you have to check. Is it causing physical harm? Is it causing mental harm? Is it causing  someone to lose social standing? How much effect will it have to the person.. if I don't fact-check the story? .. We do the ones that actually have greater harm, we give them priority.. We have to get it out before most people  see it to reduce the harm that it's causing.} - P3
\end{displayquote}
\end{small}

\subsubsection{Guiding fact-checkers} The job of news editors does not end after approving a claim. They also guide and help fact-checkers  in ``\textit{gathering sources and evidence (P11)}'' for verifying  stories, ``\textit{understanding concepts, finding best available data [for research],..and connecting with the experts given [editors'] long experience in the media ecosystem (P12)}''.


\subsection{Copy Editors---Ensuring Quality  of the Fact-checks} \label{copy_editors}
{Copy editors do quality control} of the fact-check story/document through multiple iterative copy editing cycles. {A fact-check story or document is a report  written by  fact-checkers containing the claim investigated, sources used for investigation, and verdict indicating veracity of the claim.} 
{This stakeholder group} acts like the \textit{first readers} who determine whether  fact-checkers have accurately interpreted the claim in the fact-check, used multiple primary sources as evidence  for investigation,  provided working links to the sources used for investigation and  presented the evidence in such a way that it leads to a logical and correct conclusion. 
They also work towards making the fact-checks engaging by checking the phrasing and grammatical errors. We discuss the {tasks} performed by copy editors in detail below.




\subsubsection{Performing copy editing cycles} Through our interviews we realized that a fact-check goes through two to three copy editing stages. Each stage is supervised by a different {copy} editor to ensure higher quality.
The written fact-check is provided to the {copy} editor in a shared document (e.g. Google doc) where they leave comments to provide feedback. 
At the first copy edit stage, {copy} editors check the central premise of the fact-check.

\begin{small}
\begin{displayquote}
\enquote{When I get a fact-check, I read through it three times to understand it  before I make any change on it, before I ask any questions. I look for the claim and  the debunk. Does it really hold? If there are questions about the debunk then  we send it back to the news desk.. At that stage, the fact checker will pick it up and go back and try to sort out any queries that have been raised.} - P1
\end{displayquote}
\end{small}


The second copy-edit stage focuses on refining the language and flow of the fact-check.  The final copy-edit stage focuses  on making the fact-check more engaging and interesting to read.  We discuss this aspect briefly.


\subsubsection{Making  fact-checks more engaging} {Copy} editors try to keep the fact-check short, clear, crisp and interesting. They ensure that the fact-checks are written in a language that is understood by laymen.
P1 spoke most candidly about the engagement aspect in the editorial process. They revealed that they often collaborate with social media managers to get feedback  on the engagement aspect of the fact-checks. For example, ensuring that  the country relevant to the fact-check is present ``\textit{in the title or in the blurb (P1)}''  so that people could quickly determine if its of interest to them, or making certain that the verdict on the claim is placed high up in the fact-check to get more attention from the public. Copy editors along with social media managers also suggest addition of info-graphics (engaging visuals, imagery, tables, charts, etc.)  to fact-checks in order to attract the eyes of the readers. The info-graphics are added to quickly communicate ``\textit{complex information in a visual manner} (P2)''. They are usually added in long-form fact-checks---the ones that debunk multiple claims and delve into the subject of the claim in greater depths.

{Copy editors also ensure that the fact-check contains  terms  that people are most likely to search online.} P1, P9, P12 and P18 talked about claim-review schema used by fact-checking organizations to  allow   internet companies like Google to  index their fact-checks \cite{GitHubFu84:online}. These participants believe that using popular  search terms in fact-check  helps increase its visibility since search engines then  rank it higher in the search results. We briefly touch upon this collaboration between fact-checking organizations and  internet companies later in Section \ref{advocacy}.

\subsubsection{Use of collaborative systems, labelling and color coding schemes}
{In most of the organizations that we interviewed}, both news desk and copy editors use Google docs to comment and provide feedback on the fact-checks. In addition, {fact-checking  organizations like Pesacheck, Africa Check, etc.} also use an open-source project management and collaboration tool called Trello \cite{Trello12:online}. The tool provides a dashboard with a series of columns  that contain cards. The columns are named so that they denote the current stage of the fact-check. For example, P2 showed the Trello board of their organization which had several stages such as \textit{complete fact checks, copy edit 1, copy edit 2, copy edit 3, Q/A \& final review, final review done, published live, and amplification done}. 
Trello cards are the basic functioning unit in the tool. The cards hold various information about the fact-checks including due dates, conversations,  attachments etc. The tool allows  editors to add  colors and labels to the cards.
{Copy} editors use labels and color coding schemes for different purposes. For example, P2 uses labels for prioritizing the fact checks (``\textit{Earlier this year there was a lot of interest in COVID, so I would label fact-check as COVID because we were prioritizing those}). P10 uses colors to indicate  tasks allocated to people (``\textit{It's easier for me to see what people are working on.. If its research, it will be purple, if its documentation, it will be green.}).'' 

\subsection{External Fact-checkers---Monitoring, Investigating and Publishing Fact-checks}
 \label{fact-checkers-role}
{External fact-checkers are the most evident stakeholder group supporting the fact-checking process. }
Their role is to continuously monitor the external world for potentially false claims, investigate them for veracity, assign the verdict and publish the fact-check. We discuss these roles below. We also specify the technological infrastructures supporting the roles along with the description of the roles.

\vspace{3pt}
\subsubsection{Monitoring online spaces}\label{fbq} Monitoring content is one of the most tedious steps in the fact-checking workflow. Fact checkers monitor content reactively (in response to user tips), preemptively (before events like elections, presidential speeches, etc.) and in real time (tracking current affairs via trends and listening to conversations in real time). 
They monitor the content via several tools and artifacts. First, they rely on user reports and tiplines which are useful ways to access misleading content circulating in private groups and WhatsApp that are otherwise hard to access.

\begin{small}
\begin{displayquote}
\enquote{We have our WhatsApp tipline and emails.. Public who wants to verify a particular piece of information send it on WhatsApp to us.. We verify those queries at our end and then send them the replies with the fact check story if possible. If we don't have fact check story, then we send people whatever information we have on the query. 
} - P14
\end{displayquote}
\end{small}

Second, fact-checkers create watch lists and track social media accounts, groups, pages and websites of repeat offenders--those who posted misinformative content multiple times in the past (\textit{``We've a database where we've  tracked all accounts spreading misinformation.. We go back and check these accounts, see what they've posted on their website and personal account.''---P3}).
Third, fact-checkers rely on manual searches. They follow the current events via news or Twitter trends to get updated about  topics that people are talking about and track them on all online platforms. 
Creation of relevant search queries to track these topics  is mostly a tedious ``\textit{hit-n-trial} (P7)'' method.
Searching for a query can give millions of results on search platforms. Therefore, fact-checkers rely on search query syntax to reduce the number of search results.

\begin{small}
\begin{displayquote}
\enquote{I do not want the news content. I want user generated content. So one of the simplest tools is to write minus news (search\_query -news)  so that it cancels out the major news content.} - P7
\end{displayquote}
\end{small}

\begin{small}
\begin{displayquote}
\enquote{I use keywords like ``intitle'' [on Google search]. [intitle:coronajihad site:facebook.com].. is showing me every search term, every post on Facebook, which has the title ``coronajihad''.} - P8
\end{displayquote}
\end{small}

Fourth, fact-checkers  use several tools to track content on the internet. For example, they use CrowdTangle \cite{CrowdTan34:online} to track Facebook's public pages and groups. Organizations that have partnered with Facebook have access to a Facebook proprietary tool colloquially known as the ``Facebook Queue''. The tool aggregates potentially misleading content that is accumulating engagement on the platform. 
{In some of the organizations}, fact-checkers also rely on several third party tools (e.g. Social searcher \cite{SocialSe32:online}, Influencer \cite{FindEver1:online}, BuzzSumo \cite{BuzzSumo25:online} etc.) to search and filter content on platforms since they provide them varied search filter options that the original interface of the social media platform lacks. For example, Facebook search  allows to filter content by year and not by months and dates. 
Lastly, fact-checkers also rely on a network of stringers---{``reporters who work for a publication or news agency on a part-time basis'' \mbox{\cite{Stringer71:online}}}---who inform them about the misinformation ``\textit{circulating in their region..[and] language} (P11).''

\vspace{3pt}
\subsubsection{Making a decision to fact-check and extracting claim(s)}
Once fact-checkers have identified potentially false content on the internet, they identify claim(s) in the content to verify. The fact-checkers can decide to do a short-form or a long-form fact-check depending on the number of claims they identify in the content. For their partnership with Facebook, the fact-checkers only do short-form fact-checks where they identify one claim from the body of the content. In long form fact-checks, fact-checkers will debunk multiple claims present in the content. 

\begin{small}
\begin{displayquote}
\enquote{The reason we do [short-forms] is to make sure we are clear, we're not confusing the audience. This is a largely inline with the Facebook partnership where we mainly focus on one claim.} - P2
\end{displayquote}
\end{small}


Next, fact-checkers prepare a pitch to convince news desk editors why the content needs to be fact-checked along with a plan of how they would  accumulate proofs to debunk the claim.
Once the claim is approved, they archive the content using online websites like archiveis \cite{archivet72:online}, wayback machine \cite{Internet99:online}, etc. Many times people and organizations delete the false claims made by them once they are fact-checked. Thus, archiving becomes essential to prove that the content existed. 

\begin{small}
\begin{displayquote}
\enquote{Once you identify the claim you archive it because what purveyors of misinformation do mostly, they delete it. So once they delete it you're unable to read that particular claim.} - P11
\end{displayquote}
\end{small}

After archiving, fact-checkers find the original source of the claim--``\textit{who shared [the claim], context with which it (the claim) was originally shared}  (P14)'',   without which the fact-checker would not have a complete picture of the context in which the content was originally shared. 


%
\subsubsection{Researching} {Once a claim} is identified, fact-checkers  collect multiple primary sources to prove or disprove the claim.
They use three ways of gathering sources. First, they use quotes by experts such as doctors, physicians, meteorologists and academics.
Second, they use public authoritative sources like government databases (e.g.  Kenya National Bureau of Statistics), mainstream news sources (e.g. CNN), and peer reviewed scientific research papers and journals.
Third, they rely on several tools. For example, fact-checkers use image and video verification tools such as InVid \cite{InVIDVer51:online} followed by a reverse image search on search engines to collect metadata and digital trail of the image/video in order to determine its authenticity.
{Fact-checkers from First Check, Pesacheck, and Africa Check also reported that they depend on third party tools}  such as whois \cite{WhoisLoo11:online}, Spoonbill \cite{Spoonbil76:online}, edit history functionality in Facebook, etc. to determine the veracity of a website or post.
 Through our interviews we realized the important role played by comments in fact-checking. Comments contain useful clues that help in investigating the claims.


\begin{small}
\begin{displayquote}
\enquote{What we do first is read comments before checking. We found in comments that one woman said this is not the entire video, here is the link to entire video. So that's what led us to entire video. [In comments] we see some clues, how to look for what really happened.
} - P16
\end{displayquote}
\end{small}


\subsubsection{Assigning veracity label and publishing fact-check} After the investigation, fact-checkers assign a label to the claim that reflects its veracity. Through our interviews, we realized fact-checking organizations
use a range of labelling conventions, for example, 5 point scale ranging from completely false to true, four point Pinocchio scale \cite{AboutThe91:online}, etc. There is no commonly accepted standard for labelling misinformation.  
If the verdict on the fact-check is false, fact-checkers also ask the person/organization that had posted the misinformative content for correction.
The last task is publishing the fact check.  The fact-check contains the link to the content being debunked, description of the claim, names, quotes and links to all sources used for claim verification followed by the verdict/label. 
By publishing these details, the fact-checker takes the reader through the entire investigation ``\textit{so that the readers can replicate [the process] themselves} (P18)''.

\subsection{In-house Fact-checkers---Gathering Sources and Verifying Claims} \label{inhouse}
In-house fact checkers are employed by publication houses to fact-check the stories produced by the journalists  before they are published and disseminated. They receive the script or the news story from the journalist along with all source material that they used while researching and writing the piece. The in-house fact-checker then verifies every claim present in the story and delivers a modified story along with a list of proposed changes. 
{Unlike the job of external fact-checkers, the job of this stakeholder group is to fix or remove incorrect claims without publicizing or calling attention to the inaccuracies} \cite{graves2013deciding}. {The need for in-house fact-checkers arises} in a publication house  not only to ensure that the stories published are reporting accurate facts to readers but also to \textit{``protect the publishing house from any liability or future lawsuit (P13)''}.  We discuss the roles of this stakeholder group below.

\subsubsection{Identifying and verifying claims} 
In-house fact-checkers verify every line present in the story. Each line usually has more than one fact to be checked, from how a proper noun is spelled, grammatical idiosyncrasies to  every phrase making a claim. Journalist's opinions and arguments, and  quotes from a reputable expert are the only phrases in the content that are  not verified. However,  for opinions, the in-house fact-checkers check the context surrounding the argument to ensure it is ``\textit{right and mainstream (P13)}''. 
All the  claims that are vague and without proper sources backing them are modified or removed from the text.
\begin{small}
\begin{displayquote}
\enquote{Vague and broad facts which I.. try to get people to remove..  I recently had a whole debate with someone about a line that said America is more divided today than ever. I was like, how do you check that line.. what are the sources for that kind of statement? [It] is just so broad.} - P15
\end{displayquote}
\end{small}

In-house fact-checkers employ a myriad of techniques---top down approach, prioritization and batch processing---to identify and verify claims. {The} majority of the in-house fact-checkers we interviewed  first perform a top down linear scan of the document to get the most central ideas ``\textit{which if were incorrect, the entire piece (article) would be called into question}'' (P13). Second, they prioritize the claims for verification.  Different  fact-checkers prioritize claims differently. For example, P13, P15 and P25 work with organizations that have strict timelines for publications. Thus, they first prioritize the claims that will take maximum time to investigate. On the other hand, P20 does not have strict deadlines and prioritizes claims that they are certain are false.

\begin{small}
\begin{displayquote}
\enquote{{[I prioritize] a claim that I know is going to take me a while to nail down. So for example, if there's a claim ... about someone committing a crime and I need to file a FOIA request\footnote{https://www.foia.gov/how-to.html} or
go through public records or order a document from sort of  government agency, I want to do that as early as possible so that I get myself as much time. You know, basically any claim that relies on other actors in order to meet to verify.}} - P13
\end{displayquote}
\end{small}

\begin{small}
\begin{displayquote}
\enquote{{When I prioritize I start with the things that I know that are wrong, and then I look at the things I think are right. And then I check all dates and all names. }} - P20
\end{displayquote}
\end{small}

P15 revealed that  they \textit{``process the claims in batches''}. Everyday they work on a batch of claims and send enquiries and suggestions regarding those claims to the journalist. This technique gives journalists ample time to respond and does not inundate them with several queries towards the very end of the schedule.

\subsubsection{Gathering sources for verification} 
In-house fact-checkers use ``\textit{primary reputable sources (P15})'' to verify claims. The journalist  is expected to give  the primary sources that they used while doing research and writing the article. However, if the sources are missing, in-house fact-checkers look for primary documentation (like death certificate, house deed, etc.), mainstream news sources (like CNN, New York Times, etc.), academic peer reviewed journals and relevant experts to verify the claims. In case the source used by the journalist is not reliable, the usual journalistic practice is to gather a total of three to four sources to back up the claim. 
 In-house fact-checkers heavily rely on Google search  to hunt for sources. {Some of the in-house fact-checkers} also use Nexis search \cite{SearchRe0:online}---a paid service---that gives news articles, blogs, legal documents as search results. They find latter to be more effective for searching news and documents. 

\subsubsection{Organizing corrections} After identifying and verifying claims, in-house fact-checkers organize and communicate the list of questions and suggested changes to the journalist. According to  in-house fact-checkers, there is no standard convention for organizing corrections. However, we found that all fact-checkers use color coding schemes for organization but in different ways. For example,  P13 highlights verifiable and unverifiable claims in the Google doc with different colors  and leaves  comments  containing information about the sources. P15 copies each claim from the doc into a separate row in a spreadsheet and then uses color coding to indicate the state of the claim (pending, verified, incorrect, etc). 

\subsection{Social Media Managers---Disseminating Fact-checks, Increasing Engagement} \label{engagement_team}
{Social media managers  are responsible for} leading all social media initiatives including publishing and disseminating  fact-checks on their organization's social media handles. They determine how to make fact-checks more appealing so that they attract people's eyes. They also measure the engagement received by the posted fact-check(s)  and based on the feedback continuously update their amplification strategies to attract more audience to engage with the fact-checks.
We discuss these {tasks} below.

\subsubsection{Disseminating fact-checks} 
Social media manager's primary responsibility is to post fact-checks and educational tip-sheets produced by fact-checking organizations on all major social media platforms. Few organizations also have a ``\textit{WhatsApp number where they  broadcast weekly newsletters containing the top fact-checks of the week (P12)}''. Social media managers want to make their fact-checks accessible to the people with disabilities. Since Twitter did not have captioning feature with its audio tweets, P17 usually posts fact-checking videos   with captions. They use tools like Kapwing \cite{KapwingT82:online} to add captions in the videos.
\begin{small}
\begin{displayquote}
\enquote{We wanted to see how can we reach, for example, blind and deaf audiences.. The audio tweets did not have captioning, they did not have accessibility features for disabled audiences. And so this is why we decided to use the video instead because videos allow us to caption and so people who can't hear the video can still read the information. } - P17
\end{displayquote}
\end{small}




\subsubsection{Adopting strategies to increase engagement} Social media managers   adopt several innovative ways to increase their content's  reach inorganically (via ads) and organically (via visual storytelling). We discuss some of the strategies.

\begin{figure*}
  \centering
  \begin{subfigure}{0.33\textwidth}
      \centering
      \includegraphics[width=1\textwidth,keepaspectratio]{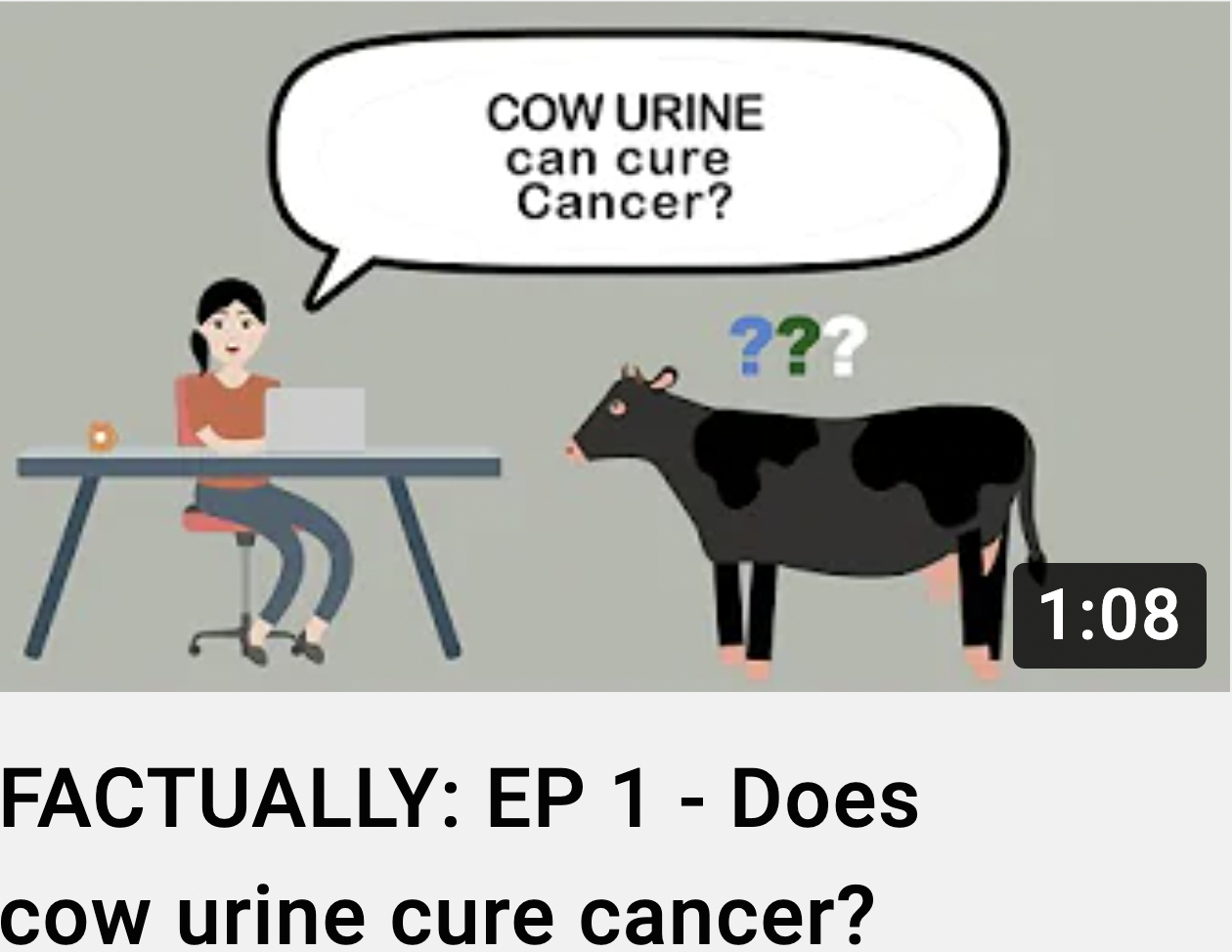}
      \caption{}
      \label{searchtopicfig}
  \end{subfigure}\hspace{5mm}
  \begin{subfigure}{0.33\textwidth}
      \centering
      \includegraphics[width=1\textwidth,keepaspectratio]{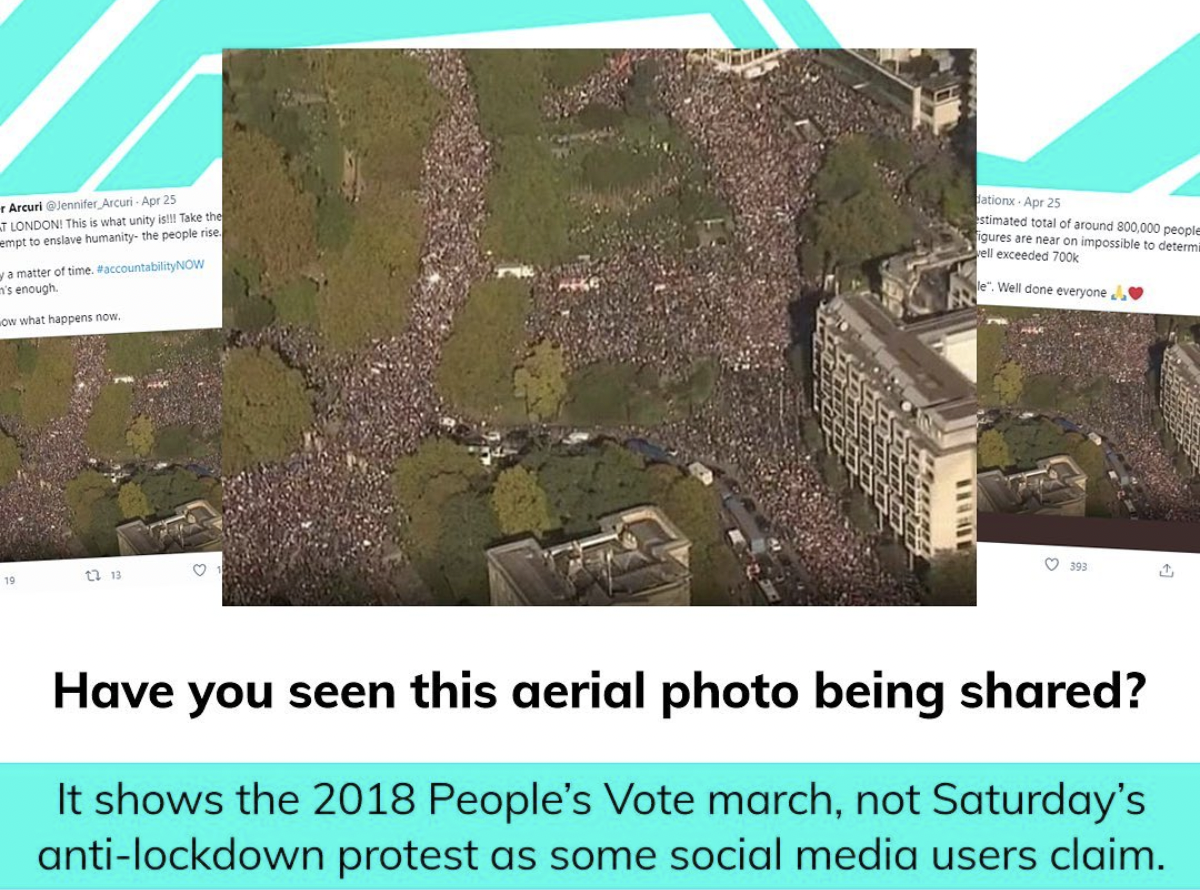}
      \caption{}
      \label{searchqueryfig}
  \end{subfigure}\hspace{5mm}
 \begin{subfigure}{0.25\textwidth}
      \centering
      \includegraphics[width=1\textwidth,height = 4cm]{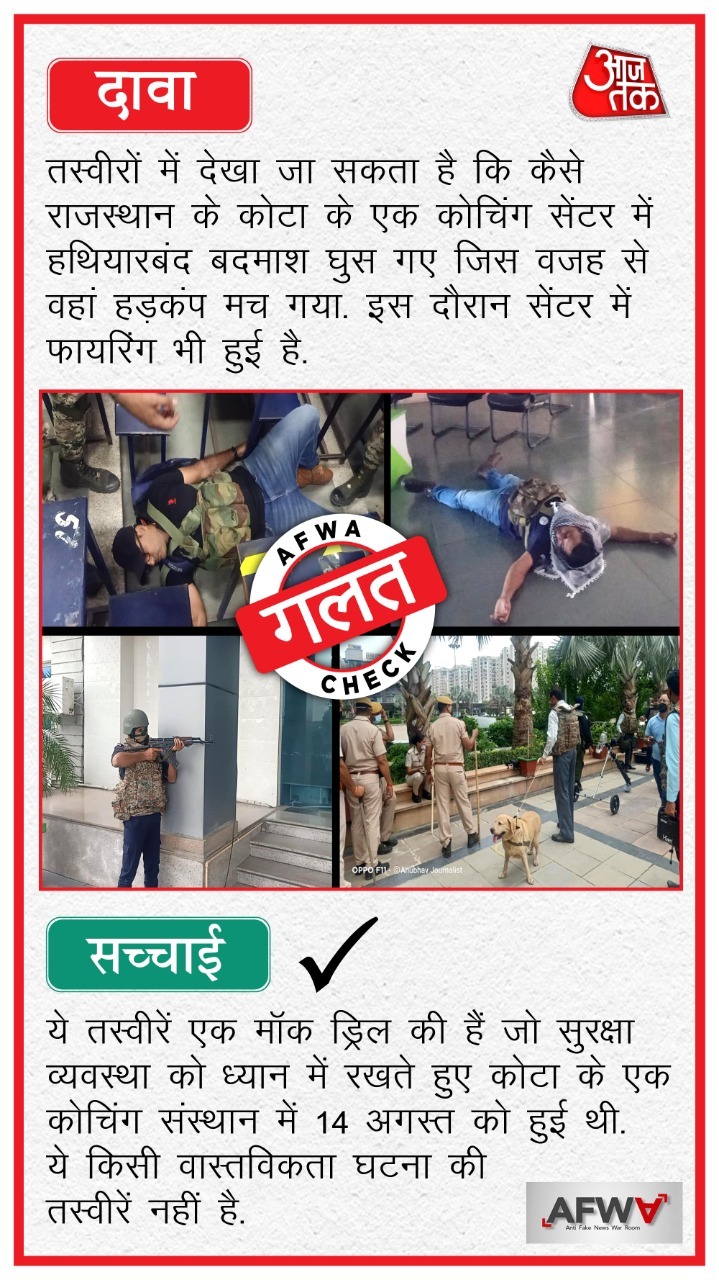}
      \caption{}
      \label{autocompletefig}
  \end{subfigure}
  \caption{{(a) A short YouTube video explaining a fact-check using comic like visuals (b) An Instagram post containing a fact-check (c) A ``postcard'' containing fact-check in Hindi language to be shared on mediums like WhatsApp. The single image contains the false-claim and the debunk. }}
  \label{visuals}
  \vspace{-0.4cm}
\end{figure*}

\setlist{nolistsep}
\begin{itemize}

\item \textit{Running advertisements:} {A few} fact-checking organizations that have partnered with Facebook receive free advertisement credits to promote their fact-checks on the platform.  P17 delved deep into the ad usage process. They advertise fact-checks that are not bounded by time since they could be promoted via ads for an extended period. To select the audience for ad targeting, they look at three attributes, namely, {the} audience's  education, relevance to the fact-check and interests. They target {audiences} who have ``\textit{graduated from a university}'' (education),  live in  ``\textit{countries that are most relevant to the fact-check}'' being promoted (relevance) and are ``\textit{interested in news, advocacy and community issues}'' (interests). 


\item  \textit{Visual storytelling: } Social media managers find visual storytelling to be a very effective way of getting people to engage with  fact-checking content. They've found that many people prefer ``\textit{watching their content over reading it (P18)}''. Therefore, they convert fact-checks into a visual narrative (images or videos) before posting them on social media platforms. To create the visual content, these stakeholder groups rely on  video and image editing tools. {For example, social media managers in few of the organizations that we interviewed use multimedia editing tools such as} Adobe illustrator \cite{Industry67:online}, Photoshop \cite{Official68:online}, etc.
Social media managers also create   comic strips  where a cartoon walks readers through the fact-checks (``\textit{I myself had started this small comic strip thing to get  more engagement.. it became quite popular and people were liking it, and sharing it''---P7}). 
To engage with the local non-English speaking audience, the comic strips, images and videos are also converted to regional languages.
\begin{small}
\begin{displayquote}
\enquote{We  translate [comics] into Swahili because there's been  this type of content, .. being done in like mainstream languages like English, French, Portuguese, But some of the more widely spoken local languages  aren't really a priority. That was a gap that we identified. } - P5
\end{displayquote}
\end{small}
{Figure \mbox{\ref{visuals}} contains three examples of fact-checks  leveraging visual storytelling techniques. }

\end{itemize}

\subsubsection{Measuring engagement and updating strategies}
Measuring social media engagement is important to determine what content is gaining more traction and which engagement strategy is working. P17 informed us how they track click through rates  to determine how many people are engaging with the content by clicking on the links posted by them. 
\begin{small}
\begin{displayquote}
\enquote{[Bitly] can tell how many people clicked on a particular link, and we can track that.. [It allowed us to] analyze how people engage with these links, how many clicks come from the newsletter, as opposed to Facebook or Twitter or  Instagram? And then that allows us to  know  whether we need to change how we present the information. } - P17
\end{displayquote}
\end{small}

The engagement statistics help social media teams to update their engagement strategies. For example, P17 described how they changed the number of stories published in their weekly WhatsApp newsletter from five to three after realizing that public only clicked on the top  URLs.

\begin{small}
\begin{displayquote}
\enquote{Earlier editions [of WhatsApp newsletter] had up to 5 stories per edition. And we were seeing that people were only clicking the top 2 or 3 links, and were ignoring the rest. So a decision was made to reduce the {number} of stories that we feature from 5 to 3 to make it shorter.} - P17
\end{displayquote}
\end{small}





\section{Infrastructures Supporting Long-term Advocacy Centric  Fact-checking } \label{labor-long-term}
 \textit{Long-term advocacy centric} fact-checking aims at improving the information landscape by conducting research about various aspects of online disinformation,  influencing policies surrounding availability and quality of  data and statistics,  conducting educational training for aspiring fact-checkers,  organizing literacy campaigns for  general public and forming coalitions with various fact-checking organizations and internet companies. This type of fact-checking is supported by two stakeholder groups---\textit{investigators and researchers} who  conduct long-term investigative projects that include data and network analysis,  and \textit{advocators} who are involved in policy and advocacy work. In this section, we elaborate on the role of these stakeholder groups.

\subsection{Investigators and Researchers---Conducting In-depth Research and Investigation} \label{inv}
Few fact-checking organizations (e.g. Full Fact, Code for Africa, etc.) have a separate team of investigators and researchers. Unlike fact-checkers who engage with individual pieces of misinformation, this stakeholder group conducts an in-depth  investigation of persistently circulating misinformation and disinformation campaigns via data and network analysis. 

\begin{small}
\begin{displayquote}
\enquote{Part of my work at the moment is creating or developing a framework for misinformation crises. So, as opposed to individual pieces of misinformation.. looking at when responses need to go over and above the day to day..everybody who works in the kind of anti misinformation space gets together and they normally introduce new responses and new policies to manage that.} - P9
\end{displayquote}
\end{small}

\begin{small}
\begin{displayquote}
\enquote{If.. we are seeing a certain type of misinformation occurring.. every single day, debunking each individual one will not help. So what they [fact-checking team] do is now they refer that case to us. And then..[we do an]  in depth investigation to try and see  where is this narrative originating from.. Data analytics team is the one that does the  sifting through the data sets that we obtained from social media.. Forensic team does profiling of key accounts..that we identified.} - P10
\end{displayquote}
\end{small}

\subsubsection{Conducting long term investigative projects}
Investigators and researchers undertake several investigative projects such as verifying the {backfire effect\footnote{https://fullfact.org/blog/2019/mar/does-backfire-effect-exist/} of fact-checking (when a claim aligns with a person's beliefs, proving that it is wrong will  make them believe it more strongly)}, studying long term effects of conspiracy theories, examining public engagement with political news, determining how to communicate fact checks effectively,  etc.

\begin{small}
\begin{displayquote}
\enquote{You've probably heard about the backfire effect which is kind of mythical idea that fact-checking.. does more damage than good. I think the original research was repeated and the same effect wasn't found.. We also did a project recently where we looked at.. how conspiracy theories affect people's beliefs in the long term.. who believes and shares misinformation, how to communicate your fact checks [while] presenting them to audience.} - P9
\end{displayquote}
\end{small}


\begin{small}
\begin{displayquote}
\enquote{Projects include doing investigations into Russian  disinformation or Russian influence in African countries.. investigating Chinese influence operations into .. African countries,.. human trafficking in four South African countries.} - P10
\end{displayquote}
\end{small}

\begin{small}
\begin{displayquote}
\enquote{There's been a lot of research that was done around.. how to design a fact check to make it more engaging, how do you phrase a headline,.. how much of this information can you put into like a video. What is the ideal video length? How to get more people to interact with them and then how to  try to make them much more long lasting and persistent in people's memory.} - P5
\end{displayquote}
\end{small}

These stakeholder groups analyze  misinformative content that is going viral on social media platforms and then trace it back to the social media accounts who started sharing it. They also conduct qualitative research by conducting surveys.

\begin{small}
\begin{displayquote}
\enquote{ {We have like a survey and we've been trying to find what kind of narratives against immigrants are more popular here in Spain. We're trying to prepare another big survey about how fact checking works in  Spain and what kind of debunk works better for us.
}} - P21
\end{displayquote}
\end{small}

{The results of the investigative projects are released as dossiers (e.g. \mbox{\cite{Presenta79}}, \mbox{\cite{Ugandain62}}). The dossier provides background of the issue investigated, data collection and analysis method(s), results of investigation, and conclusion. For example, in \mbox{\cite{Ugandain62}}, investigators study the misinformation influence operations that occurred in Uganda before the January
2021 elections. The dossier first briefly describes the political landscape in Uganda and  provides examples of misinformative tweets that acted as the starting point of investigation. These tweets leveraged trending hashtags such as   \#StopHooliganism,  \#UgandaIsBleeding \footnote{In November 2020, Robert Kyagulanyi, a presidential
candidate in Uganda was arrested on two separate occasions. The protests that broke out after his arrests were documented in Twitter posts containing hashtags such as    \#StopHooliganism,   \#UgandaIsBleeding, etc.}, etc.  to spread false narratives against the opposition party using past events from other countries. The dossier then deep dives into the methods used to collect and analyze relevant tweets and finally attributes the influence operation to the supporters of National Resistance Movement\footnote{https://en.wikipedia.org/wiki/National\_Resistance\_Movement}.}

 Investigators and researchers use  tools like python libraries for data analysis and gephi \cite{GephiThe94:online} for network visualization in addition to the tools used by fact-checkers.

\begin{small}
\begin{displayquote}
\enquote{We were actually using an open source tool called tweepy to collect tweets from Twitter.. The network analysis is [done] using a tool called gephi [which] will be able to show  relationships between tweets using the retweet function like which are the accounts that has been highly retweeted, which are the influential accounts within the network. } - P10
\end{displayquote}
\end{small}

\subsection{Advocators---Influencing Policy, Building Coalitions, Conducting Educational Workshops and Literacy Campaigns} \label{advocacy}


Several of our study participants revealed that fact-checking doesn't stop with  generation and distribution of fact-checks and is much more like a sustained campaign. It also includes influencing policymakers and information producers to improve the quality of information (and in turn the quality of fact-checking), building coalitions with other fact-checking organizations, social media companies,  and journalistic organizations as well as  providing fact-checking training to newsrooms and organizations. 
Such initiatives are led by advocators. 
This stakeholder group  identifies and realizes several ways to improve the  \textit{short-term claims centric} fact checking process. They steward multiple outreach programs, policy initiatives and advocacy projects locally and globally. All the initiatives started by these stakeholder groups could be considered actions that are performed via technological and informational infrastructures such as workshops, appeals and training programs. We present the tasks performed by this stakeholder group below.



\subsubsection{Creating new generation of fact-checkers and fact-checking organizations} The advocators  conduct training, workshops and fellowship programs for people and organizations all over the world, teach them nitty-gritty details of fact-checking along with how to setup and operate a fact-checking organization of their own.  

\begin{small}
\begin{displayquote}
\enquote{Ethiopia.. [is] a country where press freedom is very limited, online false information often leads to offline.. And most of the media there are state controlled. So we were conducting trainings on how they can set up fact checking desks and try to be independent.} - P2
\end{displayquote}
\end{small}

\begin{small}
\begin{displayquote}
\enquote{{In Germany, we only got two IFCN signatories. And that's not enough. So we try to convince traditional media outlets in the regions..to start with fact checking.. We are training them. [We have created] a community of fact checkers..and more than 600 journalists and we are doing training, encouraging them to start with fact checks.}} - P19
\end{displayquote}
\end{small}

 {Advocators conduct trainings through webinars or online platforms. Some organizations have set up their online learning platforms where they provide video tutorials to fact-checkers (e.g. \mbox{\cite{factify}}), while others have partnered with academic universities to conduct  educational trainings.}

\begin{small}
\begin{displayquote}
\enquote{{The advocacy campaigns.. on fact-checking, we conduct them virtually. We have partnerships with  Kenyan institutions of higher learning, such as Daystar and Aga Khan university, which allows us to conduct webinars on fact-checking. Those sessions are attended by media professors and their students. 
We mainly use Jitsi, Slack and Google Meet to do the training.
}} - P2
\end{displayquote}
\end{small}

\begin{small}
\begin{displayquote}
\enquote{{We are proud of digital e learning platform we have built at DPA. we have a lot of videos here, where we explain how to work with the Internet Archive.
Those are webinars [where we] train maximum 15 people, zoom webinars means it's live training.
}} - P19
\end{displayquote}
\end{small}


\subsubsection{Pitching the importance of  evidence based decision making to policymakers} The advocators are also actively trying to reach their country's policymakers and civil society organizations, informing them about their work and bidding the importance of facts and evidence based decision making. For example, P12 attended parliamentary researchers' conference in Kenya where they  pitched the importance of facts in informing the country's policies and intervention programs. Similarly, P9's organization tried to get  parliamentary support to attribute electoral imprints   to election campaigners during referendum and elections in the United Kingdom.

\begin{small}
\begin{displayquote}
\enquote{Civil society organizations generate data that then is used by government to put in place  policy intervention, say  poverty eradication.. interventions in healthcare. So we just want them to understand that as you're doing this, you also have to check your facts, don't just rely on a news story or rely on a document or rely on a statement by a public politician to define your problem.. And so I.. tell them.. their job is to tell the leaders what the data says.
} - P12
\end{displayquote}
\end{small}

\begin{small}
\begin{displayquote}
\enquote{We sort of started a project, a couple of years ago about imprints. So in the UK if you distribute companion materials on paper, you have to say who is that from whereas online that's not the case. And during the referendum and.. election.. where certain information come from online, it wasn't attributed to campaigners in the same way that it would be offline. And so we tried to get some parliamentary support to change those rules.} - P9
\end{displayquote}
\end{small}

\subsubsection{Organizing literacy campaigns}
The advocators actively organize literacy campaigns educating people how to critically examine information that they find on online platforms. For example, P5 shared how they  partner with community radio stations, design MOOC \cite{MOOCorgM86:online} courses and frequently share tip sheets  explaining people what fake news is and how  they can identify it.

\begin{small}
\begin{displayquote}
\enquote{We try to also work with community radio stations .. and talk through what like what fake news is,.. talking about information literacy and information disorder and how it manifests} - P5
\end{displayquote}
\end{small}

\subsubsection{Spearheading initiatives to improve accessibility and quality of data and information}
Better decisions are made when better data is available. Quality data and information is essential because it acts as a source to verify facts in the \textit{short-term claims centric} fact-checking. The advocators are actively working to improve the code of practices in releasing data, for example, updating it from the paper to the digital age. 

\begin{small}
\begin{displayquote}
\enquote{Improving the code of practice for official statistics and updating it to the internet age. And we've done lots of individual pieces of work trying to improve specific statistical releases because obviously they form the basis of a lot of what we do.. we try and get ..the office for statistics regulation to be a bit bolder in how they treat misuse of statistics officially.} - P9
\end{displayquote}
\end{small}
Our interviews also revealed that data in most of the African countries is either old, or is not accessible. The advocators there are  engaging with government agencies, making them aware about the problem and stressing the importance of having data publicly accessible . 

\begin{small}
\begin{displayquote}
\enquote{Before the census in 2019, the last census had been done in 2009. So while there are estimates available on the census data, we would use those estimates. But then the data is just not accurate  when it's estimated.. [So we] talk to the people at the National Statistics office and say we would like this data. [We] talk to people at the ministry, and trade unions telling them that this is how it would be better if you track unemployment. 
} - P12
\end{displayquote}
\end{small}


\subsubsection{ Building collaborations and coalitions}
Several advocators stressed on the need for all fact-checking organizations to work together, collaborate, and share resources. Such collaborations would be helpful in understanding common challenges and needs of the fact-checkers.  The coalition would also better position the organizations while making certain demands from the internet companies. IFCN has played a huge part in forming such a collaboration and several advocators are actively working to expand this network. 
\begin{small}
\begin{displayquote}
\enquote{There's not a lot of us working together. So that's what I'm working on in collaboration with  IFCN.. how different organizations and sectors can work better together to complement each other and collaborate and kind of information sharing and resources.., [understand]  common challenges. For example, when we work with internet companies, there are certain things we might all want to ask them for which, at the moment, [only] some of us asking.} - P9
\end{displayquote}
\end{small}




In addition to a coalition among themselves, few fact-checking organizations are also actively partnering with companies like Google and Facebook to fight fake news on their platform \cite{engadget} ``\textit{because they have a huge impact on the way people experience misinformation} (P23)''. P9 and P12 informed us how Google is  working with their organizations to make their fact-checks more visible by ranking them higher in Google search. One of our study participant (P18) was in-fact instrumental in starting the initiative.

\begin{small}
\begin{displayquote}
\enquote{When we want what our fact checks to rank higher or to be more visible, so there is a back end tool claim review that we use that is integrated into WordPress.. And so  the search engine looks at whatever you post as a fact check and then it ranks it higher in the matrix.} - P12
\end{displayquote}
\end{small}

\section{Needs and Challenges of Stakeholder Groups} \label{challenges}
In this section we answer RQ2 by presenting the challenges faced by various stakeholder groups categorized by emerging themes. 
\subsection{Skepticism Towards AI and Automation}

{Fact-checkers expressed skepticism and distrust towards artificial intelligence (AI) and automation of the fact-checking process.} {This finding resonates with  prior work that revealed that the black box nature of artificial intelligence techniques and machine learning algorithms make their inner workings unintelligible to humans thereby decreasing users' trust in their outputs \mbox{\cite{schmidt2020transparency,paez2019pragmatic}}}. P4 divulged that they are skeptical of Facebook's AI-based tool that aggregates potentially misinformative content for fact-checkers to verify and rate on its platform because they think that  ``\textit{algorithms hide a lot of stuff}''. {They believe that fact-checkers or independent organizations should be responsible for aggregating  content to fact-check on the social media platforms rather than ``\textit{[companies] that are running the platforms}.''}

 Fact-checkers understand that machine learning models work when there are lots of similar data for training and pattern recognition. {Thus, P18 doubts AI's capability to detect falsehood in politicians' statements which can be very diverse in language and topics. They also doubt AI's capability to  differentiate between a true and a false statement especially when there are only subtle differences between the two. P19 does not believe that fact-checking could be automated since its a complicated process that requires several humans to discuss and make decisions.} 
 

\begin{small}
\begin{displayquote}
\enquote{
I think [AI based tools are] going to be less useful for most.. politicians, because the problem is people don't repeat stuff the same way  and the addition of a word or two can make a huge difference.. I don't think a computer is ever going to be able to figure that out. } - P18
\end{displayquote}
\end{small}

\begin{small}
\begin{displayquote}
\enquote{{[Fact-checking is] such a complex process, for example, extracting a claim,- what's the underlying meaning of a certain claim, how to understand it. Even [for humans] it's a process of discussing and then deciding. So I think it still will be humans work in a way. }} - P19
\end{displayquote}
\end{small}

{Despite the skepticism towards AI,  we found that some fact-checking organizations have indeed adopted AI based tools and  a few others showed a willingness to adopt such tools. However, such tools are only acceptable for low stake scenarios of  monitoring content on social media as compared to high stake scenarios of assigning a veracity label to the content. This observation is inline with recent work that found that the use of AI is more acceptable in low stake compared to high stake decision making processes \mbox{\cite{ashoori2019ai,scharowski2020transparency}}.} 

\begin{small}
\begin{displayquote}
\enquote{{I will not trust any algorithm or any AI  to flag something as right or wrong, at least not at this stage.. I prefer AI only to give us a curated list, flag us that this is something that we should look in. Make a tool that picks up the signals that [indicate the content is] misleading and makes a curated list.  }} - P22
\end{displayquote}
\end{small}

\begin{small}
\begin{displayquote}
\enquote{{If you're serious with fact-checking, you cannot replace it by automatization or things like that. But it could be helpful in [social-media] monitoring.  }} - P19
\end{displayquote}
\end{small}

{To increase fact-checkers acceptance of AI based tools, P16 and P12  stressed} how its essential to have humans in the fact-checking loop. P12 further said how they would only trust the AI output if the tool is able to explain how it arrived at a particular conclusion. {Past studies have  also developed human-in the loop AI systems  \mbox{\cite{smith2018closing,xin2018accelerating}} and
tried to make them more explainable in order to foster trust in them \mbox{\cite{linardatos2021explainable,abdul2018trends,schmidt2020transparency,miller2019explanation,danilevsky-etal-2020-survey}}}. 

    \begin{small}
\begin{displayquote}
\enquote{I think at a certain point, AI stops and human being needs to come in to verify.. I don't think that AI can really do exactly the same what we can do. Not yet at least.} - P16
\end{displayquote}
\end{small}

    \begin{small}
\begin{displayquote}
\enquote{
The manual process would still have [to be there].
I would be willing to use it [automated tool] to see how it arrives at that conclusion. So if you say this is misinformation, and these are the sources of data that we're using to make that. And we check and find that the algorithm.. is not using them out of context. So at that point, we would be  in a position to say let's check it.
} - P12
\end{displayquote}
\end{small}

\subsection{Need For Tools and Limiting Social Media Affordances} 
We found stakeholder groups divulging several challenges related to the tools they used and the affordances provided by social media platforms. In the process, we also discovered a few needs with respect to tools and systems. We present them next.

    \subsubsection{Monitoring Social Media Platforms is Manual, Time Consuming and Difficult.} Fact checkers complained about the information overload problem. With the emergence of a variety of social media platforms, the amount of online information is increasing exponentially. However, for the fact-checkers, searching for misleading content mostly remains a manual task. In addition, generation of search queries that could lead to potentially  dubious content is still based on hit and trial method. {As discussed in Section \mbox{\ref{fact-checkers-role}}}, only Facebook has provided a tool to its partner fact-checking organizations that can aggregate potential misinformation. {While a number of social media monitoring tools  have been developed to identify and aggregate misinformation, particularly on Twitter (such as \mbox{\cite{tolmie2017supporting,alrubaian2016credibility,hunt2020monitoring, shao2016hoaxy,cerone2020watch,resnick2014rumorlens}, etc.)}, other social media platforms (e.g. YouTube, Google, Yandex, etc.) where misinformation is equally prolific also need attention. }

    
 
    \begin{small}
\begin{displayquote}
\enquote{The search is not easy to be honest.. it takes a lot of time to actually find the content because we only have a manual method to do that, hit n trial method to do that.. I have even scrolled to the point where YouTube shows me no more results. So that's how manual it gets.} - P7
\end{displayquote}
\end{small} 

    
    \begin{small}
\begin{displayquote}
\enquote{If there's some.. tool that helps filter..different types of misinformation will help because now you have  information overload and you don't know what to choose and what not to choose.} - P3
\end{displayquote}
\end{small}


    \subsubsection{Limitations of platform affordances.} Many fact-checkers complained how online social media platforms' affordances become a hindrance for searching and filtering content. For example,  inability to search for posts and comments on Facebook, lack of search trends feature on platforms (with an exception of Twitter), unavailability of fine grained search filters, inability to download content on platforms like Instagram and inability to search for same videos that were uploaded with different keywords (title, description, etc.) on YouTube are some of the limitations of platform affordances.
    
    \begin{small}
\begin{displayquote}
\enquote{Facebook is actually tricky to be honest, because  there's no one way to search content. you can only search people..pages and groups, etc.. but I need user generated content.} - P7
\end{displayquote}
\end{small}




    \begin{small}
\begin{displayquote}
\enquote{Youtube search engine is not very good.  It will just show you only videos which are popular which has no use.
Misleading videos won't have too much views, but they have a lot of uploads. What happens is that since a lot of people.. upload videos using different caption, different keywords. So.. [there might be] 10 versions of the same thing. } - P8
\end{displayquote}
\end{small}

     \subsubsection{Systems and tools needed to detect misleading claims on private message platforms. } Privacy settings on social media groups (e.g. Facebook) and 
     end-to-end encryption in messaging platforms (e.g. WhatsApp) act as a hindrance in accessing misleading content circulating on these platforms. The fact-checkers {informed that they} need tools that allow them to access and flag content on these platforms. {Recent research work has focused on building crowd sourced WhatsApp tip lines for discovering content to fact-check \mbox{\cite{Melo_Messias_Resende_Garimella_Almeida_Benevenuto_2019,kazemi2021tiplines}}. However, being able to track or report private messages via tools necessitates serious ethical considerations.  Private messaging platforms give users a false sense of security, thereby making them share sensitive information (e.g.  clinical records of patients  \mbox{\cite{mars2016whatsapp}}), without anonymizing it \mbox{\cite{guerra2021use}} and hence, expose users to privacy risks  \mbox{\cite{article_whatsapp}.}}



\subsubsection{Overload of tools.} During our interviews, fact-checkers showed and talked about several tools that they use  during fact-checking. One of the participants P8 showed us around 15 tools. There is a tool overload problem.  {P3 suggested building a single tool that could provide all functionalities that fact-checkers need.  However, since fact-checking is a complex task involving multiple steps, it is difficult for a single tool to cater to a divergent set of requirements and functionalities \mbox{\cite{tolmie2017supporting}}. Thus, building a suite of purpose-built tools that cater to specific needs of fact-checkers in the various steps involved in fact-checking would be more useful for the fact-checkers.}


         \begin{small}
\begin{displayquote}
\enquote{[If one could] put all these  tools in one, .. So I don't have to look for different tools when I'm analyzing a video or an audio. I can do it in one place instead. Like.. I can use 10 tools to analyze the video. But if we can have one..[tool that] gives you the information you need.} - P3
\end{displayquote}
\end{small} 

\subsubsection{Need for specific tools}  Despite the problem with tools overload, existing tools lack  task-specific functionalities. For instance, a lot of steps involved in video fact-checking are manual. While tools like Invid and reverse image search are available to verify a video, they are only useful to check if the video is digitally altered or used in a different context than the original. For all other videos (e.g. videos with conspiracy theories), claim extraction and claim verification process is manual. For lengthier videos, this process could get even more tedious.



\begin{small}
\begin{displayquote}
\enquote{There is no tool  to debunk [conspiracy theory videos]. For example, I cannot do a reverse image search, I cannot divide the video into key frames, because it is a narrative that is false.. We have to really go line by line and see what the person is saying, and then all we can do is search quotes from different organizations to [verify the claim] .} - P7
\end{displayquote}
\end{small} 

Fact-checkers believe that {``\textit{efforts against misinformation advance  much more and much quicker in English'' (P23)} than in other regional languages.} They need tools to transcribe  videos in local regional languages. 

\begin{small}
\begin{displayquote}
\enquote{The problem with transcribing videos online or using any other software is  languages because India has so many languages and we tend to get a video and information in every possible language. So it becomes difficult to have one dedicated tool to transcribe all our videos.} - P14
\end{displayquote}
\end{small} 

 

    

Editors revealed that editorial work is mostly manual. 
P1 spoke about  how editing work is 
``\textit{still very human}'' and the human resource management tools like Trello could be further strengthened. For example, currently, ``t\textit{here are a lot of issues of accountability}'' with the Trello board, it lacks control features because of which ``\textit{anybody can move a card anywhere}.''


\subsubsection{Getting organic engagement for fact-checks without advertisements}
P17 revealed that 
while on some platforms (e.g. Twitter) it is easier to \textit{``get organic reach and engagement''} by adopting appropriate strategies, it is 
\textit{``a big challenge to get [same].. attention without advertising''}  on other platforms (e.g. Facebook). 


\subsection{Issues around policy and information infrastructure} Stakeholder groups discussed several challenges surrounding information availability and quality. They understand that quality data is essential not only for developing AI based automated tools but also for investigating claims.


\subsubsection{Need to improve information quality before automation}    
  AI models are as good as the data available. If the information against which the claims are to be verified is missing or of low quality, the models would never work.  

        \begin{small}
\begin{displayquote}
\enquote{Automation is tricky because.. in a  place like East Africa ..information is not readily available. You cannot say that I'll go to this site and get this information so that when these numbers are presented it can easily be automated.} - P12
\end{displayquote}
\end{small}

P9 raised an interesting point along the same lines. They explained how success of automated fact-checking is dependent on the accessibility and format of statistics and information available. The data has to be in the same format for the machine to be able to understand it. Their organization is working with institutes around the world to improve statistics globally. 

\begin{small}
\begin{displayquote}
\enquote{[We want] statistics being published in a kind of open accessible and consistent format.  So, for example, like any symbols that are used to show a caveat about data needs to be the same so that our machine can understand them each time.. My colleague.. is working with the open data Institute in UK and  also globally.. to [determine] whether there are ways of improving statistics so that automated fact-checking can work.} - P9
\end{displayquote}
\end{small}

{P22 elicited how  data in their country is not available in a user friendly  or machine usable format. }

\begin{small}
\begin{displayquote}
\enquote{{The data, you will find that it is in a PDF... file or a photograph on the website..then how do we use it?...You need the data ..[to be] put into a excel sheet so  you can clean it.  Data should be in a format which is user friendly so it can be used, [it should be] machine learning friendly so that the machine can pick up and use the data.
}} - P22
\end{displayquote}
\end{small}

{The aforementioned concerns elicited by fact-checkers are in line with the prior research that has also listed absence of structured and quality data as well as lack of adherence to data standards, as few of the major challenges faced in the field of big data \mbox{\cite{adnan2019information,desouza2014big}}. Lack of harmonization between data sets makes data integration a complicated and time consuming task \mbox{\cite{3Challen73:online,Challeng38:online}}. Data integration is necessary in various scenarios related to fact-checking \mbox{\cite{leblay2018computational}}, for example, querying different databases originating from different sources to determine the veracity of content or determining whether a piece of content has already been fact-checked by searching in various fact-checking databases \mbox{\cite{leblay2018computational,cazalens2018content}}.  Thus, scholars have stressed on the need to adhere to common data standards to help facilitate data integration and reuse \mbox{\cite{10.1145/2479724.2479730}}. There have been a number of cross country initiatives to set common standards for data in various domains. For example, in 2017, European Medicines Agency held a meeting to discuss the opportunities and challenges in applying a common data model to healthcare data across the countries in Europe to support regulatory decision-making
\mbox{\cite{ACommonD38:online,Acommond78:online}}. Furthermore, several advocators (Section \mbox{\ref{advocacy}}) have also been spearheading initiatives to improve the data quality in their respective countries.}

\subsubsection{Lack of information sources.} \label{lack_of_info}{Our interviews with fact-checkers in the Global South  revealed that the information needed to investigate claims  either does not exist or is not updated periodically. }

\begin{small}
\begin{displayquote}
\enquote{{There's a lot that we don't cover because of lack of sources.}} - P2
\end{displayquote}
\end{small}

\begin{small}
\begin{displayquote}
\enquote{{In Kenya,.. demographic Health Survey, which .. shows the health situation in the country, the last one was done in 2014, this is 2020.  It's just too old, and we can't use it.}} - P12
\end{displayquote}
\end{small} 

\subsubsection{Difficulty in getting information for research from civic organizations.} In most cases, fact checkers need multiple sources to debunk a claim. To obtain these resources they rely on public data sets and information from government and civic organizations. Six fact-checkers working in African countries and the Balkan regions informed us how  information needed for research is not publicly available and  getting it from officials is a long and difficult process.

\begin{small}
\begin{displayquote}
\enquote{It's so hard to get information from the government.. because everybody wants to protect themselves. They don't want to give you the information that you actually need.} - P3
\end{displayquote}
\end{small} 

\subsubsection{Algorithmic bias against local content}
P5 complained about algorithmic bias in terms of how content indexed in search engines is skewed towards the Global North region making it very difficult to search for local content in regional languages of the Global South.  
    
    \begin{small}
\begin{displayquote}
\enquote{The way the algorithms works was very I'd say Euro-centric or like North America [centric].. Trying to find tools that would enable me to find.. stuff that's not in English, like things in local languages was quite challenging.. it would take a while for some of the stuff from from this side to be indexed on.. Google searches and other platforms .. So, there's sort of algorithmic bias when it comes to.. find stuff like that.} - P5
\end{displayquote}
\end{small}

\subsection{Emotional cost of fact-checking}
In addition to the manual labor involved in fact-checking,  fact-checkers also face significant  emotional toll and stress in their job. They are often victims of online threats and abuse from users and conspiracy theorists whose posts they were tasked to debunk. Manually scanning through misinformative content about certain topics, such as riots, conspiracy theories, also has adverse affects on their mental health.

 \begin{small}
\begin{displayquote}
\enquote{It becomes kind of very stressful job. Seeing all  violence and getting into each and every detail, kind of takes a toll on your mental health. And then again you have to listen to abuses. And the situation is worse when you like put this content online and then people attack you} - P8
\end{displayquote}
\end{small}

 \begin{small}
\begin{displayquote}
\enquote{We are exposed to threats, these conspiracy theorists are very aggressive and I worked only for like a couple weeks when I saw a CrowdTangle post with my own photo saying, this is your censor and  it was very unpleasant feeling to find that.} - P16
\end{displayquote}
\end{small}

While previous work has studied emotional labor and  psychological symptomatology in content moderation work \cite{roberts2014behind,kerr2015recruitment,arsht2018human,steiger2021psychological,dosono2019moderation}, {no study has investigated the emotional cost of fact-checking work. Studying human costs underlying the fact-checking process and determining wellness interventions to psychological effects of fact-checking are another fruitful avenues for future research. }

\section{Discussion} \label{sec:discussion}
{In this section, we discuss how our study renders visibility to the human and technological infrastructures supporting the fact-checking work and the collaborative efforts involved in the process. We also discuss the  needs of the stakeholder groups and the implications of our findings  on the future research directions in fact-checking.}

\subsection{Rendering visibility to the human infrastructure of fact-checking}
Till date the primary objective of fact-checking is considered  as   debunking misleading claims. Based on this objective,  prior work suggests that fact-checking can only influence three constituencies--- people, journalists and political operatives \mbox{\cite{amazeen2013critical}}. By identifying and examining the human infrastructure---{the stakeholder groups that need to be brought into alignment to accomplish fact-checking}, our work 
provides a means to think about  fact-checking as a multidimensional initiative which then assists in understanding 1) the invisible aspects of the process, and 2) the other ways fact-checking process can have an influence. Our work shows that fact-checking is supported by several processes, such as editorial work, social media engagement work, in-depth research and data analysis, as well as advocacy and policy work that might not be visible to the external world. Through the study of these processes, 
we establish how  fact-checking has evolved to include both \textit{short-term claims centric}  and \textit{long-term advocacy centric} fact-checking. 
{We make visible the efforts that fact-checking advocators are putting 
in to improve the availability, accessibility and quality of data and statistics by aligning the focus and interests of governments and internet companies with fact-checking organizations.}  Through these efforts, the fact-checking organizations are not only improving the information landscape of their country but in turn are also improving the quality of (\textit{short-term claims centric}) fact-checking itself.  Rendering visibility to the work of the human infrastructure and the invisible processes of fact-checking has helped in uncovering the needs---both social and technical---of the entire fact-checking ecosystem consisting of all the stakeholder groups. The knowledge of needs of the ecosystem could  enable design and development of tools and policies  to support various aspects of the  fact-checking process. 

\subsection{Collaborative efforts in the fact-checking process}


{Our study highlights  fact-checking as a distributed problem where collaboration takes place at multiple stages among  people with different skill sets within and outside the fact-checking team/organization. First,  collaboration occurs among the  stakeholder groups: editors, fact-checkers, social media managers, researchers \& investigators, and advocators (refer Figure \mbox{\ref{factcheckdiag}} for an overview).  Second,  collaboration  extends to the outside world with experts such as doctors, oncologists, academics, etc. whose expertise is needed to investigate dubious claims. Third, fact-checkers collaborate with civic and government organizations  during the investigative stage to access data and statistics related to the claims under investigation.  Fourth, in parallel, advocates reach out to policymakers to influence policy by highlighting the challenges faced by fact-checkers. Fifth, several internet companies, like Facebook collaborate with  fact-checking organizations to fact-check and debunk misleading claims on their platform. Finally, collaboration occurs between social media users and the stakeholder groups at two stages: 1) at the content monitoring stage where users report  dubious claims that they encountered online directly to fact-checkers via tip-lines, and 2)  when users engage with fact-checks disseminated by social media managers.}
Rendering visibility to the collaborative efforts in the fact-checking process has multiple  benefits. We discuss a few.

\subsubsection{Increase in efforts to foster collaborations} Making visible the collaborative efforts in the fact-checking ecosystem  can  lead to efforts and policies that foster these collaborations. There has been growing research  on  how to make users  engage with the published fact-checks \mbox{\cite{amazeen2019reinforcing,from2020communicating,young2018fact,ecker2020effectiveness,chen2021citizens}}. 
For example, fact-checking organization Pesacheck  created a Twitter bot named \textit{debunk bot} that detects tweets  containing  URL(s) to misinformative content and replies to them with the link to the fact-check that their organization has published 
\mbox{\cite{DebunkBo89:online}}.
Similar investigations and efforts can be put to support other collaborations, such as between experts and fact-checkers.  There has been only a handful of recent efforts in this direction, for example, 
Meedan's Digital Health Lab\footnote{\url{https://meedan.com/digital-health-lab}} and  Facebook's Journalism project\footnote{\url{https://www.facebook.com/journalismproject/facebook-partners-with-meedan-digital-health-lab-to-help-fact-checkers-debunk-health-misinformation?locale=pa_IN}}  support fact-checkers in debunking health related misinformation by connecting them with health experts and providing them with  resources on the health topics that they are covering. Imagine a private social media platform  consisting of separate communities (like subreddits) of fact-checkers, and experts from different fields (from doctors, journalists, meteorologists, to university librarians and professors) to facilitate easy, and targeted communication and information sharing. Fact-checkers can post questions in relevant communities, seek quotes from experts, and get suggestions for online and offline resources  to support their investigations. Fact-checkers can also share with each other their concerns or information about new tools that they discovered. Such a platform could facilitate fact-checking organizations in addressing online misinformation
effectively and in a timely manner.

\subsubsection{Revealing the value of fact-checking work to internet companies} In recent times several platforms such as Google search, YouTube, and Google images have started actively using fact-checks produced by
fact-checking organizations with their search results to help determine their validity and truthfulness  \mbox{\cite{GoogleFa29:online,Seefactc11:online,Bringing1:online}}. Our study also  revealed how fact-checking organizations are collaborating with internet companies and allowing them to index their fact-checks. This finding contributes towards the HCI scholars call of making people aware of ``the value their data brings
to intelligent technologies'' \mbox{\cite{HCIandth97:online,vincent2019measuring}}. Our work highlights  the value that fact-checking is bringing to the  social media companies who regularly  use the fact-checks to regulate the content on their platforms and provide reliable information to the users.  Previous scholarly works have raised questions on whether volunteer  created content such as
Wikipedia articles should receive more economic benefits from the internet companies \mbox{\cite{vincent2019measuring,vincent2019data}}. On similar lines, we want to raise the question  whether fact-checkers and fact-checking organizations should also receive more economic benefits for their work.


 
 

\subsubsection{Revealing power dynamics in collaborations} Our study also sheds light on  the power dynamics of  collaborations happening in the  fact-checking ecosystem. For example, our work shows how editors  have the power over fact-checkers in determining what kinds of misinformative claims should be prioritized. Additionally, in certain situations, fact-checkers depend on civic and government organizations to get data for investigation. Tools released by social media companies, such as Facebook Queue (refer Section \ref{fbq}) also dictate what kind of claims fact-checkers investigate. These companies also have the power to remove the fact-checks from their platform. For example, Facebook removed a fact-check on abortion after receiving complaints from Republican senators \mbox{\cite{Facebook19:online}}.  Further investigation of the
potential consequences of these power dynamics in the  fact-checking process is a fruitful avenue of future research.

\subsection{Implications for future research on fact-checking} \label{dissc2}
{Taking a multi-stakeholder perspective on the fact-checking process helped us learn the  needs of stakeholder groups as well as uncover the challenges that go beyond the technical aspects of fact-checking. We use our findings to discuss and propose various directions that the future research on fact-checking can take. In Section \mbox{\ref{tech}}, we start by discussing the  needs of various stakeholder groups and propose  solutions for the same. 
In Section \mbox{\ref{trust}}, we discuss 
the values that the stakeholder groups desire in the tools and systems built for them. Next, in Section \mbox{\ref{dissc2}}, we discuss how focusing on technical solutions alone is not enough and how the existing automated approaches fail to work in real life since they ignore the social aspect of fact-checking.  We also reflect on the social and civic challenges faced by fact-checking organizations by discussing the role of information infrastructure  on fact-checking. Finally, in Section \mbox{\ref{local}}, we end by stressing how the current research on  misinformation has not focused on the Global South countries and how  there is a dearth of fact-checking tools built for regional languages of the Global South.  }

\subsubsection{Technical needs of fact-checking}\label{tech}

Our study reveals that monitoring social media platforms is the most challenging aspect in the job of a fact-checker. A combination of over reliance on third party tools to discover potentially dubious content, limiting platform affordances, and a manual way of going through each search result to determine if its potentially misleading makes the process extremely tedious. Access to better search filters on social media platforms, and a community-based approach to reporting misinformation  where users of social media platforms are able to report problematic content
are some useful ways to assist fact-checkers in finding misleading claims.

Fact-checkers also agreed that its impossible for them to have access to all corners of the web. For example, they do not have access to content on private messaging platforms  like WhatsApp that have become a popular haven for groups interested in sharing misinformation \mbox{\cite{almeida2019misinformation}}. 
Given fact-checkers' skepticism towards AI and automation, user reporting via tip lines appears to be a  feasible solution to access content on such platforms.
Research on  how to motivate people to report problematic content is a fruitful avenue of future research. 
We also found that fact-checkers end up viewing long videos to extract misleading claims and reading through lengthy comments section to  get clues that would help them investigate the claims. A system that
{could utilize comments to highlight 
all the misleading claims present in the video, leaving the final decision of selecting what claims to verify to fact-checkers could be useful for the fact-checking community. Furthermore, a tool that  highlights credibility indicators (such as  comments containing useful information for investigating the claims) would significantly reduce the manual effort put in by the fact-checkers.
}

Our interviews revealed that when it comes to assigning a verdict to a claim
there is no single standard labelling system shared across fact-checking organizations.
Collective agreement on the labels could allow for greater information sharing and interoperability. 
Future efforts can focus on developing  structured ontologies for representing credibility assessment metrics that would lead to development of common labels and benchmarks for assigning veracity labels to the claims.

Our work also sheds light on the needs of stakeholder groups other than fact-checkers. There are lack of editorial and process management tools that could be used by news desk and copy editors. Social media managers need effective strategies to increase users'  engagement with the fact-checks. In order for fact-checks to really have an impact, they must be ``seen and attended to
by audiences'' \mbox{\cite{amazeen2019reinforcing}}. To accomplish this, it is essential to understand who shares fact-checks on social media platforms and what modality or visual storytelling technique is more suited for which platform. 
Tools are also needed to convert
fact-checks  to multiple languages to increase their reach. Platforms can also help in making the fact-checks more visible and accessible.
  Google's efforts (e.g. claim review) to prioritize the ranks of fact-checks in searches is one such effort. 
Technology critics  
have called structured journalism, where fact-checks  are produced in a machine readable form, as the future of fact-checking \mbox{\cite{structurejournalism}}. Recent work has tried to automate extraction of structured information---claim, claimant, and verdict from the fact-check, to allow  search engines to display it in the search results \mbox{\cite{jiang2020factoring}}. More such efforts towards structured journalism are needed to integrate fact-checks with online content. 
\subsubsection{Values desired in fact-checking tools and systems}
\label{trust}{
Our study participants  expressed  skepticism in automation and AI technology because of its black-box nature. At the same time, they also showed willingness to adopt automated solutions for low stake tasks.  Fact-checkers do not want systems that decide the veracity of information, rather they want tools that could help them with their day-to-day tasks such as monitoring online platforms, checking whether a video is digitally altered, transcribing videos in regional languages, etc.  Algorithm explainability combined with tools that have humans in the loop emerged as key values that fact-checkers desire in the systems built for them.
Familiarity with the inner workings of  algorithms along with tools that use both human and machine  capabilities for problem solving can help increase fact-checkers' trust in the automated systems.  Recent times have witnessed burgeoning interest in the field of human-centered XAI  where researchers draw from formal HCI theories to design explanations on how machines reach a particular decision \mbox{\cite{10.1145/3173574.3174156,8466590,arya2019one}}. Scholars are also studying how to design human and machine  configurations to operationalize human-in-the-loop systems \mbox{\cite{GRONSUND2020101614,xin2018accelerating}}.
Understanding specific needs for explainability with respect to fact-checkers, operationalizing those needs at the conceptual and methodological levels in tools developed for fact-checking, designing systems  that could take fact-checker's feedback and use that to modify the algorithm used by the system are few useful directions for future research.}


\subsubsection{Going beyond the technical: need for socio-technical solutions}\label{dissc2}
Would technology mediated solutions alone lead to improvement in the quality of fact-checking? While automating the entire fact-checking process and developing new tools to increase efficiency and scalability seems promising, it is not a panacea to all the problems faced by the stakeholder groups. A holistic change could only be achieved via systematic changes in the civic, political, and informational contexts. Through our study we found how accessing data from the civic and government bodies is a difficult task, especially if it portrays the government in a less-than-favorable light. 
While fact-checking is now a global endeavor, in some countries information  is either not publicly available or essential information (e.g. census data, health surveys, etc.) is outdated because of lack of periodic collection.  Our interviews revealed how several claims are left unchecked because of lack of sources. The availability of high quality up-to date information is essential for fact-checking---manual or automatic. Additionally, good quality data is precursor to having good machine learning models \cite{sessions2006effects} that would be needed to automate the investigative step in the fact-checking pipeline where  publicly available authoritative statistics are used to determine the veracity of claims. Thus, as part of \textit{long-term advocacy centric fact-checking}, advocators within the fact-checking organizations have been actively pushing for policy changes to improve the availability and quality of data and statistics. For example,  fact-checking organization Full Fact gave oral evidence to the House of Commons Public Administration and Constitutional Affairs Committee on issues of coherent and accessible health statistics in the United Kingdom \mbox{\cite{ukgov}}. {There is a growing interest in the HCI community to engage with policymakers as a way to inform policy that could benefit society \mbox{\cite{10.1145/2807916,10.1145/3290605.3300314}}.  Future work in fact-checking could focus on understanding  the opportunities and difficulties that advocators face while engaging with  civic organizations  and determining strategies  that advocators can  adopt to shape policies surrounding statistics and public data in their respective countries.}

\subsubsection{Fact-checking in the Global South}\label{local}
Recent work in the CSCW community stressed on the fact that academic research, till date, has primarily focused on misinformation in Western countries, while not addressing the phenomenon in the Global South \cite{haque2020combating}. Our study reiterates the lack of  knowledge and  context surrounding the misinformation landscape in the Global South. 
Local regional languages are under resourced both by the online platforms and search engines making it difficult for fact-checkers to gain access to local context online and for regional speakers to gain access to reliable information \cite{local_context2}. 
{The current  design-based approaches to fact-checking do not take into account the lack of fact-checking resources in regional languages  and thus, fail to account for the unevenness of the viability of fact-checking across the globe.}
Advocators recognize how access to local community-specific knowledge and culture is essential to understand the characteristics of misinformation and why it spreads in a particular region \cite{local_context1}. Thus, they are calling attention to the need to improve the information infrastructure and research in the Global South. {Our work also acts as call to action for researchers to study the misinformation landscape in the Global South region. }

\section{Conclusions and Limitations}
Our work sheds light on  how fact-checking is practised in the real world by presenting the infrastructures---both human and technological---supporting the fact-checking work. We interviewed 26 participants belonging to six primary stakeholder groups involved in the fact-checking process namely editors, external fact-checkers, in-house fact-checkers, investigators and researchers, social media managers, and advocators. By studying the various tasks performed by these stakeholder groups, we identified the role of tools, technology and policy in their work. Finally, we also identified key challenges faced by the stakeholder groups along with opportunities of advancing current tools, policies and technology for fact-checkers. 

Our work is not without limitations. 
The majority of the organizations that we interviewed are either IFCN signatories or work closely with IFCN signatories. 
Thus, the fact-checking model and tools used by the stakeholders presented by our study might not be applicable to every fact-checking organization.
The fact-checking process could also be subjected to various degrees of local and regional variabilities that our work does not capture. 
 We also acknowledge that not all stakeholder roles are present in every fact-checking organization that we interviewed.  For example, not every  organization is doing long term investigative  or advocacy work. We leave the examination of the factors influencing the fact-checking work in different regions and the variabilities in fact-checking work across regions to future work.
\bibliographystyle{ACM-Reference-Format}
\bibliography{sample-base}


\begin{thebibliography}{179}


\ifx \showCODEN    \undefined \def \showCODEN     #1{\unskip}     \fi
\ifx \showDOI      \undefined \def \showDOI       #1{#1}\fi
\ifx \showISBNx    \undefined \def \showISBNx     #1{\unskip}     \fi
\ifx \showISBNxiii \undefined \def \showISBNxiii  #1{\unskip}     \fi
\ifx \showISSN     \undefined \def \showISSN      #1{\unskip}     \fi
\ifx \showLCCN     \undefined \def \showLCCN      #1{\unskip}     \fi
\ifx \shownote     \undefined \def \shownote      #1{#1}          \fi
\ifx \showarticletitle \undefined \def \showarticletitle #1{#1}   \fi
\ifx \showURL      \undefined \def \showURL       {\relax}        \fi
\providecommand\bibfield[2]{#2}
\providecommand\bibinfo[2]{#2}
\providecommand\natexlab[1]{#1}
\providecommand\showeprint[2][]{arXiv:#2}

\bibitem[\protect\citeauthoryear{??}{3Ch}{2021}]%
        {3Challen73:online}
 \bibinfo{year}{2021}\natexlab{}.
\newblock \bibinfo{title}{3 Challenges of Integrating Heterogeneous Data
  Sources - DZone Integration}.
\newblock
  \bibinfo{howpublished}{\url{https://dzone.com/articles/3-challenges-of-integrating-heterogeneous-data-sou}}.
\newblock
\newblock
\shownote{(Accessed on 08/03/2021).}


\bibitem[\protect\citeauthoryear{??}{Cha}{2021}]%
        {Challeng38:online}
 \bibinfo{year}{2021}\natexlab{}.
\newblock \bibinfo{title}{Challenges of Integrating Heterogeneous Data Sources
  - DATAVERSITY}.
\newblock
  \bibinfo{howpublished}{\url{https://www.dataversity.net/challenges-of-integrating-heterogeneous-data-sources/}}.
\newblock
\newblock
\shownote{(Accessed on 08/03/2021).}


\bibitem[\protect\citeauthoryear{??}{ACo}{2021}]%
        {ACommonD38:online}
 \bibinfo{year}{2021}\natexlab{}.
\newblock \bibinfo{title}{A Common Data Model for Europe? - Why? Which? How? -
  workshop report}.
\newblock
  \bibinfo{howpublished}{\url{https://www.ema.europa.eu/en/documents/report/common-data-model-europe-why-which-how-workshop-report_en.pdf}}.
\newblock
\newblock
\shownote{(Accessed on 08/03/2021).}


\bibitem[\protect\citeauthoryear{??}{Aco}{2021}]%
        {Acommond78:online}
 \bibinfo{year}{2021}\natexlab{}.
\newblock \bibinfo{title}{A common data model in Europe? – Why? Which? How? |
  European Medicines Agency}.
\newblock
  \bibinfo{howpublished}{\url{https://www.ema.europa.eu/en/events/common-data-model-europe-why-which-how}}.
\newblock
\newblock
\shownote{(Accessed on 08/03/2021).}


\bibitem[\protect\citeauthoryear{??}{DER}{2021}]%
        {DERSPIEG52:online}
 \bibinfo{year}{2021}\natexlab{}.
\newblock \bibinfo{title}{DER SPIEGEL | Online-Nachrichten}.
\newblock
  \bibinfo{howpublished}{\url{https://www.spiegel.de/consent-a-?targetUrl=https\%3A\%2F\%2Fwww.spiegel.de\%2Finternational\%2F&ref=https\%3A\%2F\%2Fwww.google.com\%2F}}.
\newblock
\newblock
\shownote{(Accessed on 09/14/2021).}


\bibitem[\protect\citeauthoryear{??}{dpa}{2021}]%
        {dpaen59}
 \bibinfo{year}{2021}\natexlab{}.
\newblock \bibinfo{title}{dpa: en}.
\newblock \bibinfo{howpublished}{\url{https://www.dpa.com/en/}}.
\newblock
\newblock
\shownote{(Accessed on 08/17/2021).}


\bibitem[\protect\citeauthoryear{??}{Fac}{2021}]%
        {FactChec52}
 \bibinfo{year}{2021}\natexlab{}.
\newblock \bibinfo{title}{Fact Check}.
\newblock \bibinfo{howpublished}{\url{https://www.indiatoday.in/fact-check}}.
\newblock
\newblock
\shownote{(Accessed on 08/17/2021).}


\bibitem[\protect\citeauthoryear{??}{ort}{2021}]%
        {ortada}
 \bibinfo{year}{2021}\natexlab{}.
\newblock \bibinfo{title}{Portada · Maldita.es - Periodismo para que no te la
  cuelen}.
\newblock \bibinfo{howpublished}{\url{https://maldita.es/}}.
\newblock
\newblock
\shownote{(Accessed on 08/17/2021).}


\bibitem[\protect\citeauthoryear{??}{Pre}{2021}]%
        {Presenta79}
 \bibinfo{year}{2021}\natexlab{}.
\newblock \bibinfo{title}{Presentación de PowerPoint}.
\newblock
  \bibinfo{howpublished}{\url{https://maldita.es/uploads/public/docs/barometro_desinformacion_parte_1.pdf}}.
\newblock
\newblock
\shownote{(Accessed on 08/25/2021).}


\bibitem[\protect\citeauthoryear{??}{Sea}{2021}]%
        {SearchRe0:online}
 \bibinfo{year}{2021}\natexlab{}.
\newblock \bibinfo{title}{Search Results - LexisNexis}.
\newblock
  \bibinfo{howpublished}{\url{https://www.lexisnexis.com/en-us/search.page}}.
\newblock
\newblock
\shownote{(Accessed on 04/15/2021).}


\bibitem[\protect\citeauthoryear{??}{Uga}{2021}]%
        {Ugandain62}
 \bibinfo{year}{2021}\natexlab{}.
\newblock \bibinfo{title}{Uganda in Crisis – ANCIR's iLAB}.
\newblock
  \bibinfo{howpublished}{\url{https://investigate.africa/reports/uganda-in-crisis/}}.
\newblock
\newblock
\shownote{(Accessed on 08/25/2021).}


\bibitem[\protect\citeauthoryear{??}{Bri}{2022}]%
        {Bringing1:online}
 \bibinfo{year}{2022}\natexlab{}.
\newblock \bibinfo{title}{Bringing fact check information to Google Images}.
\newblock
  \bibinfo{howpublished}{\url{https://blog.google/products/search/bringing-fact-check-information-google-images/}}.
\newblock
\newblock
\shownote{(Accessed on 01/13/2022).}


\bibitem[\protect\citeauthoryear{??}{Deb}{2022}]%
        {DebunkBo89:online}
 \bibinfo{year}{2022}\natexlab{}.
\newblock \bibinfo{title}{Debunk Bot (@DebunkBotAfrica) / Twitter}.
\newblock \bibinfo{howpublished}{\url{https://twitter.com/debunkbotafrica}}.
\newblock
\newblock
\shownote{(Accessed on 01/12/2022).}


\bibitem[\protect\citeauthoryear{??}{Fac}{2022}]%
        {Facebook19:online}
 \bibinfo{year}{2022}\natexlab{}.
\newblock \bibinfo{title}{Facebook Takes Down Fact-Check Of Live Action, Lila
  Rose Anti-Abortion Videos}.
\newblock
  \bibinfo{howpublished}{\url{https://www.buzzfeednews.com/article/claudiakoerner/facebook-fact-check-abortion-video-doctors-medical}}.
\newblock
\newblock
\shownote{(Accessed on 01/05/2022).}


\bibitem[\protect\citeauthoryear{??}{Gep}{2022}]%
        {GephiThe94:online}
 \bibinfo{year}{2022}\natexlab{}.
\newblock \bibinfo{title}{Gephi - The Open Graph Viz Platform}.
\newblock \bibinfo{howpublished}{\url{https://gephi.org/}}.
\newblock
\newblock
\shownote{(Accessed on 01/08/2022).}


\bibitem[\protect\citeauthoryear{??}{Goo}{2022}]%
        {GoogleFa29:online}
 \bibinfo{year}{2022}\natexlab{}.
\newblock \bibinfo{title}{Google Fact Check Feature: What It Means for Your
  Online Efforts - Act-On}.
\newblock
  \bibinfo{howpublished}{\url{https://act-on.com/blog/google-fact-check-feature-what-it-means-for-your-online-efforts/}}.
\newblock
\newblock
\shownote{(Accessed on 01/05/2022).}


\bibitem[\protect\citeauthoryear{??}{HCI}{2022}]%
        {HCIandth97:online}
 \bibinfo{year}{2022}\natexlab{}.
\newblock \bibinfo{title}{HCI and the U.S. Presidential Election: A Few
  Thoughts on a Research Agenda | by Brent Hecht | Medium}.
\newblock
  \bibinfo{howpublished}{\url{https://brenthecht.medium.com/hci-and-the-u-s-presidential-election-a-few-thoughts-on-a-research-agenda-7c1a0a04986}}.
\newblock
\newblock
\shownote{(Accessed on 01/05/2022).}


\bibitem[\protect\citeauthoryear{??}{Ind}{2022}]%
        {Industry67:online}
 \bibinfo{year}{2022}\natexlab{}.
\newblock \bibinfo{title}{Industry-leading vector graphics software | Adobe
  Illustrator}.
\newblock
  \bibinfo{howpublished}{\url{https://www.adobe.com/products/illustrator.html}}.
\newblock
\newblock
\shownote{(Accessed on 01/07/2022).}


\bibitem[\protect\citeauthoryear{??}{Kap}{2022}]%
        {KapwingT82:online}
 \bibinfo{year}{2022}\natexlab{}.
\newblock \bibinfo{title}{Kapwing: The Collaborative Online Video Editor}.
\newblock \bibinfo{howpublished}{\url{https://www.kapwing.com/}}.
\newblock
\newblock
\shownote{(Accessed on 01/07/2022).}


\bibitem[\protect\citeauthoryear{??}{MOO}{2022}]%
        {MOOCorgM86:online}
 \bibinfo{year}{2022}\natexlab{}.
\newblock \bibinfo{title}{MOOC.org | Massive Open Online Courses | An edX
  Site}.
\newblock \bibinfo{howpublished}{\url{https://www.mooc.org/}}.
\newblock
\newblock
\shownote{(Accessed on 01/08/2022).}


\bibitem[\protect\citeauthoryear{??}{Off}{2022}]%
        {Official68:online}
 \bibinfo{year}{2022}\natexlab{}.
\newblock \bibinfo{title}{Official Adobe Photoshop | Photo and design
  software}.
\newblock
  \bibinfo{howpublished}{\url{https://www.adobe.com/products/photoshop.html}}.
\newblock
\newblock
\shownote{(Accessed on 01/07/2022).}


\bibitem[\protect\citeauthoryear{??}{See}{2022}]%
        {Seefactc11:online}
 \bibinfo{year}{2022}\natexlab{}.
\newblock \bibinfo{title}{See fact checks in YouTube search results - YouTube
  Help}.
\newblock
  \bibinfo{howpublished}{\url{https://support.google.com/youtube/answer/9229632?hl=en}}.
\newblock
\newblock
\shownote{(Accessed on 01/05/2022).}


\bibitem[\protect\citeauthoryear{??}{fnc}{2021}]%
        {fnc}
 \bibinfo{year}{accessed in 2021}\natexlab{}.
\newblock \bibinfo{title}{The African Network of Centers for
  InvestigativeReporting’s Investigative Lab}.
\newblock
\newblock
\urldef\tempurl%
\url{https://investigate.africa/}
\showURL{%
\tempurl}


\bibitem[\protect\citeauthoryear{Abdul, Vermeulen, Wang, Lim, and
  Kankanhalli}{Abdul et~al\mbox{.}}{2018a}]%
        {abdul2018trends}
\bibfield{author}{\bibinfo{person}{Ashraf Abdul}, \bibinfo{person}{Jo
  Vermeulen}, \bibinfo{person}{Danding Wang}, \bibinfo{person}{Brian~Y Lim},
  {and} \bibinfo{person}{Mohan Kankanhalli}.} \bibinfo{year}{2018}\natexlab{a}.
\newblock \showarticletitle{Trends and trajectories for explainable,
  accountable and intelligible systems: An hci research agenda}. In
  \bibinfo{booktitle}{\emph{Proceedings of the 2018 CHI conference on human
  factors in computing systems}}. \bibinfo{pages}{1--18}.
\newblock


\bibitem[\protect\citeauthoryear{Abdul, Vermeulen, Wang, Lim, and
  Kankanhalli}{Abdul et~al\mbox{.}}{2018b}]%
        {10.1145/3173574.3174156}
\bibfield{author}{\bibinfo{person}{Ashraf Abdul}, \bibinfo{person}{Jo
  Vermeulen}, \bibinfo{person}{Danding Wang}, \bibinfo{person}{Brian~Y. Lim},
  {and} \bibinfo{person}{Mohan Kankanhalli}.} \bibinfo{year}{2018}\natexlab{b}.
\newblock \bibinfo{booktitle}{\emph{Trends and Trajectories for Explainable,
  Accountable and Intelligible Systems: An HCI Research Agenda}}.
\newblock \bibinfo{publisher}{Association for Computing Machinery},
  \bibinfo{address}{New York, NY, USA}, \bibinfo{pages}{1–18}.
\newblock
\showISBNx{9781450356206}
\urldef\tempurl%
\url{https://doi.org/10.1145/3173574.3174156}
\showURL{%
\tempurl}


\bibitem[\protect\citeauthoryear{Abraham and Reddy}{Abraham and Reddy}{2013}]%
        {abraham2013re}
\bibfield{author}{\bibinfo{person}{Joanna Abraham} {and}
  \bibinfo{person}{Madhu~C Reddy}.} \bibinfo{year}{2013}\natexlab{}.
\newblock \showarticletitle{Re-coordinating activities: an investigation of
  articulation work in patient transfers}. In
  \bibinfo{booktitle}{\emph{Proceedings of the 2013 conference on Computer
  supported cooperative work}}. \bibinfo{pages}{67--78}.
\newblock


\bibitem[\protect\citeauthoryear{Ackerman}{Ackerman}{2000}]%
        {ackerman2000intellectual}
\bibfield{author}{\bibinfo{person}{Mark~S Ackerman}.}
  \bibinfo{year}{2000}\natexlab{}.
\newblock \showarticletitle{The intellectual challenge of CSCW: The gap between
  social requirements and technical feasibility}.
\newblock \bibinfo{journal}{\emph{Human--Computer Interaction}}
  \bibinfo{volume}{15}, \bibinfo{number}{2-3} (\bibinfo{year}{2000}),
  \bibinfo{pages}{179--203}.
\newblock


\bibitem[\protect\citeauthoryear{Adadi and Berrada}{Adadi and Berrada}{2018}]%
        {8466590}
\bibfield{author}{\bibinfo{person}{Amina Adadi} {and} \bibinfo{person}{Mohammed
  Berrada}.} \bibinfo{year}{2018}\natexlab{}.
\newblock \showarticletitle{Peeking Inside the Black-Box: A Survey on
  Explainable Artificial Intelligence (XAI)}.
\newblock \bibinfo{journal}{\emph{IEEE Access}}  \bibinfo{volume}{6}
  (\bibinfo{year}{2018}), \bibinfo{pages}{52138--52160}.
\newblock
\urldef\tempurl%
\url{https://doi.org/10.1109/ACCESS.2018.2870052}
\showDOI{\tempurl}


\bibitem[\protect\citeauthoryear{Adair}{Adair}{2021}]%
        {structurejournalism}
\bibfield{author}{\bibinfo{person}{Bill Adair}.}
  \bibinfo{year}{2021}\natexlab{}.
\newblock \bibinfo{title}{The future of fact-checking is all about structured
  data}.
\newblock
\newblock
\urldef\tempurl%
\url{https://www.niemanlab.org/2020/12/the-future-of-fact-checking-is-all-about-structured-data/}
\showURL{%
\tempurl}


\bibitem[\protect\citeauthoryear{Adnan, Akbar, and Wang}{Adnan
  et~al\mbox{.}}{2019}]%
        {adnan2019information}
\bibfield{author}{\bibinfo{person}{Kiran Adnan}, \bibinfo{person}{Rehan Akbar},
  {and} \bibinfo{person}{Khor~Siak Wang}.} \bibinfo{year}{2019}\natexlab{}.
\newblock \showarticletitle{Information Extraction from Multifaceted
  Unstructured Big Data}.
\newblock \bibinfo{journal}{\emph{International Journal of Recent Technology
  and Engineering (IJRTE)}}  \bibinfo{volume}{8} (\bibinfo{year}{2019}),
  \bibinfo{pages}{1398--1404}.
\newblock


\bibitem[\protect\citeauthoryear{AFP}{AFP}{2021}]%
        {afp}
\bibfield{author}{\bibinfo{person}{AFP}.} \bibinfo{year}{2021}\natexlab{}.
\newblock \bibinfo{title}{Fact Check |}.
\newblock \bibinfo{howpublished}{\url{https://factcheck.afp.com//}}.
\newblock
\newblock
\shownote{(Accessed on 04/15/2021).}


\bibitem[\protect\citeauthoryear{Agadjanian, Bakhru, Chi, Greenberg, Hollander,
  Hurt, Kind, Lu, Ma, Nyhan, et~al\mbox{.}}{Agadjanian et~al\mbox{.}}{2019}]%
        {agadjanian2019counting}
\bibfield{author}{\bibinfo{person}{Alexander Agadjanian},
  \bibinfo{person}{Nikita Bakhru}, \bibinfo{person}{Victoria Chi},
  \bibinfo{person}{Devyn Greenberg}, \bibinfo{person}{Byrne Hollander},
  \bibinfo{person}{Alexander Hurt}, \bibinfo{person}{Joseph Kind},
  \bibinfo{person}{Ray Lu}, \bibinfo{person}{Annie Ma},
  \bibinfo{person}{Brendan Nyhan}, {et~al\mbox{.}}}
  \bibinfo{year}{2019}\natexlab{}.
\newblock \showarticletitle{Counting the Pinocchios: The effect of summary
  fact-checking data on perceived accuracy and favorability of politicians}.
\newblock \bibinfo{journal}{\emph{Research \& Politics}} \bibinfo{volume}{6},
  \bibinfo{number}{3} (\bibinfo{year}{2019}),
  \bibinfo{pages}{2053168019870351}.
\newblock


\bibitem[\protect\citeauthoryear{Almeida}{Almeida}{2019}]%
        {almeida2019misinformation}
\bibfield{author}{\bibinfo{person}{Jussara Almeida}.}
  \bibinfo{year}{2019}\natexlab{}.
\newblock \showarticletitle{Misinformation dissemination on the web}. In
  \bibinfo{booktitle}{\emph{Companion Proceedings of the 2019 World Wide Web
  Conference}}. \bibinfo{pages}{740--740}.
\newblock


\bibitem[\protect\citeauthoryear{Alrubaian, Al-Qurishi, Hassan, and
  Alamri}{Alrubaian et~al\mbox{.}}{2016}]%
        {alrubaian2016credibility}
\bibfield{author}{\bibinfo{person}{Majed Alrubaian}, \bibinfo{person}{Muhammad
  Al-Qurishi}, \bibinfo{person}{Mohammad~Mehedi Hassan}, {and}
  \bibinfo{person}{Atif Alamri}.} \bibinfo{year}{2016}\natexlab{}.
\newblock \showarticletitle{A credibility analysis system for assessing
  information on twitter}.
\newblock \bibinfo{journal}{\emph{IEEE Transactions on Dependable and Secure
  Computing}} \bibinfo{volume}{15}, \bibinfo{number}{4} (\bibinfo{year}{2016}),
  \bibinfo{pages}{661--674}.
\newblock


\bibitem[\protect\citeauthoryear{Amazeen}{Amazeen}{2013}]%
        {amazeen2013critical}
\bibfield{author}{\bibinfo{person}{Michelle~A Amazeen}.}
  \bibinfo{year}{2013}\natexlab{}.
\newblock \bibinfo{title}{A Critical Assessment of Fact-checking in 2012}.
\newblock
\newblock


\bibitem[\protect\citeauthoryear{Amazeen, Vargo, and Hopp}{Amazeen
  et~al\mbox{.}}{2019}]%
        {amazeen2019reinforcing}
\bibfield{author}{\bibinfo{person}{Michelle~A Amazeen},
  \bibinfo{person}{Chris~J Vargo}, {and} \bibinfo{person}{Toby Hopp}.}
  \bibinfo{year}{2019}\natexlab{}.
\newblock \showarticletitle{Reinforcing attitudes in a gatewatching news era:
  Individual-level antecedents to sharing fact-checks on social media}.
\newblock \bibinfo{journal}{\emph{Communication Monographs}}
  \bibinfo{volume}{86}, \bibinfo{number}{1} (\bibinfo{year}{2019}),
  \bibinfo{pages}{112--132}.
\newblock


\bibitem[\protect\citeauthoryear{Archive}{Archive}{2021}]%
        {Internet99:online}
\bibfield{author}{\bibinfo{person}{Internet Archive}.}
  \bibinfo{year}{2021}\natexlab{}.
\newblock \bibinfo{title}{Internet Archive: Wayback Machine}.
\newblock \bibinfo{howpublished}{\url{https://archive.org/web/}}.
\newblock
\newblock
\shownote{(Accessed on 04/15/2021).}


\bibitem[\protect\citeauthoryear{Archive.today}{Archive.today}{2021}]%
        {archivet72:online}
\bibfield{author}{\bibinfo{person}{Archive.today}.}
  \bibinfo{year}{2021}\natexlab{}.
\newblock \bibinfo{title}{archive.today - Wikipedia}.
\newblock
  \bibinfo{howpublished}{\url{https://en.wikipedia.org/wiki/Archive.today}}.
\newblock
\newblock
\shownote{(Accessed on 04/15/2021).}


\bibitem[\protect\citeauthoryear{Arsht and Etcovitch}{Arsht and
  Etcovitch}{2018}]%
        {arsht2018human}
\bibfield{author}{\bibinfo{person}{Andrew Arsht} {and} \bibinfo{person}{Daniel
  Etcovitch}.} \bibinfo{year}{2018}\natexlab{}.
\newblock \showarticletitle{The human cost of online content moderation}.
\newblock \bibinfo{journal}{\emph{Harvard Law Review Online, Harvard
  University, Cambridge, MA, USA. Retrieved from https://jolt. law. harvard.
  edu/digest/the-human-cost-ofonline-content-moderation}}
  (\bibinfo{year}{2018}).
\newblock


\bibitem[\protect\citeauthoryear{Arya, Bellamy, Chen, Dhurandhar, Hind,
  Hoffman, Houde, Liao, Luss, Mojsilovi{\'c}, et~al\mbox{.}}{Arya
  et~al\mbox{.}}{2019}]%
        {arya2019one}
\bibfield{author}{\bibinfo{person}{Vijay Arya}, \bibinfo{person}{Rachel~KE
  Bellamy}, \bibinfo{person}{Pin-Yu Chen}, \bibinfo{person}{Amit Dhurandhar},
  \bibinfo{person}{Michael Hind}, \bibinfo{person}{Samuel~C Hoffman},
  \bibinfo{person}{Stephanie Houde}, \bibinfo{person}{Q~Vera Liao},
  \bibinfo{person}{Ronny Luss}, \bibinfo{person}{Aleksandra Mojsilovi{\'c}},
  {et~al\mbox{.}}} \bibinfo{year}{2019}\natexlab{}.
\newblock \showarticletitle{One explanation does not fit all: A toolkit and
  taxonomy of ai explainability techniques}.
\newblock \bibinfo{journal}{\emph{arXiv preprint arXiv:1909.03012}}
  (\bibinfo{year}{2019}).
\newblock


\bibitem[\protect\citeauthoryear{Ashoori and Weisz}{Ashoori and Weisz}{2019}]%
        {ashoori2019ai}
\bibfield{author}{\bibinfo{person}{Maryam Ashoori} {and}
  \bibinfo{person}{Justin~D Weisz}.} \bibinfo{year}{2019}\natexlab{}.
\newblock \showarticletitle{In AI we trust? Factors that influence
  trustworthiness of AI-infused decision-making processes}.
\newblock \bibinfo{journal}{\emph{arXiv preprint arXiv:1912.02675}}
  (\bibinfo{year}{2019}).
\newblock


\bibitem[\protect\citeauthoryear{Ballotpedia}{Ballotpedia}{2021}]%
        {Ballotpedia}
\bibfield{author}{\bibinfo{person}{Ballotpedia}.} \bibinfo{year}{accessed in
  March 2021}\natexlab{}.
\newblock \bibinfo{title}{The methodologies of fact-checking}.
\newblock
\newblock
\urldef\tempurl%
\url{https://ballotpedia.org/The_methodologies_of_fact-checking}
\showURL{%
\tempurl}


\bibitem[\protect\citeauthoryear{Barbosa and Milan}{Barbosa and Milan}{2019}]%
        {article_whatsapp}
\bibfield{author}{\bibinfo{person}{Sérgio Barbosa} {and}
  \bibinfo{person}{Stefania Milan}.} \bibinfo{year}{2019}\natexlab{}.
\newblock \showarticletitle{Do Not Harm in Private Chat Apps: Ethical Issues
  for Research on and with WhatsApp}.
\newblock \bibinfo{journal}{\emph{Westminster Papers in Communication and
  Culture}}  \bibinfo{volume}{14} (\bibinfo{date}{08} \bibinfo{year}{2019}),
  \bibinfo{pages}{49--65}.
\newblock
\urldef\tempurl%
\url{https://doi.org/10.16997/wpcc.313}
\showDOI{\tempurl}


\bibitem[\protect\citeauthoryear{Bertot and Choi}{Bertot and Choi}{2013}]%
        {10.1145/2479724.2479730}
\bibfield{author}{\bibinfo{person}{John~Carlo Bertot} {and}
  \bibinfo{person}{Heeyoon Choi}.} \bibinfo{year}{2013}\natexlab{}.
\newblock \showarticletitle{Big Data and E-Government: Issues, Policies, and
  Recommendations}. In \bibinfo{booktitle}{\emph{Proceedings of the 14th Annual
  International Conference on Digital Government Research}} (Quebec, Canada)
  \emph{(\bibinfo{series}{dg.o '13})}. \bibinfo{publisher}{Association for
  Computing Machinery}, \bibinfo{address}{New York, NY, USA},
  \bibinfo{pages}{1–10}.
\newblock
\showISBNx{9781450320573}
\urldef\tempurl%
\url{https://doi.org/10.1145/2479724.2479730}
\showDOI{\tempurl}


\bibitem[\protect\citeauthoryear{Bondielli and Marcelloni}{Bondielli and
  Marcelloni}{2019}]%
        {bondielli2019survey}
\bibfield{author}{\bibinfo{person}{Alessandro Bondielli} {and}
  \bibinfo{person}{Francesco Marcelloni}.} \bibinfo{year}{2019}\natexlab{}.
\newblock \showarticletitle{A survey on fake news and rumour detection
  techniques}.
\newblock \bibinfo{journal}{\emph{Information Sciences}}  \bibinfo{volume}{497}
  (\bibinfo{year}{2019}), \bibinfo{pages}{38--55}.
\newblock


\bibitem[\protect\citeauthoryear{Braun and Clarke}{Braun and Clarke}{2006}]%
        {braun2006using}
\bibfield{author}{\bibinfo{person}{Virginia Braun} {and}
  \bibinfo{person}{Victoria Clarke}.} \bibinfo{year}{2006}\natexlab{}.
\newblock \showarticletitle{Using thematic analysis in psychology}.
\newblock \bibinfo{journal}{\emph{Qualitative research in psychology}}
  \bibinfo{volume}{3}, \bibinfo{number}{2} (\bibinfo{year}{2006}),
  \bibinfo{pages}{77--101}.
\newblock


\bibitem[\protect\citeauthoryear{BuzzSumo}{BuzzSumo}{2021}]%
        {BuzzSumo25:online}
\bibfield{author}{\bibinfo{person}{BuzzSumo}.} \bibinfo{year}{2021}\natexlab{}.
\newblock \bibinfo{title}{BuzzSumo.com}.
\newblock \bibinfo{howpublished}{\url{https://buzzsumo.com/}}.
\newblock
\newblock
\shownote{(Accessed on 04/15/2021).}


\bibitem[\protect\citeauthoryear{Cass, Schwanen, and Shove}{Cass
  et~al\mbox{.}}{2018}]%
        {CASS2018160}
\bibfield{author}{\bibinfo{person}{Noel Cass}, \bibinfo{person}{Tim Schwanen},
  {and} \bibinfo{person}{Elizabeth Shove}.} \bibinfo{year}{2018}\natexlab{}.
\newblock \showarticletitle{Infrastructures, intersections and societal
  transformations}.
\newblock \bibinfo{journal}{\emph{Technological Forecasting and Social Change}}
   \bibinfo{volume}{137} (\bibinfo{year}{2018}), \bibinfo{pages}{160--167}.
\newblock
\showISSN{0040-1625}
\urldef\tempurl%
\url{https://doi.org/10.1016/j.techfore.2018.07.039}
\showDOI{\tempurl}


\bibitem[\protect\citeauthoryear{Cazalens, Lamarre, Leblay, Manolescu, and
  Tannier}{Cazalens et~al\mbox{.}}{2018}]%
        {cazalens2018content}
\bibfield{author}{\bibinfo{person}{Sylvie Cazalens}, \bibinfo{person}{Philippe
  Lamarre}, \bibinfo{person}{Julien Leblay}, \bibinfo{person}{Ioana Manolescu},
  {and} \bibinfo{person}{Xavier Tannier}.} \bibinfo{year}{2018}\natexlab{}.
\newblock \showarticletitle{A content management perspective on fact-checking}.
  In \bibinfo{booktitle}{\emph{Companion Proceedings of the The Web Conference
  2018}}. \bibinfo{pages}{565--574}.
\newblock


\bibitem[\protect\citeauthoryear{Cerone, Naghizade, Scholer, Mallal, Skelton,
  and Spina}{Cerone et~al\mbox{.}}{2020}]%
        {cerone2020watch}
\bibfield{author}{\bibinfo{person}{Assunta Cerone}, \bibinfo{person}{Elham
  Naghizade}, \bibinfo{person}{Falk Scholer}, \bibinfo{person}{Devi Mallal},
  \bibinfo{person}{Russell Skelton}, {and} \bibinfo{person}{Damiano Spina}.}
  \bibinfo{year}{2020}\natexlab{}.
\newblock \showarticletitle{Watch’n’Check: Towards a social media
  monitoring tool to assist fact-checking experts}. In
  \bibinfo{booktitle}{\emph{2020 IEEE 7th International Conference on Data
  Science and Advanced Analytics (DSAA)}}. IEEE, \bibinfo{pages}{607--613}.
\newblock


\bibitem[\protect\citeauthoryear{Chaudhuri}{Chaudhuri}{2019}]%
        {chaudhuri2019paradoxes}
\bibfield{author}{\bibinfo{person}{Bidisha Chaudhuri}.}
  \bibinfo{year}{2019}\natexlab{}.
\newblock \showarticletitle{Paradoxes of Intermediation in Aadhaar: Human
  making of a digital infrastructure}.
\newblock \bibinfo{journal}{\emph{South Asia: Journal of South Asian Studies}}
  \bibinfo{volume}{42}, \bibinfo{number}{3} (\bibinfo{year}{2019}),
  \bibinfo{pages}{572--587}.
\newblock


\bibitem[\protect\citeauthoryear{Check}{Check}{2021}]%
        {africacheck}
\bibfield{author}{\bibinfo{person}{Africa Check}.}
  \bibinfo{year}{2021}\natexlab{}.
\newblock \bibinfo{title}{Africa Check | Sorting fact from fiction}.
\newblock \bibinfo{howpublished}{\url{https://africacheck.org/}}.
\newblock
\newblock
\shownote{(Accessed on 04/15/2021).}


\bibitem[\protect\citeauthoryear{Chen, Zhang, Evans, and Min}{Chen
  et~al\mbox{.}}{2021}]%
        {chen2021citizens}
\bibfield{author}{\bibinfo{person}{Qiang Chen}, \bibinfo{person}{Yangyi Zhang},
  \bibinfo{person}{Richard Evans}, {and} \bibinfo{person}{Chen Min}.}
  \bibinfo{year}{2021}\natexlab{}.
\newblock \showarticletitle{Why Do Citizens Share COVID-19 Fact-Checks Posted
  by Chinese Government Social Media Accounts? The Elaboration Likelihood
  Model}.
\newblock \bibinfo{journal}{\emph{International Journal of Environmental
  Research and Public Health}} \bibinfo{volume}{18}, \bibinfo{number}{19}
  (\bibinfo{year}{2021}), \bibinfo{pages}{10058}.
\newblock


\bibitem[\protect\citeauthoryear{Cheruiyot and Ferrer-Conill}{Cheruiyot and
  Ferrer-Conill}{2018}]%
        {doi:10.1080/21670811.2018.1493940}
\bibfield{author}{\bibinfo{person}{David Cheruiyot} {and} \bibinfo{person}{Raul
  Ferrer-Conill}.} \bibinfo{year}{2018}\natexlab{}.
\newblock \showarticletitle{“Fact-Checking Africa”}.
\newblock \bibinfo{journal}{\emph{Digital Journalism}} \bibinfo{volume}{6},
  \bibinfo{number}{8} (\bibinfo{year}{2018}), \bibinfo{pages}{964--975}.
\newblock
\urldef\tempurl%
\url{https://doi.org/10.1080/21670811.2018.1493940}
\showDOI{\tempurl}
\showeprint{https://doi.org/10.1080/21670811.2018.1493940}


\bibitem[\protect\citeauthoryear{Conditt}{Conditt}{2017}]%
        {engadget}
\bibfield{author}{\bibinfo{person}{J. Conditt}.}
  \bibinfo{year}{2017}\natexlab{}.
\newblock \bibinfo{title}{Google partners with fact-checking network to fight
  fake news}.
\newblock
\newblock
\urldef\tempurl%
\url{https://www.engadget.com/2017-10-26-google-fake-news-international-fact-checking-network.html}
\showURL{%
\tempurl}


\bibitem[\protect\citeauthoryear{COOPERATIVE}{COOPERATIVE}{2021}]%
        {TechChec3:online}
\bibfield{author}{\bibinfo{person}{TECH \&~CHECK COOPERATIVE}.}
  \bibinfo{year}{2021}\natexlab{}.
\newblock \bibinfo{title}{Tech \& Check — Duke Reporters' Lab}.
\newblock
  \bibinfo{howpublished}{\url{https://reporterslab.org/tech-and-check/}}.
\newblock
\newblock
\shownote{(Accessed on 04/15/2021).}


\bibitem[\protect\citeauthoryear{CrowdTanle}{CrowdTanle}{2021}]%
        {CrowdTan34:online}
\bibfield{author}{\bibinfo{person}{CrowdTanle}.}
  \bibinfo{year}{2021}\natexlab{}.
\newblock \bibinfo{title}{CrowdTangle | Content Discovery and Social Monitoring
  Made Easy}.
\newblock \bibinfo{howpublished}{\url{https://www.crowdtangle.com/}}.
\newblock
\newblock
\shownote{(Accessed on 04/15/2021).}


\bibitem[\protect\citeauthoryear{Danilevsky, Qian, Aharonov, Katsis, Kawas, and
  Sen}{Danilevsky et~al\mbox{.}}{2020}]%
        {danilevsky-etal-2020-survey}
\bibfield{author}{\bibinfo{person}{Marina Danilevsky}, \bibinfo{person}{Kun
  Qian}, \bibinfo{person}{Ranit Aharonov}, \bibinfo{person}{Yannis Katsis},
  \bibinfo{person}{Ban Kawas}, {and} \bibinfo{person}{Prithviraj Sen}.}
  \bibinfo{year}{2020}\natexlab{}.
\newblock \showarticletitle{A Survey of the State of Explainable {AI} for
  Natural Language Processing}. In \bibinfo{booktitle}{\emph{Proceedings of the
  1st Conference of the Asia-Pacific Chapter of the Association for
  Computational Linguistics and the 10th International Joint Conference on
  Natural Language Processing}}. \bibinfo{publisher}{Association for
  Computational Linguistics}, \bibinfo{address}{Suzhou, China},
  \bibinfo{pages}{447--459}.
\newblock
\urldef\tempurl%
\url{https://aclanthology.org/2020.aacl-main.46}
\showURL{%
\tempurl}


\bibitem[\protect\citeauthoryear{Desouza and Smith}{Desouza and Smith}{2014}]%
        {desouza2014big}
\bibfield{author}{\bibinfo{person}{Kevin~C Desouza} {and}
  \bibinfo{person}{Kendra~L Smith}.} \bibinfo{year}{2014}\natexlab{}.
\newblock \showarticletitle{Big data for social innovation}.
\newblock \bibinfo{journal}{\emph{Stanford Social Innovation Review}}
  \bibinfo{volume}{12}, \bibinfo{number}{3} (\bibinfo{year}{2014}),
  \bibinfo{pages}{38--43}.
\newblock


\bibitem[\protect\citeauthoryear{Dias and Sippitt}{Dias and Sippitt}{2020}]%
        {dias2020researching}
\bibfield{author}{\bibinfo{person}{Nicholas Dias} {and} \bibinfo{person}{Amy
  Sippitt}.} \bibinfo{year}{2020}\natexlab{}.
\newblock \showarticletitle{Researching fact checking: Present limitations and
  future opportunities}.
\newblock \bibinfo{journal}{\emph{The Political Quarterly}}
  \bibinfo{volume}{91}, \bibinfo{number}{3} (\bibinfo{year}{2020}),
  \bibinfo{pages}{605--613}.
\newblock


\bibitem[\protect\citeauthoryear{Dickey}{Dickey}{2019}]%
        {cjr}
\bibfield{author}{\bibinfo{person}{Colin Dickey}.}
  \bibinfo{year}{2019}\natexlab{}.
\newblock \bibinfo{title}{The Rise and Fall of Facts}.
\newblock
\newblock
\urldef\tempurl%
\url{https://www.cjr.org/special_report/rise-and-fall-of-fact-checking.php}
\showURL{%
\tempurl}


\bibitem[\protect\citeauthoryear{Dosono and Semaan}{Dosono and Semaan}{2019}]%
        {dosono2019moderation}
\bibfield{author}{\bibinfo{person}{Bryan Dosono} {and} \bibinfo{person}{Bryan
  Semaan}.} \bibinfo{year}{2019}\natexlab{}.
\newblock \showarticletitle{Moderation practices as emotional labor in
  sustaining online communities: The case of AAPI identity work on Reddit}. In
  \bibinfo{booktitle}{\emph{Proceedings of the 2019 CHI conference on human
  factors in computing systems}}. \bibinfo{pages}{1--13}.
\newblock


\bibitem[\protect\citeauthoryear{DPA}{DPA}{2021}]%
        {factify}
\bibfield{author}{\bibinfo{person}{DPA}.} \bibinfo{year}{2021}\natexlab{}.
\newblock \bibinfo{title}{Fact Check}.
\newblock \bibinfo{howpublished}{\url{https://dps-factify.com}}.
\newblock


\bibitem[\protect\citeauthoryear{Draft}{Draft}{2020a}]%
        {local_context2}
\bibfield{author}{\bibinfo{person}{First Draft}.}
  \bibinfo{year}{2020}\natexlab{a}.
\newblock \bibinfo{title}{Combating misinformation in under-resourced
  languages: lessons from around the world}.
\newblock
\newblock
\urldef\tempurl%
\url{https://firstdraftnews.org/latest/combating-misinformation-in-under-resourced-languages-lessons-from-around-the-world/}
\showURL{%
\tempurl}


\bibitem[\protect\citeauthoryear{Draft}{Draft}{2020b}]%
        {local_context1}
\bibfield{author}{\bibinfo{person}{First Draft}.}
  \bibinfo{year}{2020}\natexlab{b}.
\newblock \bibinfo{title}{The importance of local context in taking on
  misinformation: Lessons from AfricaCheck}.
\newblock
\newblock
\urldef\tempurl%
\url{https://firstdraftnews.org/latest/the-importance-of-local-context-to-taking-on-misinformation-lessons-from-africacheck/}
\showURL{%
\tempurl}


\bibitem[\protect\citeauthoryear{Dwyer}{Dwyer}{2007}]%
        {dwyer2007task}
\bibfield{author}{\bibinfo{person}{Catherine Dwyer}.}
  \bibinfo{year}{2007}\natexlab{}.
\newblock \showarticletitle{Task Technology Fit, the social technical gap and
  social networking sites}.
\newblock \bibinfo{journal}{\emph{AMCIS 2007 Proceedings}}
  (\bibinfo{year}{2007}), \bibinfo{pages}{374}.
\newblock


\bibitem[\protect\citeauthoryear{Dye, Nemer, Mangiameli, Bruckman, and
  Kumar}{Dye et~al\mbox{.}}{2018}]%
        {dye2018paquete}
\bibfield{author}{\bibinfo{person}{Michaelanne Dye}, \bibinfo{person}{David
  Nemer}, \bibinfo{person}{Josiah Mangiameli}, \bibinfo{person}{Amy~S
  Bruckman}, {and} \bibinfo{person}{Neha Kumar}.}
  \bibinfo{year}{2018}\natexlab{}.
\newblock \showarticletitle{El Paquete Semanal: The Week's Internet in Havana}.
  In \bibinfo{booktitle}{\emph{Proceedings of the 2018 CHI Conference on Human
  Factors in Computing Systems}}. \bibinfo{pages}{1--12}.
\newblock


\bibitem[\protect\citeauthoryear{Ecker, O'Reilly, Reid, and Chang}{Ecker
  et~al\mbox{.}}{2020}]%
        {ecker2020effectiveness}
\bibfield{author}{\bibinfo{person}{Ullrich~KH Ecker}, \bibinfo{person}{Ziggy
  O'Reilly}, \bibinfo{person}{Jesse~S Reid}, {and} \bibinfo{person}{Ee~Pin
  Chang}.} \bibinfo{year}{2020}\natexlab{}.
\newblock \showarticletitle{The effectiveness of short-format refutational
  fact-checks}.
\newblock \bibinfo{journal}{\emph{British Journal of Psychology}}
  \bibinfo{volume}{111}, \bibinfo{number}{1} (\bibinfo{year}{2020}),
  \bibinfo{pages}{36--54}.
\newblock


\bibitem[\protect\citeauthoryear{Elizabeth}{Elizabeth}{2014}]%
        {americanpressinstitute}
\bibfield{author}{\bibinfo{person}{Jane Elizabeth}.}
  \bibinfo{year}{2014}\natexlab{}.
\newblock \bibinfo{title}{Who are you calling a fact checker?}
\newblock
\newblock
\urldef\tempurl%
\url{https://www.americanpressinstitute.org/fact-checking-project/fact-checker-definition/}
\showURL{%
\tempurl}


\bibitem[\protect\citeauthoryear{Etikan, Musa, and Alkassim}{Etikan
  et~al\mbox{.}}{2016}]%
        {etikan2016comparison}
\bibfield{author}{\bibinfo{person}{Ilker Etikan},
  \bibinfo{person}{Sulaiman~Abubakar Musa}, {and}
  \bibinfo{person}{Rukayya~Sunusi Alkassim}.} \bibinfo{year}{2016}\natexlab{}.
\newblock \showarticletitle{Comparison of convenience sampling and purposive
  sampling}.
\newblock \bibinfo{journal}{\emph{American journal of theoretical and applied
  statistics}} \bibinfo{volume}{5}, \bibinfo{number}{1} (\bibinfo{year}{2016}),
  \bibinfo{pages}{1--4}.
\newblock


\bibitem[\protect\citeauthoryear{Fact}{Fact}{2021}]%
        {fullfact}
\bibfield{author}{\bibinfo{person}{Full Fact}.}
  \bibinfo{year}{2021}\natexlab{}.
\newblock \bibinfo{title}{Full Fact}.
\newblock \bibinfo{howpublished}{\url{https://fullfact.org/}}.
\newblock
\newblock
\shownote{(Accessed on 04/15/2021).}


\bibitem[\protect\citeauthoryear{Ferguson}{Ferguson}{2005}]%
        {ferguson2005impact}
\bibfield{author}{\bibinfo{person}{Kaethe~Post Ferguson}.}
  \bibinfo{year}{2005}\natexlab{}.
\newblock \showarticletitle{Impact of technology on rural Appalachian health
  care providers: Assessment of technological infrastructure, behaviors, and
  attitudes.}
\newblock  (\bibinfo{year}{2005}).
\newblock


\bibitem[\protect\citeauthoryear{Ferreira, Sharp, and Robinson}{Ferreira
  et~al\mbox{.}}{2011}]%
        {ferreira2011user}
\bibfield{author}{\bibinfo{person}{Jennifer Ferreira}, \bibinfo{person}{Helen
  Sharp}, {and} \bibinfo{person}{Hugh Robinson}.}
  \bibinfo{year}{2011}\natexlab{}.
\newblock \showarticletitle{User experience design and agile development:
  managing cooperation through articulation work}.
\newblock \bibinfo{journal}{\emph{Software: Practice and Experience}}
  \bibinfo{volume}{41}, \bibinfo{number}{9} (\bibinfo{year}{2011}),
  \bibinfo{pages}{963--974}.
\newblock


\bibitem[\protect\citeauthoryear{Fish, Kraut, and Leland}{Fish
  et~al\mbox{.}}{1988}]%
        {fish1988quilt}
\bibfield{author}{\bibinfo{person}{Robert~S Fish}, \bibinfo{person}{Robert~E
  Kraut}, {and} \bibinfo{person}{Mary~DP Leland}.}
  \bibinfo{year}{1988}\natexlab{}.
\newblock \showarticletitle{Quilt: A collaborative tool for cooperative
  writing}. In \bibinfo{booktitle}{\emph{Proceedings of the ACM SIGOIS and
  IEEECS TC-OA 1988 conference on Office information systems}}.
  \bibinfo{pages}{30--37}.
\newblock


\bibitem[\protect\citeauthoryear{Fridkin, Kenney, and Wintersieck}{Fridkin
  et~al\mbox{.}}{2015}]%
        {fridkin2015liar}
\bibfield{author}{\bibinfo{person}{Kim Fridkin}, \bibinfo{person}{Patrick~J
  Kenney}, {and} \bibinfo{person}{Amanda Wintersieck}.}
  \bibinfo{year}{2015}\natexlab{}.
\newblock \showarticletitle{Liar, liar, pants on fire: How fact-checking
  influences citizens’ reactions to negative advertising}.
\newblock \bibinfo{journal}{\emph{Political Communication}}
  \bibinfo{volume}{32}, \bibinfo{number}{1} (\bibinfo{year}{2015}),
  \bibinfo{pages}{127--151}.
\newblock


\bibitem[\protect\citeauthoryear{FROM}{FROM}{2020}]%
        {from2020communicating}
\bibfield{author}{\bibinfo{person}{A~JOINT~BRIEFING FROM}.}
  \bibinfo{year}{2020}\natexlab{}.
\newblock \showarticletitle{Communicating fact checks online}.
\newblock  (\bibinfo{year}{2020}).
\newblock


\bibitem[\protect\citeauthoryear{FullFact}{FullFact}{2021}]%
        {GitHubFu84:online}
\bibfield{author}{\bibinfo{person}{FullFact}.} \bibinfo{year}{2021}\natexlab{}.
\newblock \bibinfo{title}{GitHub -
  FullFact/claim-review-schema-wordpress-plugin: An open source project to
  create a Wordpress plugin for claim review schema.}
\newblock
  \bibinfo{howpublished}{\url{https://github.com/FullFact/claim-review-schema-wordpress-plugin}}.
\newblock
\newblock
\shownote{(Accessed on 04/15/2021).}


\bibitem[\protect\citeauthoryear{Gao, Wang, Barbier, and Liu}{Gao
  et~al\mbox{.}}{2011}]%
        {gao2011promoting}
\bibfield{author}{\bibinfo{person}{Huiji Gao}, \bibinfo{person}{Xufei Wang},
  \bibinfo{person}{Geoffrey Barbier}, {and} \bibinfo{person}{Huan Liu}.}
  \bibinfo{year}{2011}\natexlab{}.
\newblock \showarticletitle{Promoting coordination for disaster relief--from
  crowdsourcing to coordination}. In \bibinfo{booktitle}{\emph{International
  Conference on Social Computing, Behavioral-Cultural Modeling, and
  Prediction}}. Springer, \bibinfo{pages}{197--204}.
\newblock


\bibitem[\protect\citeauthoryear{Ghenai and Mejova}{Ghenai and Mejova}{2017}]%
        {ghenai2017catching}
\bibfield{author}{\bibinfo{person}{Amira Ghenai} {and} \bibinfo{person}{Yelena
  Mejova}.} \bibinfo{year}{2017}\natexlab{}.
\newblock \showarticletitle{Catching Zika fever: Application of crowdsourcing
  and machine learning for tracking health misinformation on Twitter}.
\newblock \bibinfo{journal}{\emph{arXiv preprint arXiv:1707.03778}}
  (\bibinfo{year}{2017}).
\newblock


\bibitem[\protect\citeauthoryear{Girardin}{Girardin}{2007}]%
        {girardin2007towards}
\bibfield{author}{\bibinfo{person}{Fabien Girardin}.}
  \bibinfo{year}{2007}\natexlab{}.
\newblock \emph{\bibinfo{title}{Towards Reducing the Social-Technical Gap in
  Location-Aware Computing}}.
\newblock \bibinfo{thesistype}{Ph.D. Dissertation}. \bibinfo{school}{Citeseer}.
\newblock


\bibitem[\protect\citeauthoryear{Goodman}{Goodman}{1961}]%
        {goodman1961snowball}
\bibfield{author}{\bibinfo{person}{Leo~A Goodman}.}
  \bibinfo{year}{1961}\natexlab{}.
\newblock \showarticletitle{Snowball sampling}.
\newblock \bibinfo{journal}{\emph{The annals of mathematical statistics}}
  (\bibinfo{year}{1961}), \bibinfo{pages}{148--170}.
\newblock


\bibitem[\protect\citeauthoryear{Graves}{Graves}{2018}]%
        {graves2018understanding}
\bibfield{author}{\bibinfo{person}{D Graves}.} \bibinfo{year}{2018}\natexlab{}.
\newblock \showarticletitle{Understanding the promise and limits of automated
  fact-checking}.
\newblock  (\bibinfo{year}{2018}).
\newblock


\bibitem[\protect\citeauthoryear{Graves}{Graves}{2013}]%
        {graves2013deciding}
\bibfield{author}{\bibinfo{person}{Lucas Graves}.}
  \bibinfo{year}{2013}\natexlab{}.
\newblock \emph{\bibinfo{title}{Deciding what's true: Fact-checking journalism
  and the new ecology of news}}.
\newblock \bibinfo{thesistype}{Ph.D. Dissertation}. \bibinfo{school}{Columbia
  University}.
\newblock


\bibitem[\protect\citeauthoryear{Graves}{Graves}{2017}]%
        {graves2017anatomy}
\bibfield{author}{\bibinfo{person}{Lucas Graves}.}
  \bibinfo{year}{2017}\natexlab{}.
\newblock \showarticletitle{Anatomy of a fact check: Objective practice and the
  contested epistemology of fact checking}.
\newblock \bibinfo{journal}{\emph{Communication, Culture \& Critique}}
  \bibinfo{volume}{10}, \bibinfo{number}{3} (\bibinfo{year}{2017}),
  \bibinfo{pages}{518--537}.
\newblock


\bibitem[\protect\citeauthoryear{Graves and Amazeen}{Graves and
  Amazeen}{2019}]%
        {graves2019fact}
\bibfield{author}{\bibinfo{person}{Lucas Graves} {and}
  \bibinfo{person}{Michelle~A Amazeen}.} \bibinfo{year}{2019}\natexlab{}.
\newblock \showarticletitle{Fact-checking as idea and practice in journalism}.
\newblock In \bibinfo{booktitle}{\emph{Oxford Research Encyclopedia of
  Communication}}.
\newblock


\bibitem[\protect\citeauthoryear{Graves and Anderson}{Graves and
  Anderson}{2020}]%
        {graves2020discipline}
\bibfield{author}{\bibinfo{person}{Lucas Graves} {and}
  \bibinfo{person}{Charles~W Anderson}.} \bibinfo{year}{2020}\natexlab{}.
\newblock \showarticletitle{Discipline and promote: Building infrastructure and
  managing algorithms in a “structured journalism” project by professional
  fact-checking groups}.
\newblock \bibinfo{journal}{\emph{New Media \& Society}} \bibinfo{volume}{22},
  \bibinfo{number}{2} (\bibinfo{year}{2020}), \bibinfo{pages}{342--360}.
\newblock


\bibitem[\protect\citeauthoryear{Graves and Cherubini}{Graves and
  Cherubini}{2016}]%
        {graves2016rise}
\bibfield{author}{\bibinfo{person}{Lucas Graves} {and}
  \bibinfo{person}{Federica Cherubini}.} \bibinfo{year}{2016}\natexlab{}.
\newblock \showarticletitle{The rise of fact-checking sites in Europe}.
\newblock  (\bibinfo{year}{2016}).
\newblock


\bibitem[\protect\citeauthoryear{Graves and Glaisyer}{Graves and
  Glaisyer}{2012}]%
        {graves2012fact}
\bibfield{author}{\bibinfo{person}{Lucas Graves} {and} \bibinfo{person}{Tom
  Glaisyer}.} \bibinfo{year}{2012}\natexlab{}.
\newblock \showarticletitle{The fact-checking universe in Spring 2012}.
\newblock \bibinfo{journal}{\emph{New America}} (\bibinfo{year}{2012}).
\newblock


\bibitem[\protect\citeauthoryear{Grinter}{Grinter}{1995}]%
        {grinter1995using}
\bibfield{author}{\bibinfo{person}{Rebecca~E Grinter}.}
  \bibinfo{year}{1995}\natexlab{}.
\newblock \showarticletitle{Using a configuration management tool to coordinate
  software development}. In \bibinfo{booktitle}{\emph{Proceedings of conference
  on Organizational computing systems}}. \bibinfo{pages}{168--177}.
\newblock


\bibitem[\protect\citeauthoryear{Grønsund and Aanestad}{Grønsund and
  Aanestad}{2020}]%
        {GRONSUND2020101614}
\bibfield{author}{\bibinfo{person}{Tor Grønsund} {and}
  \bibinfo{person}{Margunn Aanestad}.} \bibinfo{year}{2020}\natexlab{}.
\newblock \showarticletitle{Augmenting the algorithm: Emerging
  human-in-the-loop work configurations}.
\newblock \bibinfo{journal}{\emph{The Journal of Strategic Information
  Systems}} \bibinfo{volume}{29}, \bibinfo{number}{2} (\bibinfo{year}{2020}),
  \bibinfo{pages}{101614}.
\newblock
\showISSN{0963-8687}
\urldef\tempurl%
\url{https://doi.org/10.1016/j.jsis.2020.101614}
\showDOI{\tempurl}
\newblock
\shownote{Strategic Perspectives on Digital Work and Organizational
  Transformation.}


\bibitem[\protect\citeauthoryear{Guerra, Linz, Garcia, Kommata, Kosiuk, Chun,
  Boveda, and Duncker}{Guerra et~al\mbox{.}}{2021}]%
        {guerra2021use}
\bibfield{author}{\bibinfo{person}{F Guerra}, \bibinfo{person}{D Linz},
  \bibinfo{person}{R Garcia}, \bibinfo{person}{B Kommata}, \bibinfo{person}{J
  Kosiuk}, \bibinfo{person}{J Chun}, \bibinfo{person}{S Boveda}, {and}
  \bibinfo{person}{D Duncker}.} \bibinfo{year}{2021}\natexlab{}.
\newblock \showarticletitle{The use of instant messaging in clinical data
  sharing: the EHRA SMS survey}.
\newblock \bibinfo{journal}{\emph{EP Europace}} \bibinfo{volume}{23},
  \bibinfo{number}{Supplement\_3} (\bibinfo{year}{2021}),
  \bibinfo{pages}{euab116--515}.
\newblock


\bibitem[\protect\citeauthoryear{{H. Carstensen}}{{H. Carstensen}}{1995}]%
        {HCARSTENSEN1995327}
\bibfield{author}{\bibinfo{person}{Peter {H. Carstensen}}.}
  \bibinfo{year}{1995}\natexlab{}.
\newblock \showarticletitle{Modeling coordination work: Lessons learned from
  analyzing a cooperative work setting}.
\newblock In \bibinfo{booktitle}{\emph{Symbiosis of Human and Artifact}},
  \bibfield{editor}{\bibinfo{person}{Yuichiro Anzai},
  \bibinfo{person}{Katsuhiko Ogawa}, {and} \bibinfo{person}{Hirohiko Mori}}
  (Eds.). \bibinfo{series}{Advances in Human Factors/Ergonomics},
  Vol.~\bibinfo{volume}{20}. \bibinfo{publisher}{Elsevier},
  \bibinfo{pages}{327--332}.
\newblock
\showISSN{0921-2647}
\urldef\tempurl%
\url{https://doi.org/10.1016/S0921-2647(06)80054-7}
\showDOI{\tempurl}


\bibitem[\protect\citeauthoryear{Haque, Yousuf, Alam, Saha, Ahmed, and
  Hassan}{Haque et~al\mbox{.}}{2020}]%
        {haque2020combating}
\bibfield{author}{\bibinfo{person}{Md~Mahfuzul Haque},
  \bibinfo{person}{Mohammad Yousuf}, \bibinfo{person}{Ahmed~Shatil Alam},
  \bibinfo{person}{Pratyasha Saha}, \bibinfo{person}{Syed~Ishtiaque Ahmed},
  {and} \bibinfo{person}{Naeemul Hassan}.} \bibinfo{year}{2020}\natexlab{}.
\newblock \showarticletitle{Combating Misinformation in Bangladesh: Roles and
  Responsibilities as Perceived by Journalists, Fact-checkers, and Users}.
\newblock \bibinfo{journal}{\emph{Proceedings of the ACM on Human-Computer
  Interaction}} \bibinfo{volume}{4}, \bibinfo{number}{CSCW2}
  (\bibinfo{year}{2020}), \bibinfo{pages}{1--32}.
\newblock


\bibitem[\protect\citeauthoryear{Hassan, Arslan, Li, and Tremayne}{Hassan
  et~al\mbox{.}}{2017}]%
        {hassan2017toward}
\bibfield{author}{\bibinfo{person}{Naeemul Hassan}, \bibinfo{person}{Fatma
  Arslan}, \bibinfo{person}{Chengkai Li}, {and} \bibinfo{person}{Mark
  Tremayne}.} \bibinfo{year}{2017}\natexlab{}.
\newblock \showarticletitle{Toward automated fact-checking: Detecting
  check-worthy factual claims by claimbuster}. In
  \bibinfo{booktitle}{\emph{Proceedings of the 23rd ACM SIGKDD International
  Conference on Knowledge Discovery and Data Mining}}.
  \bibinfo{pages}{1803--1812}.
\newblock


\bibitem[\protect\citeauthoryear{Hassan, Yousuf, Mahfuzul~Haque,
  A.~Suarez~Rivas, and Khadimul~Islam}{Hassan et~al\mbox{.}}{2019}]%
        {hassan2019examining}
\bibfield{author}{\bibinfo{person}{Naeemul Hassan}, \bibinfo{person}{Mohammad
  Yousuf}, \bibinfo{person}{Md Mahfuzul~Haque}, \bibinfo{person}{Javier
  A.~Suarez~Rivas}, {and} \bibinfo{person}{Md Khadimul~Islam}.}
  \bibinfo{year}{2019}\natexlab{}.
\newblock \showarticletitle{Examining the roles of automation, crowds and
  professionals towards sustainable fact-checking}. In
  \bibinfo{booktitle}{\emph{Companion Proceedings of The 2019 World Wide Web
  Conference}}. \bibinfo{pages}{1001--1006}.
\newblock


\bibitem[\protect\citeauthoryear{He and Han}{He and Han}{2006}]%
        {he2006method}
\bibfield{author}{\bibinfo{person}{Fazhi He} {and} \bibinfo{person}{Soonhung
  Han}.} \bibinfo{year}{2006}\natexlab{}.
\newblock \showarticletitle{A method and tool for human--human interaction and
  instant collaboration in CSCW-based CAD}.
\newblock \bibinfo{journal}{\emph{Computers in Industry}} \bibinfo{volume}{57},
  \bibinfo{number}{8-9} (\bibinfo{year}{2006}), \bibinfo{pages}{740--751}.
\newblock


\bibitem[\protect\citeauthoryear{Heeks}{Heeks}{2003}]%
        {heeks2003most}
\bibfield{author}{\bibinfo{person}{Richard Heeks}.}
  \bibinfo{year}{2003}\natexlab{}.
\newblock \showarticletitle{Most eGovernment-for-development projects fail: how
  can risks be reduced?}
\newblock  (\bibinfo{year}{2003}).
\newblock


\bibitem[\protect\citeauthoryear{Hunt, Agarwal, and Zhuang}{Hunt
  et~al\mbox{.}}{2020}]%
        {hunt2020monitoring}
\bibfield{author}{\bibinfo{person}{Kyle Hunt}, \bibinfo{person}{Puneet
  Agarwal}, {and} \bibinfo{person}{Jun Zhuang}.}
  \bibinfo{year}{2020}\natexlab{}.
\newblock \showarticletitle{Monitoring misinformation on Twitter during crisis
  events: a machine learning approach}.
\newblock \bibinfo{journal}{\emph{Risk analysis}} (\bibinfo{year}{2020}).
\newblock


\bibitem[\protect\citeauthoryear{Influencer}{Influencer}{2021}]%
        {FindEver1:online}
\bibfield{author}{\bibinfo{person}{Influencer}.}
  \bibinfo{year}{2021}\natexlab{}.
\newblock \bibinfo{title}{Find Everything About YouTube on NoxInfluencer}.
\newblock \bibinfo{howpublished}{\url{https://www.noxinfluencer.com/}}.
\newblock
\newblock
\shownote{(Accessed on 04/15/2021).}


\bibitem[\protect\citeauthoryear{InVID}{InVID}{2021}]%
        {InVIDVer51:online}
\bibfield{author}{\bibinfo{person}{InVID}.} \bibinfo{year}{2021}\natexlab{}.
\newblock \bibinfo{title}{InVID Verification Plugin - InVID project}.
\newblock
  \bibinfo{howpublished}{\url{https://www.invid-project.eu/tools-and-services/invid-verification-plugin/}}.
\newblock
\newblock
\shownote{(Accessed on 04/15/2021).}


\bibitem[\protect\citeauthoryear{Jack, Chen, and Jackson}{Jack
  et~al\mbox{.}}{2017}]%
        {jack2017infrastructure}
\bibfield{author}{\bibinfo{person}{Margaret Jack}, \bibinfo{person}{Jay Chen},
  {and} \bibinfo{person}{Steven~J Jackson}.} \bibinfo{year}{2017}\natexlab{}.
\newblock \showarticletitle{Infrastructure as creative action: Online buying,
  selling, and delivery in Phnom Penh}. In
  \bibinfo{booktitle}{\emph{Proceedings of the 2017 CHI Conference on Human
  Factors in Computing Systems}}. \bibinfo{pages}{6511--6522}.
\newblock


\bibitem[\protect\citeauthoryear{Jazeera}{Jazeera}{2021}]%
        {aj}
\bibfield{author}{\bibinfo{person}{Al Jazeera}.}
  \bibinfo{year}{2021}\natexlab{}.
\newblock \bibinfo{title}{Breaking News, World News and Video from Al Jazeera |
  Today's latest from Al Jazeera}.
\newblock \bibinfo{howpublished}{\url{https://www.aljazeera.com/}}.
\newblock
\newblock
\shownote{(Accessed on 04/15/2021).}


\bibitem[\protect\citeauthoryear{Jiang, Baumgartner, Ittycheriah, and Yu}{Jiang
  et~al\mbox{.}}{2020}]%
        {jiang2020factoring}
\bibfield{author}{\bibinfo{person}{Shan Jiang}, \bibinfo{person}{Simon
  Baumgartner}, \bibinfo{person}{Abe Ittycheriah}, {and} \bibinfo{person}{Cong
  Yu}.} \bibinfo{year}{2020}\natexlab{}.
\newblock \showarticletitle{Factoring fact-checks: Structured information
  extraction from fact-checking articles}. In
  \bibinfo{booktitle}{\emph{Proceedings of The Web Conference 2020}}.
  \bibinfo{pages}{1592--1603}.
\newblock


\bibitem[\protect\citeauthoryear{Jirotka, Procter, Rodden, and Bowker}{Jirotka
  et~al\mbox{.}}{2006}]%
        {jirotka2006collaboration}
\bibfield{author}{\bibinfo{person}{Marina Jirotka}, \bibinfo{person}{Rob
  Procter}, \bibinfo{person}{Tom Rodden}, {and} \bibinfo{person}{Geoffrey~C
  Bowker}.} \bibinfo{year}{2006}\natexlab{}.
\newblock \showarticletitle{Collaboration in e-Research}.
\newblock \bibinfo{journal}{\emph{Computer Supported Cooperative Work (CSCW)}}
  \bibinfo{volume}{15}, \bibinfo{number}{4} (\bibinfo{year}{2006}),
  \bibinfo{pages}{251--255}.
\newblock


\bibitem[\protect\citeauthoryear{Karadzhov, Nakov, M{\`a}rquez,
  Barr{\'o}n-Cede{\~n}o, and Koychev}{Karadzhov et~al\mbox{.}}{2017}]%
        {karadzhov2017fully}
\bibfield{author}{\bibinfo{person}{Georgi Karadzhov}, \bibinfo{person}{Preslav
  Nakov}, \bibinfo{person}{Llu{\'\i}s M{\`a}rquez}, \bibinfo{person}{Alberto
  Barr{\'o}n-Cede{\~n}o}, {and} \bibinfo{person}{Ivan Koychev}.}
  \bibinfo{year}{2017}\natexlab{}.
\newblock \showarticletitle{Fully automated fact checking using external
  sources}.
\newblock \bibinfo{journal}{\emph{arXiv preprint arXiv:1710.00341}}
  (\bibinfo{year}{2017}).
\newblock


\bibitem[\protect\citeauthoryear{Karasti and Blomberg}{Karasti and
  Blomberg}{2018}]%
        {karasti2018studying}
\bibfield{author}{\bibinfo{person}{Helena Karasti} {and}
  \bibinfo{person}{Jeanette Blomberg}.} \bibinfo{year}{2018}\natexlab{}.
\newblock \showarticletitle{Studying infrastructuring ethnographically}.
\newblock \bibinfo{journal}{\emph{Computer Supported Cooperative Work (CSCW)}}
  \bibinfo{volume}{27}, \bibinfo{number}{2} (\bibinfo{year}{2018}),
  \bibinfo{pages}{233--265}.
\newblock


\bibitem[\protect\citeauthoryear{Karp and Pardo}{Karp and Pardo}{2017}]%
        {karp2017hapteq}
\bibfield{author}{\bibinfo{person}{Aaron Karp} {and} \bibinfo{person}{Bryan
  Pardo}.} \bibinfo{year}{2017}\natexlab{}.
\newblock \showarticletitle{HaptEQ: A collaborative tool for visually impaired
  audio producers}. In \bibinfo{booktitle}{\emph{Proceedings of the 12th
  International Audio Mostly Conference on Augmented and Participatory Sound
  and Music Experiences}}. \bibinfo{pages}{1--4}.
\newblock


\bibitem[\protect\citeauthoryear{Kartal, Guvenen, and Kutlu}{Kartal
  et~al\mbox{.}}{2020}]%
        {kartal2020too}
\bibfield{author}{\bibinfo{person}{Yavuz~Selim Kartal}, \bibinfo{person}{Busra
  Guvenen}, {and} \bibinfo{person}{Mucahid Kutlu}.}
  \bibinfo{year}{2020}\natexlab{}.
\newblock \showarticletitle{Too many claims to fact-check: Prioritizing
  political claims based on check-worthiness}.
\newblock \bibinfo{journal}{\emph{arXiv preprint arXiv:2004.08166}}
  (\bibinfo{year}{2020}).
\newblock


\bibitem[\protect\citeauthoryear{Kazemi, Garimella, Shahi, Gaffney, and
  Hale}{Kazemi et~al\mbox{.}}{2021}]%
        {kazemi2021tiplines}
\bibfield{author}{\bibinfo{person}{Ashkan Kazemi}, \bibinfo{person}{Kiran
  Garimella}, \bibinfo{person}{Gautam~Kishore Shahi}, \bibinfo{person}{Devin
  Gaffney}, {and} \bibinfo{person}{Scott~A Hale}.}
  \bibinfo{year}{2021}\natexlab{}.
\newblock \showarticletitle{Tiplines to Combat Misinformation on Encrypted
  Platforms: A Case Study of the 2019 Indian Election on WhatsApp}.
\newblock \bibinfo{journal}{\emph{arXiv preprint arXiv:2106.04726}}
  (\bibinfo{year}{2021}).
\newblock


\bibitem[\protect\citeauthoryear{Kerr and Kelleher}{Kerr and Kelleher}{2015}]%
        {kerr2015recruitment}
\bibfield{author}{\bibinfo{person}{Aphra Kerr} {and} \bibinfo{person}{John~D
  Kelleher}.} \bibinfo{year}{2015}\natexlab{}.
\newblock \showarticletitle{The recruitment of passion and community in the
  service of capital: Community managers in the digital games industry}.
\newblock \bibinfo{journal}{\emph{Critical studies in media communication}}
  \bibinfo{volume}{32}, \bibinfo{number}{3} (\bibinfo{year}{2015}),
  \bibinfo{pages}{177--192}.
\newblock


\bibitem[\protect\citeauthoryear{Kim, Tabibian, Oh, Sch{\"o}lkopf, and
  Gomez-Rodriguez}{Kim et~al\mbox{.}}{2018}]%
        {kim2018leveraging}
\bibfield{author}{\bibinfo{person}{Jooyeon Kim}, \bibinfo{person}{Behzad
  Tabibian}, \bibinfo{person}{Alice Oh}, \bibinfo{person}{Bernhard
  Sch{\"o}lkopf}, {and} \bibinfo{person}{Manuel Gomez-Rodriguez}.}
  \bibinfo{year}{2018}\natexlab{}.
\newblock \showarticletitle{Leveraging the crowd to detect and reduce the
  spread of fake news and misinformation}. In
  \bibinfo{booktitle}{\emph{Proceedings of the eleventh ACM international
  conference on web search and data mining}}. \bibinfo{pages}{324--332}.
\newblock


\bibitem[\protect\citeauthoryear{Kotonya and Toni}{Kotonya and Toni}{2020}]%
        {kotonya2020explainable}
\bibfield{author}{\bibinfo{person}{Neema Kotonya} {and}
  \bibinfo{person}{Francesca Toni}.} \bibinfo{year}{2020}\natexlab{}.
\newblock \showarticletitle{Explainable automated fact-checking for public
  health claims}.
\newblock \bibinfo{journal}{\emph{arXiv preprint arXiv:2010.09926}}
  (\bibinfo{year}{2020}).
\newblock


\bibitem[\protect\citeauthoryear{Lampinen, Bellotti, Cheshire, and
  Gray}{Lampinen et~al\mbox{.}}{2016}]%
        {lampinen2016cscw}
\bibfield{author}{\bibinfo{person}{Airi Lampinen}, \bibinfo{person}{Victoria
  Bellotti}, \bibinfo{person}{Coye Cheshire}, {and} \bibinfo{person}{Mary
  Gray}.} \bibinfo{year}{2016}\natexlab{}.
\newblock \showarticletitle{CSCW and theSharing Economy: The Future of
  Platforms as Sites of Work Collaboration and Trust}. In
  \bibinfo{booktitle}{\emph{Proceedings of the 19th ACM Conference on Computer
  Supported Cooperative Work and Social Computing Companion}}.
  \bibinfo{pages}{491--497}.
\newblock


\bibitem[\protect\citeauthoryear{Lazar}{Lazar}{2015}]%
        {10.1145/2807916}
\bibfield{author}{\bibinfo{person}{Jonathan Lazar}.}
  \bibinfo{year}{2015}\natexlab{}.
\newblock \showarticletitle{Public Policy and HCI: Making an Impact in the
  Future}.
\newblock \bibinfo{journal}{\emph{Interactions}} \bibinfo{volume}{22},
  \bibinfo{number}{5} (\bibinfo{date}{Aug.} \bibinfo{year}{2015}),
  \bibinfo{pages}{69–71}.
\newblock
\showISSN{1072-5520}
\urldef\tempurl%
\url{https://doi.org/10.1145/2807916}
\showDOI{\tempurl}


\bibitem[\protect\citeauthoryear{Leads}{Leads}{2021}]%
        {dl}
\bibfield{author}{\bibinfo{person}{Data Leads}.}
  \bibinfo{year}{2021}\natexlab{}.
\newblock \bibinfo{title}{dataleads}.
\newblock \bibinfo{howpublished}{\url{https://dataleads.co.in/}}.
\newblock
\newblock
\shownote{(Accessed on 04/15/2021).}


\bibitem[\protect\citeauthoryear{Leblay, Manolescu, and Tannier}{Leblay
  et~al\mbox{.}}{2018}]%
        {leblay2018computational}
\bibfield{author}{\bibinfo{person}{Julien Leblay}, \bibinfo{person}{Ioana
  Manolescu}, {and} \bibinfo{person}{Xavier Tannier}.}
  \bibinfo{year}{2018}\natexlab{}.
\newblock \showarticletitle{Computational fact-checking: Problems, state of the
  art, and perspectives}. In \bibinfo{booktitle}{\emph{The Web Conference}}.
\newblock


\bibitem[\protect\citeauthoryear{Lee, Dourish, and Mark}{Lee
  et~al\mbox{.}}{2006}]%
        {lee2006human}
\bibfield{author}{\bibinfo{person}{Charlotte~P Lee}, \bibinfo{person}{Paul
  Dourish}, {and} \bibinfo{person}{Gloria Mark}.}
  \bibinfo{year}{2006}\natexlab{}.
\newblock \showarticletitle{The human infrastructure of cyberinfrastructure}.
  In \bibinfo{booktitle}{\emph{Proceedings of the 2006 20th anniversary
  conference on Computer supported cooperative work}}.
  \bibinfo{pages}{483--492}.
\newblock


\bibitem[\protect\citeauthoryear{Linardatos, Papastefanopoulos, and
  Kotsiantis}{Linardatos et~al\mbox{.}}{2021}]%
        {linardatos2021explainable}
\bibfield{author}{\bibinfo{person}{Pantelis Linardatos},
  \bibinfo{person}{Vasilis Papastefanopoulos}, {and} \bibinfo{person}{Sotiris
  Kotsiantis}.} \bibinfo{year}{2021}\natexlab{}.
\newblock \showarticletitle{Explainable ai: A review of machine learning
  interpretability methods}.
\newblock \bibinfo{journal}{\emph{Entropy}} \bibinfo{volume}{23},
  \bibinfo{number}{1} (\bibinfo{year}{2021}), \bibinfo{pages}{18}.
\newblock


\bibitem[\protect\citeauthoryear{Lundberg and Tellio{\u{g}}lu}{Lundberg and
  Tellio{\u{g}}lu}{1999}]%
        {lundberg1999understanding}
\bibfield{author}{\bibinfo{person}{Nina Lundberg} {and} \bibinfo{person}{Hilda
  Tellio{\u{g}}lu}.} \bibinfo{year}{1999}\natexlab{}.
\newblock \showarticletitle{Understanding complex coordination processes in
  health care}.
\newblock \bibinfo{journal}{\emph{Scandinavian Journal of Information Systems}}
  \bibinfo{volume}{11}, \bibinfo{number}{1} (\bibinfo{year}{1999}),
  \bibinfo{pages}{5}.
\newblock


\bibitem[\protect\citeauthoryear{Mark, Al-Ani, and Semaan}{Mark
  et~al\mbox{.}}{2009}]%
        {mark2009repairing}
\bibfield{author}{\bibinfo{person}{Gloria Mark}, \bibinfo{person}{Ban Al-Ani},
  {and} \bibinfo{person}{Bryan Semaan}.} \bibinfo{year}{2009}\natexlab{}.
\newblock \showarticletitle{Repairing human infrastructure in war zones}.
\newblock \bibinfo{journal}{\emph{Proceedings of ISCRAM}}
  (\bibinfo{year}{2009}), \bibinfo{pages}{10--13}.
\newblock


\bibitem[\protect\citeauthoryear{Markussen}{Markussen}{1996}]%
        {markussen1996politics}
\bibfield{author}{\bibinfo{person}{Randi Markussen}.}
  \bibinfo{year}{1996}\natexlab{}.
\newblock \showarticletitle{Politics of intervention in design: Feminist
  reflections on the Scandinavian tradition}.
\newblock \bibinfo{journal}{\emph{ai \& Society}} \bibinfo{volume}{10},
  \bibinfo{number}{2} (\bibinfo{year}{1996}), \bibinfo{pages}{127--141}.
\newblock


\bibitem[\protect\citeauthoryear{Mars and Scott}{Mars and Scott}{2016}]%
        {mars2016whatsapp}
\bibfield{author}{\bibinfo{person}{Maurice Mars} {and}
  \bibinfo{person}{Richard~E Scott}.} \bibinfo{year}{2016}\natexlab{}.
\newblock \showarticletitle{Whatsapp in clinical practice: A literature}.
\newblock \bibinfo{journal}{\emph{The Promise of New Technologies in an Age of
  New Health Challenges}} (\bibinfo{year}{2016}), \bibinfo{pages}{82}.
\newblock


\bibitem[\protect\citeauthoryear{Meedan}{Meedan}{2021}]%
        {meedan}
\bibfield{author}{\bibinfo{person}{Meedan}.} \bibinfo{year}{2021}\natexlab{}.
\newblock \bibinfo{title}{Meedan}.
\newblock \bibinfo{howpublished}{\url{https://meedan.com/}}.
\newblock
\newblock
\shownote{(Accessed on 04/15/2021).}


\bibitem[\protect\citeauthoryear{Melo, Messias, Resende, Garimella, Almeida,
  and Benevenuto}{Melo et~al\mbox{.}}{2019}]%
        {Melo_Messias_Resende_Garimella_Almeida_Benevenuto_2019}
\bibfield{author}{\bibinfo{person}{Philipe Melo}, \bibinfo{person}{Johnnatan
  Messias}, \bibinfo{person}{Gustavo Resende}, \bibinfo{person}{Kiran
  Garimella}, \bibinfo{person}{Jussara Almeida}, {and}
  \bibinfo{person}{Fabrício Benevenuto}.} \bibinfo{year}{2019}\natexlab{}.
\newblock \showarticletitle{WhatsApp Monitor: A Fact-Checking System for
  WhatsApp}.
\newblock \bibinfo{journal}{\emph{Proceedings of the International AAAI
  Conference on Web and Social Media}} \bibinfo{volume}{13},
  \bibinfo{number}{01} (\bibinfo{date}{Jul.} \bibinfo{year}{2019}),
  \bibinfo{pages}{676--677}.
\newblock
\urldef\tempurl%
\url{https://ojs.aaai.org/index.php/ICWSM/article/view/3271}
\showURL{%
\tempurl}


\bibitem[\protect\citeauthoryear{Miller}{Miller}{2019}]%
        {miller2019explanation}
\bibfield{author}{\bibinfo{person}{Tim Miller}.}
  \bibinfo{year}{2019}\natexlab{}.
\newblock \showarticletitle{Explanation in artificial intelligence: Insights
  from the social sciences}.
\newblock \bibinfo{journal}{\emph{Artificial intelligence}}
  \bibinfo{volume}{267} (\bibinfo{year}{2019}), \bibinfo{pages}{1--38}.
\newblock


\bibitem[\protect\citeauthoryear{Morgan, Gilbert, McDonald, and Zachry}{Morgan
  et~al\mbox{.}}{2013}]%
        {morgan2013project}
\bibfield{author}{\bibinfo{person}{Jonathan~T Morgan}, \bibinfo{person}{Michael
  Gilbert}, \bibinfo{person}{David~W McDonald}, {and} \bibinfo{person}{Mark
  Zachry}.} \bibinfo{year}{2013}\natexlab{}.
\newblock \showarticletitle{Project talk: Coordination work and group
  membership in WikiProjects}. In \bibinfo{booktitle}{\emph{Proceedings of the
  9th International Symposium on Open Collaboration}}. \bibinfo{pages}{1--10}.
\newblock


\bibitem[\protect\citeauthoryear{Mota, de~Carvalho, and Reis}{Mota
  et~al\mbox{.}}{2011}]%
        {mota2011fostering}
\bibfield{author}{\bibinfo{person}{Dulce Mota}, \bibinfo{person}{Carlos~Vaz de
  Carvalho}, {and} \bibinfo{person}{Luis~Paulo Reis}.}
  \bibinfo{year}{2011}\natexlab{}.
\newblock \showarticletitle{Fostering Collaborative Work between educators in
  higher education}. In \bibinfo{booktitle}{\emph{2011 IEEE International
  Conference on Systems, Man, and Cybernetics}}. IEEE,
  \bibinfo{pages}{1286--1291}.
\newblock


\bibitem[\protect\citeauthoryear{Nardi and Engestr{\"o}m}{Nardi and
  Engestr{\"o}m}{1999}]%
        {nardi1999web}
\bibfield{author}{\bibinfo{person}{Bonnie~A Nardi} {and}
  \bibinfo{person}{Yrj{\"o} Engestr{\"o}m}.} \bibinfo{year}{1999}\natexlab{}.
\newblock \showarticletitle{A web on the wind: The structure of invisible
  work}.
\newblock \bibinfo{journal}{\emph{Computer supported cooperative work}}
  \bibinfo{volume}{8}, \bibinfo{number}{1-2} (\bibinfo{year}{1999}),
  \bibinfo{pages}{1--8}.
\newblock


\bibitem[\protect\citeauthoryear{Nguyen, Kharosekar, Krishnan, Krishnan, Tate,
  Wallace, and Lease}{Nguyen et~al\mbox{.}}{2018}]%
        {nguyen2018believe}
\bibfield{author}{\bibinfo{person}{An~T Nguyen}, \bibinfo{person}{Aditya
  Kharosekar}, \bibinfo{person}{Saumyaa Krishnan}, \bibinfo{person}{Siddhesh
  Krishnan}, \bibinfo{person}{Elizabeth Tate}, \bibinfo{person}{Byron~C
  Wallace}, {and} \bibinfo{person}{Matthew Lease}.}
  \bibinfo{year}{2018}\natexlab{}.
\newblock \showarticletitle{Believe it or not: Designing a human-ai partnership
  for mixed-initiative fact-checking}. In \bibinfo{booktitle}{\emph{Proceedings
  of the 31st Annual ACM Symposium on User Interface Software and Technology}}.
  \bibinfo{pages}{189--199}.
\newblock


\bibitem[\protect\citeauthoryear{Nguyen}{Nguyen}{2016}]%
        {nguyen2016infrastructural}
\bibfield{author}{\bibinfo{person}{Lilly~U Nguyen}.}
  \bibinfo{year}{2016}\natexlab{}.
\newblock \showarticletitle{Infrastructural action in Vietnam: Inverting the
  techno-politics of hacking in the global South}.
\newblock \bibinfo{journal}{\emph{New Media \& Society}} \bibinfo{volume}{18},
  \bibinfo{number}{4} (\bibinfo{year}{2016}), \bibinfo{pages}{637--652}.
\newblock


\bibitem[\protect\citeauthoryear{Oeldorf-Hirsch, Schmierbach, Appelman, and
  Boyle}{Oeldorf-Hirsch et~al\mbox{.}}{2020}]%
        {oeldorf2020ineffectiveness}
\bibfield{author}{\bibinfo{person}{Anne Oeldorf-Hirsch}, \bibinfo{person}{Mike
  Schmierbach}, \bibinfo{person}{Alyssa Appelman}, {and}
  \bibinfo{person}{Michael~P Boyle}.} \bibinfo{year}{2020}\natexlab{}.
\newblock \showarticletitle{The ineffectiveness of fact-checking labels on news
  memes and articles}.
\newblock \bibinfo{journal}{\emph{Mass Communication and Society}}
  \bibinfo{volume}{23}, \bibinfo{number}{5} (\bibinfo{year}{2020}),
  \bibinfo{pages}{682--704}.
\newblock


\bibitem[\protect\citeauthoryear{of~Commons}{of~Commons}{2019}]%
        {ukgov}
\bibfield{author}{\bibinfo{person}{House of Commons}.}
  \bibinfo{year}{2019}\natexlab{}.
\newblock \bibinfo{title}{Public Administration and Constitutional Affairs
  Committee, Oral evidence: Governance of Statistics}.
\newblock
\newblock
\urldef\tempurl%
\url{http://data.parliament.uk/writtenevidence/committeeevidence.svc/evidencedocument/public-administration-and-constitutional-affairs-committee/governance-of-official-statistics/oral/97624.html}
\showURL{%
\tempurl}


\bibitem[\protect\citeauthoryear{P{\'a}ez}{P{\'a}ez}{2019}]%
        {paez2019pragmatic}
\bibfield{author}{\bibinfo{person}{Andr{\'e}s P{\'a}ez}.}
  \bibinfo{year}{2019}\natexlab{}.
\newblock \showarticletitle{The pragmatic turn in explainable artificial
  intelligence (XAI)}.
\newblock \bibinfo{journal}{\emph{Minds and Machines}} \bibinfo{volume}{29},
  \bibinfo{number}{3} (\bibinfo{year}{2019}), \bibinfo{pages}{441--459}.
\newblock


\bibitem[\protect\citeauthoryear{Pendse, Lalani, De~Choudhury, Sharma, and
  Kumar}{Pendse et~al\mbox{.}}{2020}]%
        {pendse2020like}
\bibfield{author}{\bibinfo{person}{Sachin~R Pendse}, \bibinfo{person}{Faisal~M
  Lalani}, \bibinfo{person}{Munmun De~Choudhury}, \bibinfo{person}{Amit
  Sharma}, {and} \bibinfo{person}{Neha Kumar}.}
  \bibinfo{year}{2020}\natexlab{}.
\newblock \showarticletitle{" Like Shock Absorbers": Understanding the Human
  Infrastructures of Technology-Mediated Mental Health Support}. In
  \bibinfo{booktitle}{\emph{Proceedings of the 2020 CHI Conference on Human
  Factors in Computing Systems}}. \bibinfo{pages}{1--14}.
\newblock


\bibitem[\protect\citeauthoryear{Pesacheck}{Pesacheck}{2021}]%
        {pesa}
\bibfield{author}{\bibinfo{person}{Pesacheck}.}
  \bibinfo{year}{2021}\natexlab{}.
\newblock \bibinfo{title}{PesaCheck}.
\newblock \bibinfo{howpublished}{\url{https://pesacheck.org/}}.
\newblock
\newblock
\shownote{(Accessed on 04/15/2021).}


\bibitem[\protect\citeauthoryear{Post}{Post}{2021a}]%
        {AboutThe91:online}
\bibfield{author}{\bibinfo{person}{The~Washington Post}.}
  \bibinfo{year}{2021}\natexlab{a}.
\newblock \bibinfo{title}{About The Fact Checker - The Washington Post}.
\newblock
  \bibinfo{howpublished}{\url{https://www.washingtonpost.com/politics/2019/01/07/about-fact-checker/}}.
\newblock
\newblock
\shownote{(Accessed on 04/15/2021).}


\bibitem[\protect\citeauthoryear{Post}{Post}{2021b}]%
        {wp}
\bibfield{author}{\bibinfo{person}{The~Washington Post}.}
  \bibinfo{year}{2021}\natexlab{b}.
\newblock \bibinfo{title}{Fact Checker - The Washington Post}.
\newblock
  \bibinfo{howpublished}{\url{https://www.washingtonpost.com/news/fact-checker/}}.
\newblock
\newblock
\shownote{(Accessed on 04/15/2021).}


\bibitem[\protect\citeauthoryear{Poynter}{Poynter}{2021}]%
        {poynter2}
\bibfield{author}{\bibinfo{person}{Poynter}.} \bibinfo{year}{accessed in March,
  2021}\natexlab{}.
\newblock \bibinfo{title}{The International Fact-Checking Network}.
\newblock
\newblock
\urldef\tempurl%
\url{https://www.poynter.org/ifcn/}
\showURL{%
\tempurl}


\bibitem[\protect\citeauthoryear{Quint}{Quint}{2021}]%
        {quint}
\bibfield{author}{\bibinfo{person}{The Quint}.}
  \bibinfo{year}{2021}\natexlab{}.
\newblock \bibinfo{title}{Latest News, Breaking News LIVE, Top News Headlines,
  Viral Videos News Updates - The Quint}.
\newblock \bibinfo{howpublished}{\url{https://www.thequint.com/}}.
\newblock
\newblock
\shownote{(Accessed on 04/15/2021).}


\bibitem[\protect\citeauthoryear{Rasmussen}{Rasmussen}{2007}]%
        {rasmussen2007human}
\bibfield{author}{\bibinfo{person}{Lauge~Baungaard Rasmussen}.}
  \bibinfo{year}{2007}\natexlab{}.
\newblock \showarticletitle{From human-centred to human-context centred
  approach: looking back over ‘the hills’, what has been gained and lost?}
\newblock \bibinfo{journal}{\emph{Ai \& Society}} \bibinfo{volume}{21},
  \bibinfo{number}{4} (\bibinfo{year}{2007}), \bibinfo{pages}{471--495}.
\newblock


\bibitem[\protect\citeauthoryear{Republic}{Republic}{2021}]%
        {nr}
\bibfield{author}{\bibinfo{person}{The~New Republic}.}
  \bibinfo{year}{2021}\natexlab{}.
\newblock \bibinfo{title}{The New Republic}.
\newblock \bibinfo{howpublished}{\url{https://newrepublic.com/}}.
\newblock
\newblock
\shownote{(Accessed on 04/15/2021).}


\bibitem[\protect\citeauthoryear{Resnick, Carton, Park, Shen, and
  Zeffer}{Resnick et~al\mbox{.}}{2014}]%
        {resnick2014rumorlens}
\bibfield{author}{\bibinfo{person}{Paul Resnick}, \bibinfo{person}{Samuel
  Carton}, \bibinfo{person}{Souneil Park}, \bibinfo{person}{Yuncheng Shen},
  {and} \bibinfo{person}{Nicole Zeffer}.} \bibinfo{year}{2014}\natexlab{}.
\newblock \showarticletitle{Rumorlens: A system for analyzing the impact of
  rumors and corrections in social media}. In \bibinfo{booktitle}{\emph{Proc.
  Computational Journalism Conference}}, Vol.~\bibinfo{volume}{5}.
\newblock


\bibitem[\protect\citeauthoryear{Roberts}{Roberts}{2014}]%
        {roberts2014behind}
\bibfield{author}{\bibinfo{person}{Sarah~T Roberts}.}
  \bibinfo{year}{2014}\natexlab{}.
\newblock \emph{\bibinfo{title}{Behind the screen: The hidden digital labor of
  commercial content moderation}}.
\newblock \bibinfo{thesistype}{Ph.D. Dissertation}. \bibinfo{school}{University
  of Illinois at Urbana-Champaign}.
\newblock


\bibitem[\protect\citeauthoryear{Robinson, Maddock, and Starbird}{Robinson
  et~al\mbox{.}}{2015}]%
        {robinson2015examining}
\bibfield{author}{\bibinfo{person}{John~J Robinson}, \bibinfo{person}{Jim
  Maddock}, {and} \bibinfo{person}{Kate Starbird}.}
  \bibinfo{year}{2015}\natexlab{}.
\newblock \showarticletitle{Examining the Role of Human and Technical
  Infrastructure during Emergency Response.}. In
  \bibinfo{booktitle}{\emph{ISCRAM}}.
\newblock


\bibitem[\protect\citeauthoryear{Rummel, Spada, Hermann, Caspar, and
  Schornstein}{Rummel et~al\mbox{.}}{2002}]%
        {rummel2002promoting}
\bibfield{author}{\bibinfo{person}{Nikol Rummel}, \bibinfo{person}{Hans Spada},
  \bibinfo{person}{Fabian Hermann}, \bibinfo{person}{Franz Caspar}, {and}
  \bibinfo{person}{Katrin Schornstein}.} \bibinfo{year}{2002}\natexlab{}.
\newblock \showarticletitle{Promoting the coordination of computer-mediated
  interdisciplinary collaboration}.
\newblock  (\bibinfo{year}{2002}).
\newblock


\bibitem[\protect\citeauthoryear{Sambasivan and Smyth}{Sambasivan and
  Smyth}{2010}]%
        {sambasivan2010human}
\bibfield{author}{\bibinfo{person}{Nithya Sambasivan} {and}
  \bibinfo{person}{Thomas Smyth}.} \bibinfo{year}{2010}\natexlab{}.
\newblock \showarticletitle{The human infrastructure of ICTD}. In
  \bibinfo{booktitle}{\emph{Proceedings of the 4th ACM/IEEE international
  conference on information and communication technologies and development}}.
  \bibinfo{pages}{1--9}.
\newblock


\bibitem[\protect\citeauthoryear{Sawyer and Tapia}{Sawyer and Tapia}{2006}]%
        {sawyer2006always}
\bibfield{author}{\bibinfo{person}{Steve Sawyer} {and} \bibinfo{person}{Andrea
  Tapia}.} \bibinfo{year}{2006}\natexlab{}.
\newblock \showarticletitle{Always articulating: Theorizing on mobile and
  wireless technologies}.
\newblock \bibinfo{journal}{\emph{The Information Society}}
  \bibinfo{volume}{22}, \bibinfo{number}{5} (\bibinfo{year}{2006}),
  \bibinfo{pages}{311--323}.
\newblock


\bibitem[\protect\citeauthoryear{Scharowski}{Scharowski}{2020}]%
        {scharowski2020transparency}
\bibfield{author}{\bibinfo{person}{Nicolas Scharowski}.}
  \bibinfo{year}{2020}\natexlab{}.
\newblock \emph{\bibinfo{title}{Transparency and Trust in AI}}.
\newblock \bibinfo{thesistype}{Ph.D. Dissertation}. \bibinfo{school}{Institute
  of Psychology}.
\newblock


\bibitem[\protect\citeauthoryear{Schmidt, Biessmann, and Teubner}{Schmidt
  et~al\mbox{.}}{2020}]%
        {schmidt2020transparency}
\bibfield{author}{\bibinfo{person}{Philipp Schmidt}, \bibinfo{person}{Felix
  Biessmann}, {and} \bibinfo{person}{Timm Teubner}.}
  \bibinfo{year}{2020}\natexlab{}.
\newblock \showarticletitle{Transparency and trust in artificial intelligence
  systems}.
\newblock \bibinfo{journal}{\emph{Journal of Decision Systems}}
  \bibinfo{volume}{29}, \bibinfo{number}{4} (\bibinfo{year}{2020}),
  \bibinfo{pages}{260--278}.
\newblock


\bibitem[\protect\citeauthoryear{Searcher}{Searcher}{2021}]%
        {SocialSe32:online}
\bibfield{author}{\bibinfo{person}{Social Searcher}.}
  \bibinfo{year}{2021}\natexlab{}.
\newblock \bibinfo{title}{Social Searcher - Free Social Media Search Engine}.
\newblock \bibinfo{howpublished}{\url{https://www.social-searcher.com/}}.
\newblock
\newblock
\shownote{(Accessed on 04/15/2021).}


\bibitem[\protect\citeauthoryear{Sessions and Valtorta}{Sessions and
  Valtorta}{2006}]%
        {sessions2006effects}
\bibfield{author}{\bibinfo{person}{Valerie Sessions} {and}
  \bibinfo{person}{Marco Valtorta}.} \bibinfo{year}{2006}\natexlab{}.
\newblock \showarticletitle{The Effects of Data Quality on Machine Learning
  Algorithms.}
\newblock \bibinfo{journal}{\emph{ICIQ}}  \bibinfo{volume}{6}
  (\bibinfo{year}{2006}), \bibinfo{pages}{485--498}.
\newblock


\bibitem[\protect\citeauthoryear{Shao, Ciampaglia, Flammini, and Menczer}{Shao
  et~al\mbox{.}}{2016}]%
        {shao2016hoaxy}
\bibfield{author}{\bibinfo{person}{Chengcheng Shao},
  \bibinfo{person}{Giovanni~Luca Ciampaglia}, \bibinfo{person}{Alessandro
  Flammini}, {and} \bibinfo{person}{Filippo Menczer}.}
  \bibinfo{year}{2016}\natexlab{}.
\newblock \showarticletitle{Hoaxy: A platform for tracking online
  misinformation}. In \bibinfo{booktitle}{\emph{Proceedings of the 25th
  international conference companion on world wide web}}.
  \bibinfo{pages}{745--750}.
\newblock


\bibitem[\protect\citeauthoryear{Shiralkar, Flammini, Menczer, and
  Ciampaglia}{Shiralkar et~al\mbox{.}}{2017}]%
        {shiralkar2017finding}
\bibfield{author}{\bibinfo{person}{Prashant Shiralkar},
  \bibinfo{person}{Alessandro Flammini}, \bibinfo{person}{Filippo Menczer},
  {and} \bibinfo{person}{Giovanni~Luca Ciampaglia}.}
  \bibinfo{year}{2017}\natexlab{}.
\newblock \showarticletitle{Finding streams in knowledge graphs to support fact
  checking}. In \bibinfo{booktitle}{\emph{2017 IEEE International Conference on
  Data Mining (ICDM)}}. IEEE, \bibinfo{pages}{859--864}.
\newblock


\bibitem[\protect\citeauthoryear{Sivek and Bloyd-Peshkin}{Sivek and
  Bloyd-Peshkin}{2019}]%
        {doi:10.1080/17512786.2019.1643767}
\bibfield{author}{\bibinfo{person}{Susan~Currie Sivek} {and}
  \bibinfo{person}{Sharon Bloyd-Peshkin}.} \bibinfo{year}{2019}\natexlab{}.
\newblock \showarticletitle{Where Do Facts Matter? The Digital Paradox in
  Magazines’ Fact-checking Processes}.
\newblock \bibinfo{journal}{\emph{Journalism Practice}} \bibinfo{volume}{13},
  \bibinfo{number}{8} (\bibinfo{year}{2019}), \bibinfo{pages}{998--1002}.
\newblock
\urldef\tempurl%
\url{https://doi.org/10.1080/17512786.2019.1643767}
\showDOI{\tempurl}
\showeprint{https://doi.org/10.1080/17512786.2019.1643767}


\bibitem[\protect\citeauthoryear{Smith, Kumar, Boyd-Graber, Seppi, and
  Findlater}{Smith et~al\mbox{.}}{2018}]%
        {smith2018closing}
\bibfield{author}{\bibinfo{person}{Alison Smith}, \bibinfo{person}{Varun
  Kumar}, \bibinfo{person}{Jordan Boyd-Graber}, \bibinfo{person}{Kevin Seppi},
  {and} \bibinfo{person}{Leah Findlater}.} \bibinfo{year}{2018}\natexlab{}.
\newblock \showarticletitle{Closing the loop: User-centered design and
  evaluation of a human-in-the-loop topic modeling system}. In
  \bibinfo{booktitle}{\emph{23rd International Conference on Intelligent User
  Interfaces}}. \bibinfo{pages}{293--304}.
\newblock


\bibitem[\protect\citeauthoryear{Spaa, Durrant, Elsden, and Vines}{Spaa
  et~al\mbox{.}}{2019}]%
        {10.1145/3290605.3300314}
\bibfield{author}{\bibinfo{person}{Anne Spaa}, \bibinfo{person}{Abigail
  Durrant}, \bibinfo{person}{Chris Elsden}, {and} \bibinfo{person}{John
  Vines}.} \bibinfo{year}{2019}\natexlab{}.
\newblock \showarticletitle{Understanding the Boundaries between Policymaking
  and HCI}. In \bibinfo{booktitle}{\emph{Proceedings of the 2019 CHI Conference
  on Human Factors in Computing Systems}} (Glasgow, Scotland Uk)
  \emph{(\bibinfo{series}{CHI '19})}. \bibinfo{publisher}{Association for
  Computing Machinery}, \bibinfo{address}{New York, NY, USA},
  \bibinfo{pages}{1–15}.
\newblock
\showISBNx{9781450359702}
\urldef\tempurl%
\url{https://doi.org/10.1145/3290605.3300314}
\showDOI{\tempurl}


\bibitem[\protect\citeauthoryear{Spoonbill}{Spoonbill}{2021}]%
        {Spoonbil76:online}
\bibfield{author}{\bibinfo{person}{Spoonbill}.}
  \bibinfo{year}{2021}\natexlab{}.
\newblock \bibinfo{title}{Spoonbill}.
\newblock \bibinfo{howpublished}{\url{https://spoonbill.io/}}.
\newblock
\newblock
\shownote{(Accessed on 04/15/2021).}


\bibitem[\protect\citeauthoryear{Star and Ruhleder}{Star and Ruhleder}{1996}]%
        {star1996steps}
\bibfield{author}{\bibinfo{person}{Susan~Leigh Star} {and}
  \bibinfo{person}{Karen Ruhleder}.} \bibinfo{year}{1996}\natexlab{}.
\newblock \showarticletitle{Steps toward an ecology of infrastructure: Design
  and access for large information spaces}.
\newblock \bibinfo{journal}{\emph{Information systems research}}
  \bibinfo{volume}{7}, \bibinfo{number}{1} (\bibinfo{year}{1996}),
  \bibinfo{pages}{111--134}.
\newblock


\bibitem[\protect\citeauthoryear{Star and Strauss}{Star and Strauss}{1999}]%
        {star1999layers}
\bibfield{author}{\bibinfo{person}{Susan~Leigh Star} {and}
  \bibinfo{person}{Anselm Strauss}.} \bibinfo{year}{1999}\natexlab{}.
\newblock \showarticletitle{Layers of silence, arenas of voice: The ecology of
  visible and invisible work}.
\newblock \bibinfo{journal}{\emph{Computer supported cooperative work (CSCW)}}
  \bibinfo{volume}{8}, \bibinfo{number}{1} (\bibinfo{year}{1999}),
  \bibinfo{pages}{9--30}.
\newblock


\bibitem[\protect\citeauthoryear{Steiger, Bharucha, Venkatagiri, Riedl, and
  Lease}{Steiger et~al\mbox{.}}{2021}]%
        {steiger2021psychological}
\bibfield{author}{\bibinfo{person}{Miriah Steiger}, \bibinfo{person}{Timir~J
  Bharucha}, \bibinfo{person}{Sukrit Venkatagiri}, \bibinfo{person}{Martin~J
  Riedl}, {and} \bibinfo{person}{Matthew Lease}.}
  \bibinfo{year}{2021}\natexlab{}.
\newblock \showarticletitle{The Psychological Well-Being of Content
  Moderators}.
\newblock  (\bibinfo{year}{2021}).
\newblock


\bibitem[\protect\citeauthoryear{Stencel}{Stencel}{2019}]%
        {poynter1}
\bibfield{author}{\bibinfo{person}{Mark Stencel}.}
  \bibinfo{year}{2019}\natexlab{}.
\newblock \bibinfo{title}{Number of fact-checking outlets surges to 188 in more
  than 60 countries}.
\newblock
\newblock
\urldef\tempurl%
\url{https://www.poynter.org/fact-checking/2019/number-of-fact-checking-outlets-surges-to-188-in-more-than-60-countries/}
\showURL{%
\tempurl}


\bibitem[\protect\citeauthoryear{Stisen, Verdezoto, Blunck, Kj{\ae}rgaard, and
  Gr{\o}nb{\ae}k}{Stisen et~al\mbox{.}}{2016}]%
        {stisen2016accounting}
\bibfield{author}{\bibinfo{person}{Allan Stisen}, \bibinfo{person}{Nervo
  Verdezoto}, \bibinfo{person}{Henrik Blunck}, \bibinfo{person}{Mikkel~Baun
  Kj{\ae}rgaard}, {and} \bibinfo{person}{Kaj Gr{\o}nb{\ae}k}.}
  \bibinfo{year}{2016}\natexlab{}.
\newblock \showarticletitle{Accounting for the invisible work of hospital
  orderlies: Designing for local and global coordination}. In
  \bibinfo{booktitle}{\emph{Proceedings of the 19th ACM Conference on
  Computer-Supported Cooperative Work \& Social Computing}}.
  \bibinfo{pages}{980--992}.
\newblock


\bibitem[\protect\citeauthoryear{Tang, Chen, Semaan, and Roberson}{Tang
  et~al\mbox{.}}{2015}]%
        {tang2015restructuring}
\bibfield{author}{\bibinfo{person}{Charlotte Tang}, \bibinfo{person}{Yunan
  Chen}, \bibinfo{person}{Bryan~C Semaan}, {and} \bibinfo{person}{Jahmeilah~A
  Roberson}.} \bibinfo{year}{2015}\natexlab{}.
\newblock \showarticletitle{Restructuring human infrastructure: The impact of
  EHR deployment in a volunteer-dependent clinic}. In
  \bibinfo{booktitle}{\emph{Proceedings of the 18th ACM Conference on Computer
  Supported Cooperative Work \& Social Computing}}. \bibinfo{pages}{649--661}.
\newblock


\bibitem[\protect\citeauthoryear{Teyssou, Leung, Apostolidis, Apostolidis,
  Papadopoulos, Zampoglou, Papadopoulou, and Mezaris}{Teyssou
  et~al\mbox{.}}{2017}]%
        {teyssou2017invid}
\bibfield{author}{\bibinfo{person}{Denis Teyssou}, \bibinfo{person}{Jean-Michel
  Leung}, \bibinfo{person}{Evlampios Apostolidis},
  \bibinfo{person}{Konstantinos Apostolidis}, \bibinfo{person}{Symeon
  Papadopoulos}, \bibinfo{person}{Markos Zampoglou}, \bibinfo{person}{Olga
  Papadopoulou}, {and} \bibinfo{person}{Vasileios Mezaris}.}
  \bibinfo{year}{2017}\natexlab{}.
\newblock \showarticletitle{The InVID plug-in: web video verification on the
  browser}. In \bibinfo{booktitle}{\emph{Proceedings of the first international
  workshop on multimedia verification}}. \bibinfo{pages}{23--30}.
\newblock


\bibitem[\protect\citeauthoryear{Thorne, Vlachos, Christodoulopoulos, and
  Mittal}{Thorne et~al\mbox{.}}{2018}]%
        {thorne2018fever}
\bibfield{author}{\bibinfo{person}{James Thorne}, \bibinfo{person}{Andreas
  Vlachos}, \bibinfo{person}{Christos Christodoulopoulos}, {and}
  \bibinfo{person}{Arpit Mittal}.} \bibinfo{year}{2018}\natexlab{}.
\newblock \showarticletitle{Fever: a large-scale dataset for fact extraction
  and verification}.
\newblock \bibinfo{journal}{\emph{arXiv preprint arXiv:1803.05355}}
  (\bibinfo{year}{2018}).
\newblock


\bibitem[\protect\citeauthoryear{Tolmie, Procter, Randall, Rouncefield, Burger,
  Wong Sak~Hoi, Zubiaga, and Liakata}{Tolmie et~al\mbox{.}}{2017}]%
        {tolmie2017supporting}
\bibfield{author}{\bibinfo{person}{Peter Tolmie}, \bibinfo{person}{Rob
  Procter}, \bibinfo{person}{David~William Randall}, \bibinfo{person}{Mark
  Rouncefield}, \bibinfo{person}{Christian Burger}, \bibinfo{person}{Geraldine
  Wong Sak~Hoi}, \bibinfo{person}{Arkaitz Zubiaga}, {and}
  \bibinfo{person}{Maria Liakata}.} \bibinfo{year}{2017}\natexlab{}.
\newblock \showarticletitle{Supporting the use of user generated content in
  journalistic practice}. In \bibinfo{booktitle}{\emph{Proceedings of the 2017
  chi conference on human factors in computing systems}}.
  \bibinfo{pages}{3632--3644}.
\newblock


\bibitem[\protect\citeauthoryear{Trello}{Trello}{2021}]%
        {Trello12:online}
\bibfield{author}{\bibinfo{person}{Trello}.} \bibinfo{year}{2021}\natexlab{}.
\newblock \bibinfo{title}{Trello}.
\newblock \bibinfo{howpublished}{\url{https://trello.com/en-US}}.
\newblock
\newblock
\shownote{(Accessed on 04/15/2021).}


\bibitem[\protect\citeauthoryear{Tuffley}{Tuffley}{2009}]%
        {tuffley2009mind}
\bibfield{author}{\bibinfo{person}{David Tuffley}.}
  \bibinfo{year}{2009}\natexlab{}.
\newblock \showarticletitle{Mind the gap}.
\newblock \bibinfo{journal}{\emph{International Journal of Sociotechnology and
  Knowledge Development (IJSKD)}} \bibinfo{volume}{1}, \bibinfo{number}{1}
  (\bibinfo{year}{2009}), \bibinfo{pages}{58--69}.
\newblock


\bibitem[\protect\citeauthoryear{Unruh and Pratt}{Unruh and Pratt}{2008}]%
        {unruh2008invisible}
\bibfield{author}{\bibinfo{person}{Kenton~T Unruh} {and} \bibinfo{person}{Wanda
  Pratt}.} \bibinfo{year}{2008}\natexlab{}.
\newblock \showarticletitle{The Invisible Work of Being a Patient and
  Implications for Health Care:“[the doctor is] my business partner in the
  most important business in my life, staying alive.”}. In
  \bibinfo{booktitle}{\emph{Ethnographic Praxis in Industry Conference
  Proceedings}}, Vol.~\bibinfo{volume}{2008}. Wiley Online Library,
  \bibinfo{pages}{40--50}.
\newblock


\bibitem[\protect\citeauthoryear{van~der Meulen and Reijnierse}{van~der Meulen
  and Reijnierse}{2020}]%
        {van2020factcorp}
\bibfield{author}{\bibinfo{person}{Marten van~der Meulen} {and}
  \bibinfo{person}{W~Gudrun Reijnierse}.} \bibinfo{year}{2020}\natexlab{}.
\newblock \showarticletitle{FactCorp: A Corpus of Dutch Fact-checks and its
  Multiple Usages}. In \bibinfo{booktitle}{\emph{Proceedings of The 12th
  Language Resources and Evaluation Conference}}. \bibinfo{pages}{1286--1292}.
\newblock


\bibitem[\protect\citeauthoryear{Verdezoto, Bagalkot, Akbar, Sharma,
  Mackintosh, Harrington, and Griffiths}{Verdezoto et~al\mbox{.}}{2021}]%
        {verdezoto2021invisible}
\bibfield{author}{\bibinfo{person}{Nervo Verdezoto}, \bibinfo{person}{Naveen
  Bagalkot}, \bibinfo{person}{Syeda~Zainab Akbar}, \bibinfo{person}{Swati
  Sharma}, \bibinfo{person}{Nicola Mackintosh}, \bibinfo{person}{Deirdre
  Harrington}, {and} \bibinfo{person}{Paula Griffiths}.}
  \bibinfo{year}{2021}\natexlab{}.
\newblock \showarticletitle{The Invisible Work of Maintenance in Community
  Health: Challenges and Opportunities for Digital Health to Support Frontline
  Health Workers in Karnataka, South India}.
\newblock \bibinfo{journal}{\emph{Proceedings of the ACM on Human-Computer
  Interaction}} \bibinfo{volume}{5}, \bibinfo{number}{CSCW1}
  (\bibinfo{year}{2021}), \bibinfo{pages}{1--31}.
\newblock


\bibitem[\protect\citeauthoryear{Vincent, Hecht, and Sen}{Vincent
  et~al\mbox{.}}{2019a}]%
        {vincent2019data}
\bibfield{author}{\bibinfo{person}{Nicholas Vincent}, \bibinfo{person}{Brent
  Hecht}, {and} \bibinfo{person}{Shilad Sen}.}
  \bibinfo{year}{2019}\natexlab{a}.
\newblock \showarticletitle{“Data Strikes”: Evaluating the Effectiveness of
  a New Form of Collective Action Against Technology Companies}. In
  \bibinfo{booktitle}{\emph{The World Wide Web Conference}}.
  \bibinfo{pages}{1931--1943}.
\newblock


\bibitem[\protect\citeauthoryear{Vincent, Johnson, Sheehan, and Hecht}{Vincent
  et~al\mbox{.}}{2019b}]%
        {vincent2019measuring}
\bibfield{author}{\bibinfo{person}{Nicholas Vincent}, \bibinfo{person}{Isaac
  Johnson}, \bibinfo{person}{Patrick Sheehan}, {and} \bibinfo{person}{Brent
  Hecht}.} \bibinfo{year}{2019}\natexlab{b}.
\newblock \showarticletitle{Measuring the importance of user-generated content
  to search engines}. In \bibinfo{booktitle}{\emph{Proceedings of the
  International AAAI Conference on Web and Social Media}},
  Vol.~\bibinfo{volume}{13}. \bibinfo{pages}{505--516}.
\newblock


\bibitem[\protect\citeauthoryear{Wang}{Wang}{2017}]%
        {wang2017liar}
\bibfield{author}{\bibinfo{person}{William~Yang Wang}.}
  \bibinfo{year}{2017}\natexlab{}.
\newblock \showarticletitle{" liar, liar pants on fire": A new benchmark
  dataset for fake news detection}.
\newblock \bibinfo{journal}{\emph{arXiv preprint arXiv:1705.00648}}
  (\bibinfo{year}{2017}).
\newblock


\bibitem[\protect\citeauthoryear{Wasson}{Wasson}{1998}]%
        {wasson1998identifying}
\bibfield{author}{\bibinfo{person}{Barbara Wasson}.}
  \bibinfo{year}{1998}\natexlab{}.
\newblock \showarticletitle{Identifying coordination agents for collaborative
  telelearning}.
\newblock \bibinfo{journal}{\emph{International Journal of Artificial
  Intelligence in Education (IJAIED)}}  \bibinfo{volume}{9}
  (\bibinfo{year}{1998}), \bibinfo{pages}{275--299}.
\newblock


\bibitem[\protect\citeauthoryear{Webster}{Webster}{2022}]%
        {Stringer71:online}
\bibfield{author}{\bibinfo{person}{Merriam Webster}.}
  \bibinfo{year}{2022}\natexlab{}.
\newblock \bibinfo{title}{Stringer Definition \& Meaning - Merriam-Webster}.
\newblock
  \bibinfo{howpublished}{\url{https://www.merriam-webster.com/dictionary/stringer}}.
\newblock
\newblock
\shownote{(Accessed on 04/17/2022).}


\bibitem[\protect\citeauthoryear{Whois.net}{Whois.net}{2021}]%
        {WhoisLoo11:online}
\bibfield{author}{\bibinfo{person}{Whois.net}.}
  \bibinfo{year}{2021}\natexlab{}.
\newblock \bibinfo{title}{Whois Lookup \& IP | Whois.net}.
\newblock \bibinfo{howpublished}{\url{https://www.whois.net/}}.
\newblock
\newblock
\shownote{(Accessed on 04/15/2021).}


\bibitem[\protect\citeauthoryear{Xin, Ma, Liu, Macke, Song, and
  Parameswaran}{Xin et~al\mbox{.}}{2018}]%
        {xin2018accelerating}
\bibfield{author}{\bibinfo{person}{Doris Xin}, \bibinfo{person}{Litian Ma},
  \bibinfo{person}{Jialin Liu}, \bibinfo{person}{Stephen Macke},
  \bibinfo{person}{Shuchen Song}, {and} \bibinfo{person}{Aditya Parameswaran}.}
  \bibinfo{year}{2018}\natexlab{}.
\newblock \showarticletitle{Accelerating human-in-the-loop machine learning:
  Challenges and opportunities}. In \bibinfo{booktitle}{\emph{Proceedings of
  the second workshop on data management for end-to-end machine learning}}.
  \bibinfo{pages}{1--4}.
\newblock


\bibitem[\protect\citeauthoryear{Young, Jamieson, Poulsen, and Goldring}{Young
  et~al\mbox{.}}{2018}]%
        {young2018fact}
\bibfield{author}{\bibinfo{person}{Dannagal~G Young},
  \bibinfo{person}{Kathleen~Hall Jamieson}, \bibinfo{person}{Shannon Poulsen},
  {and} \bibinfo{person}{Abigail Goldring}.} \bibinfo{year}{2018}\natexlab{}.
\newblock \showarticletitle{Fact-checking effectiveness as a function of format
  and tone: Evaluating FactCheck. org and FlackCheck. org}.
\newblock \bibinfo{journal}{\emph{Journalism \& Mass Communication Quarterly}}
  \bibinfo{volume}{95}, \bibinfo{number}{1} (\bibinfo{year}{2018}),
  \bibinfo{pages}{49--75}.
\newblock


\end{thebibliography}

\end{document}